\documentclass[fleqn,usenatbib]{mnras}

\usepackage{amsmath,amssymb}
\usepackage{graphicx}
\usepackage{algorithm}
\usepackage{algorithmic}
\usepackage{booktabs}
\usepackage{multirow}
\usepackage{xcolor}
\usepackage{hyperref}

\newcommand{\ramses}{\textsc{ramses}}
\newcommand{\curamses}{\textsc{cuRAMSES}}
\newcommand{\ncpu}{N_{\rm cpu}}
\newcommand{\nrank}{N_{\rm rank}}
\newcommand{\nthread}{N_{\rm thread}}
\newcommand{\nfile}{N_{\rm file}}
\newcommand{\ngridmax}{N_{\rm gridmax}}
\newcommand{\nlevelmax}{N_{\rm levelmax}}
\newcommand{\twotondim}{2^{N_{\rm dim}}}
\newcommand{\ksec}{\ensuremath{k}\text{-section}}
\newcommand{\Olog}[1]{\mathcal{O}(#1)}
\newcommand{\code}[1]{\texttt{#1}}
\newcommand{\mpi}[0]{\texttt{MPI}}
\newcommand{\mpiio}[0]{\texttt{MPI-IO}}
\newcommand{\dd}[0]{{domain decomposition}}

\newcommand{\rev}[1]{#1}

\title[cuRAMSES: Scalable AMR Optimizations for Cosmological Simulations]
{cuRAMSES: Scalable AMR Optimizations for Large-Scale Cosmological Simulations}

\author[J.~Kim]{
Juhan Kim$^{1}$\thanks{E-mail: kjhan@kias.re.kr}
\\
$^{1}$Center for Advanced Computation, Korea Institute for Advanced Study,
85 Hoegiro, Dongdaemun-gu, Seoul 02455, Republic of Korea
}

\date{Accepted XXX. Received YYY; in original form ZZZ}

\pubyear{2026}

\begin{document}

\label{firstpage}
\pagerange{\pageref{firstpage}--\pageref{lastpage}}

\maketitle

\begin{abstract}
We present \curamses, a suite of advanced domain decomposition strategies and 
algorithmic optimizations for the \ramses\ adaptive mesh refinement (AMR) code, 
designed to overcome the communication, memory, and solver bottlenecks inherent 
in massive cosmological simulations. 
The central innovation is a recursive \ksec\ domain decomposition that replaces the traditional Hilbert curve ordering with a hierarchical spatial partitioning. This approach substitutes global all-to-all communications with neighbour-only point-to-point communications. By maintaining a constant number of communication partners regardless of the total rank count, it significantly improves strong scaling at high concurrency. 
To address critical memory constraints at scale, we introduce a Morton-key hash 
table for octree-neighbour lookup alongside on-demand array allocation, drastically reducing the per-rank memory footprint. Furthermore, a novel spatial hash-binning algorithm in box-type local domains accelerates supernova and AGN feedback routines by over two orders of magnitude (a ${\sim}260\times$ speedup). 
For hybrid architectures, an automatic CPU/GPU dispatch model with GPU-resident
mesh data is implemented and benchmarked.  The multigrid Poisson solver achieves a $1.7\times$ GPU speedup on H100 and A100 GPUs, although the Godunov solver is currently PCIe-bandwidth-limited. The net improvement is ${\sim}20$\,per\,cent on current PCIe-connected hardware, and a performance model predicts ${\sim}2\times$ on tightly coupled architectures such as the NVIDIA GH200. Additionally, a variable-$\nrank$ restart capability enables flexible I/O workflows. 
Extensive diagnostics verify that all modifications preserve mass, momentum, and
energy conservation, matching the reference Hilbert-ordering run to within
0.5\,per\,cent in the total energy diagnostic.
\end{abstract}

\begin{keywords}
methods: numerical -- cosmology: simulations -- hydrodynamics --
software: development
\end{keywords}

\section{INTRODUCTION}
\label{sec:intro}

Cosmological hydrodynamic simulations play a central role in modern
astrophysics, connecting the predictions of the $\Lambda$CDM paradigm
to the observable properties of galaxies, the intergalactic medium,
and the large-scale structures of the Universe. Over the past decade,
landmark galaxy formation simulations such as IllustrisTNG
\citep{Pillepich2018,Nelson2019} with the moving-mesh code \textsc{arepo}
\citep{Springel2010}, EAGLE \citep{Schaye2015} with \textsc{gadget}
\citep{Springel2005}, and the Horizon-AGN/Horzion-Run 5/NewHorizon/NewCluster simulations
\citep{Dubois2014,Lee2021,Dubois2021,Han2026} with the adaptive mesh refinement (AMR) code
\ramses\ \citep{Teyssier2002}, among others
\citep{Hopkins2018,Dave2019,Tremmel2017}, have advanced our
understanding of cosmic structure.

These achievements, however, highlight the need for simulations of
even greater scale and dynamic range.  Capturing large-scale
statistical properties with percent-level precision
\citep{Kim2011,Kim2015,Ishiyama2021,Maksimova2021,Schaye2015,Libeskind2018}
requires volumes exceeding $(1\,h^{-1}\,\mathrm{Gpc})^3$ \citep{Klypin2018, Klypin2019}.
At the same time, resolving the physics that governs individual
galaxies including supermassive black hole (SMBH) feedback
\citep{Dubois2012,Weinberger2017}, the multi-phase interstellar medium
(ISM) \citep{Springel2003}, and star formation on sub-parsec scales
\citep{Hopkins2014}, demands deep AMR hierarchies with dynamic ranges of
$10^5$ or more.
Achieving both goals simultaneously, large volume \emph{and} high
resolution requires an unprecedented computational effort that
approaches and will ultimately exploit exascale facilities
\citep{Habib2016,Frontiere2025}.
The principal barriers to reaching this regime are not only physical
modelling limitations but also \emph{algorithmic} challenges including
communication overhead that grows with the number of \mpi\ ranks, memory
consumption that limits the refinement depth per node, and solver
inefficiencies that dominate the wallclock time.

Among the numerical approaches employed by these simulations, AMR
codes such as \ramses\ and Enzo \citep{Bryan2014} provide a
particularly attractive framework, in which the
computational mesh is refined only where the physics demands it,
concentrating resources on collapsing haloes and star-forming regions
while keeping the cost of smooth low-density regions manageable.

A key design choice in parallel AMR codes is the domain decomposition
strategy for distributed-memory systems. Space-filling curves (SFCs), particularly the Hilbert
(Peano--Hilbert) curve, have widely and successfully served this purpose
\citep{Warren1993,Springel2005}. The Hilbert curve maps the
three-dimensional computational domain to a one-dimensional index
preserving spatial locality so that cells close in physical space
remain close along the curve.  For equal-volume partitioning, each
rank's subdomain typically borders 6--8 neighbours in three dimensions.
This enables a simple and effective
partitioning where the one-dimensional index range is divided into
$\nrank$ contiguous segments, each being assigned to an \mpi\ rank. The resulting
decomposition naturally produces compact subdomains with relatively
small surface-to-volume ratios, minimising the ghost-zone boundary
between neighbouring ranks.

However, scaling this approach to the regime of $10^{10}$--$10^{11}$
particles and $10^4$--$10^5$ \mpi\ ranks reveals fundamental
limitations. The one-dimensional nature of the SFC means that
\emph{every} rank may, in principle, border \emph{any other} rank in worst cases,
forcing communication patterns that scale poorly with $\nrank$. In the
standard \ramses\ implementation, ghost-zone exchange, grid and particle migration, and sink particle
synchronisation all rely on \code{MPI\_ALLTOALL} to communicate
counts among all ranks, leading to $\Olog{\nrank^2}$ communication
complexity and $\Olog{\nrank}$ per-rank buffer memory, which becomes a severe
bottleneck when $\nrank$ exceeds ${\sim}10^3$
\citep{Teyssier2002}. Furthermore, the Hilbert load balancer assigns each
cell a cost of $(80 + N_{\rm part})$, where $N_{\rm part}$ is the
particle count per oct.  While this does account for particles, the
implied 80 times more weight to grid with respect to particle, severely underweights
particle-dominated cells. For example, a grid hosting $10^4$ particles in a dense
halo carries a cost only ${\sim}125\times$ that of an empty cell,
whereas its actual memory footprint is about 450 times larger
(taking $w_{\rm grid}=2256$\,B (Bytes) for \code{nvar}=14 and $w_{\rm part}=12$\,B per particle, see
Section~\ref{sec:ksec_membal}).  The result is that particle-rich
ranks are systematically overloaded in both memory and compute time
\citep{Springel2005}. The problem is particularly acute in
cosmological zoom-in simulations \citep{Lee2021,Dubois2021}, where the
high-resolution region occupies a small fraction of the total volume.
The Hilbert curve concentrates nearly all refinement on a few ranks
while the remaining ranks are left with low-resolution void cells,
leading to severe load imbalance that worsens with increasing zoom
factor.  In the Horizon Run~5 simulation \citep{Lee2021}, for example,
the Hilbert-based load rebalancing required more than 10 hours per
invocation below redshift $z \approx 2$, and had to be abandoned entirely from the redshift.

Beyond communication and load balancing, per-rank memory overhead
(e.g.\ the \code{nbor} and Hilbert key arrays, $>$1\,GB for
$\ngridmax = 5\,\mathrm{M}$, see Section~\ref{sec:morton}) limits the
refinement depth.  The multigrid Poisson solver \citep{Guillet2011}
dominates runtime (${\sim}50$\,per\,cent), and \ramses' one-file-per-rank
I/O prevents restarting with a different rank count.

The original \ramses\ code relies exclusively on \mpi\ for
parallelism, assigning one \mpi\ rank per processor core. To exploit the
shared-memory bandwidth of modern multi-core nodes, hybrid \mpi+OpenMP
implementations have been developed. The Horizon Run~5 simulation
\citep{Lee2021} employs OMP-RAMSES, and the NewCluster zoom-in
simulation \citep{Han2026} employs RAMSES-yOMP, both adding OpenMP
threading within each \mpi\ rank for improved intra-node scalability. However,
\mpi+OpenMP alone still leaves the GPU compute capability of modern
heterogeneous nodes untapped.  As current and upcoming supercomputers
derive an increasing fraction of their floating-point throughput from
GPU accelerators, a three-level (\mpi+OpenMP+CUDA) parallelism model
becomes desirable, although the achievable speedup depends critically
on the CPU--GPU interconnect bandwidth, as we quantify in
Section~\ref{sec:hybrid}.

In this paper we describe \curamses, a comprehensive set of
modifications to \ramses\ that addresses each of these challenges. We
introduce a recursive \ksec\ domain decomposition
(Section~\ref{sec:ksection}), adaptive \mpi\ exchange backends
(Section~\ref{sec:ksec_exchange}), a Morton key hash table
(Section~\ref{sec:morton}), multigrid and FFTW3 Poisson solver
optimizations (Section~\ref{sec:multigrid}), spatial hash binning for
feedback (Section~\ref{sec:feedback}), variable-$\nrank$ restart
(Section~\ref{sec:varcpu}), and GPU-accelerated solvers
(Section~\ref{sec:hybrid}).  Performance benchmarks are presented in
Section~\ref{sec:performance}, and we conclude in
Section~\ref{sec:conclusions}.

Throughout this paper, we use the notation of \citet{Teyssier2002}, where
$\nlevelmax$ is the maximum AMR level, $\ngridmax$ is the maximum
number of grids per rank, and
$\nrank$ is the number of \mpi\ ranks,
$\nthread$ is the number of OpenMP threads per rank, and
$\ncpu = \nrank \times \nthread$ is the total number of CPU cores.

\section{RECURSIVE K-SECTION DOMAIN DECOMPOSITION}
\label{sec:ksection}

We replace the Hilbert ordering with a recursive $k$-ary spatial
partitioning that not only eliminates global \mpi\ collectives but also encodes the
communication pattern directly in the tree structure.

\subsection{Hierarchical Partitioning of Spatial and Communication Domain}
\label{sec:ksec_tree}

Given $\nrank$ \mpi\ ranks, we first compute the prime factorization as
\begin{equation}
\nrank = p^{m_1}  q^{m_2}  \cdots  r^{m_r},
\quad p > q > \cdots > r,
\label{eq:prime_factorization}
\end{equation}
where one may note the order of factors.
The splitting sequence is then,
\begin{equation}
\mathbf{k} = (\underbrace{p, \ldots, p}_{m_1},
              \underbrace{q, \ldots, q}_{m_2}, \ldots,
              \underbrace{r, \ldots, r}_{m_r}),
\label{eq:split_sequence}
\end{equation}
which yields $L\,(= \sum_i m_i)$ levels in the tree structure encoding both
the domain hierarchy and the communication pattern.
The tree is a $k$-ary structure where each \emph{node} represents a
contiguous group of \mpi\ ranks sharing a spatial sub-domain. The
\emph{root} node at level 0 spans all $\nrank$ ranks and each
\emph{leaf} node at level~$L$ corresponds to a single rank.
At each level $\ell$, the
domain of a node is split into $k_\ell$ child nodes along the longest
axis of the current bounding box. This longest-axis selection ensures
roughly isotropic sub-domains minimising the surface-to-volume ratio
and hence the ghost-zone count.

\begin{figure}
\centering
\includegraphics[width=\columnwidth]{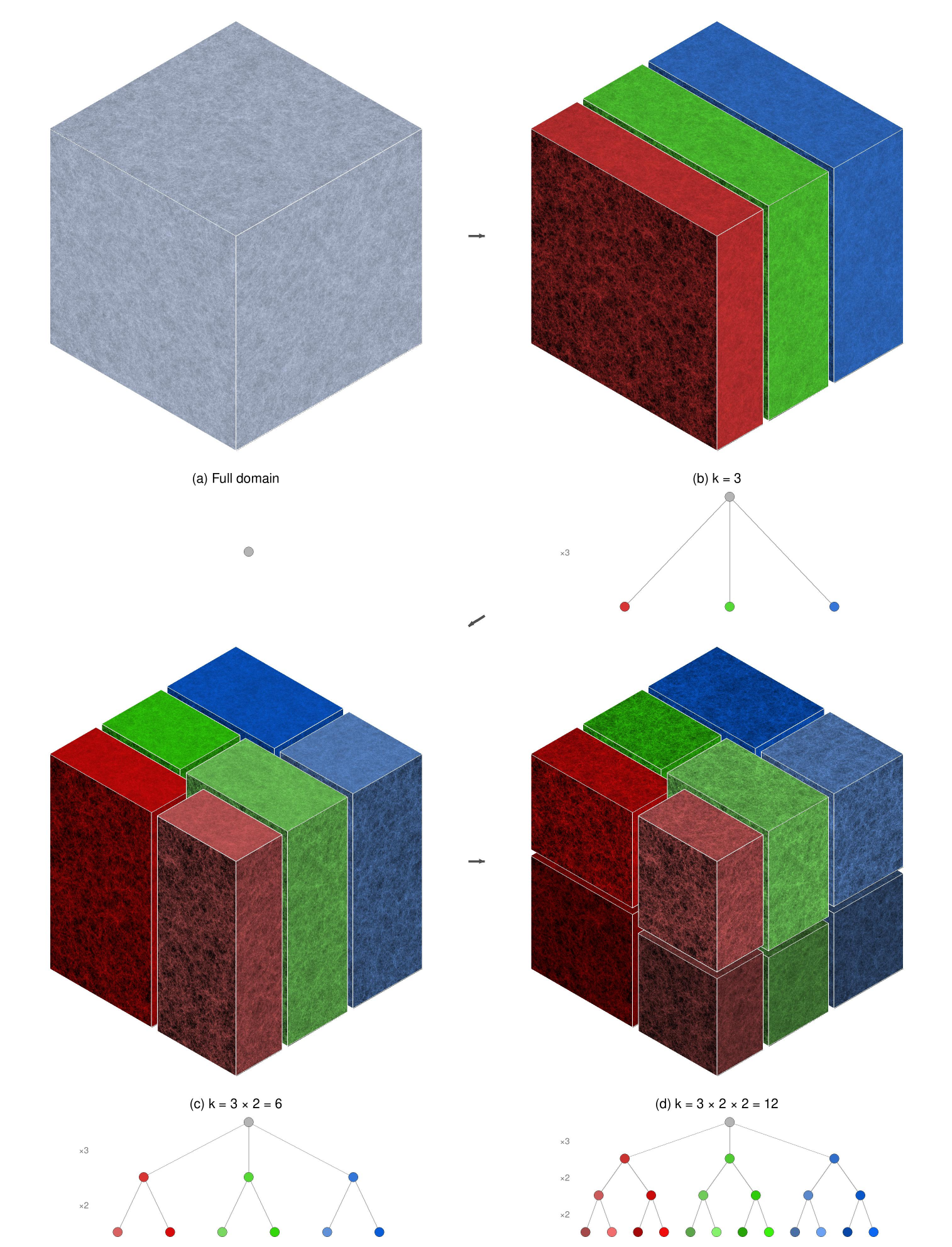}
\caption{Progressive recursive \ksec\ decomposition for $\nrank = 12 =
3 \times 2 \times 2$. (a)~The full and undivided simulation domain. (b)~First
split into $k = 3$ slabs along the longest axis. (c)~Each slab bisected
along the second axis ($3 \times 2 = 6$ sub-domains). (d)~Final bisection
along the third axis ($3 \times 2 \times 2 = 12$ leaf domains). Below
each panel, the corresponding \ksec\ tree is shown. Sub-domain volumes
vary by approximately 10--30\,per\,cent, reflecting load-balanced wall placement.}
\label{fig:ksec_progressive}
\end{figure}

For example, $\nrank = 12 = 3 \cdot 2 \cdot 2$ produces the
splitting sequence $(3, 2, 2)$ with $L = 3$ tree levels, where the root is
split into 3 slabs along the longest axis, each slab is bisected, and
each half is bisected again, yielding 12 leaf nodes, one per rank.
Fig.~\ref{fig:ksec_progressive} illustrates this progressive
decomposition for $\nrank = 12$, from the undivided domain (the upper left panel) through
three successive levels of splitting with the corresponding
\ksec\ tree shown beneath each panel.

The tree is stored as a set of arrays indexed by node identifier.
For each internal node, the child indices for each of the $k_\ell$
partitions and the spatial coordinates of the partition boundaries
are recorded. And for each leaf node, the assigned \mpi\ rank is stored.
Every node also carries the minimum and maximum rank indices of all
leaves in its subtree, enabling rapid range queries during the
hierarchical exchange.

The total number of tree nodes,
$N_{\rm nodes} = 1 + \sum_{\ell=1}^{L} \prod_{i=1}^{\ell} k_i$,
is at most a few hundreds for practical values of $\nrank$
($\nrank \leq 10^5$), representing negligible overhead.

\subsection{Load-Balanced Wall Placement}
\label{sec:ksec_loadbal}

When the tree is updated during load balancing (at every \code{nremap}
coarse steps or by constraints on the acceptable inhomogeneity), the wall positions at a certain level are adjusted by iterative
binary search so that each partition receives a load proportional to the
number of ranks it contains. The procedure operates level by level and top to bottom. At level $\ell$,
the following steps are performed.

\begin{enumerate}
\item A histogram is built over the splitting coordinate. For each
  node at level $\ell$, the cells within that node are projected onto the
  splitting axis and binned into a cumulative cost histogram with
  resolution $\Delta x_{\rm hist}$.

\item For each of the $k_\ell - 1$ walls within each node, a binary
  search adjusts the wall position until the cumulative
  load on the left side matches the target fraction. The target
  cumulative fraction for wall $j$ in a node spanning ranks
  $[i_{\rm min}, i_{\rm max}]$ with total count
  $n = i_{\rm max} - i_{\rm min} + 1$ is
  \begin{equation}
  f_j = n^{-1}\sum_{m=1}^{j} n_m,
  \label{eq:target_frac}
  \end{equation}
  where $n_m$ is the number of ranks assigned to partition $m$.

\item An \code{MPI\_ALLREDUCE} aggregates the local histograms across
  all $\nrank$ ranks to obtain the global cumulative load at each wall
  position. The binary search converges when the relative load imbalance
 falls below a
  tolerance $\epsilon_{\rm tol}$ (typically 5\,per\,cent), or when the
  wall position can no longer be resolved at the histogram resolution.

\item After wall convergence, the cells are repartitioned (sorted)
  according to the new wall positions and the histogram bounds are
  updated for the next level.
\end{enumerate}

\subsection{Memory-Weighted Cost Function}
\label{sec:ksec_membal}

The default \ramses\ load balancer uses a cost of $80 + N_{\rm part}$
per oct grid, which underweights particle-dominated cells by a factor of
${\sim}3$--$4$ relative to their true memory footprint.
\curamses\ replaces this with a memory-weighted cost function
\begin{equation}
C_{\rm cell} = {{w_{\rm grid} +
n_{\rm part}({\code{igrid}}) \cdot w_{\rm part} +
n_{\rm sink}({\code{igrid}}) \cdot w_{\rm sink}
\label{eq:cost_function}} \over 2^{N_{\rm dim}}},
\end{equation}
where $w_{\rm grid}$ is the memory cost per grid slot, computed
automatically as 
\begin{equation}
	w_{\rm grid} = 2^{N_{\rm dim}} \times (2\,\code{nvar}\times 8 + 52) + 48
\end{equation}
in bytes, for example,  1232 B(ytes) for \code{nvar}=6 or 2256\,B for \code{nvar}=14,
accounting for hydro (\code{uold}+\code{unew}), gravity, and AMR bookkeeping arrays.
A user-specified value may override the automatic formula. And
$w_{\rm part}$ is the memory cost per particle slot (default 12 bytes
for position, velocity, mass, and linked-list pointers),
$n_{\rm part}({\code{igrid}})$ is the number of dark matter and star
particles (both sharing the same memory layout) attached to grid,
\code{igrid}, and $w_{\rm sink}$ is a computational weight per sink
particle (default 500) that accounts for the expensive feedback and
accretion operations associated with each sink. The sink count
$n_{\rm sink}({\code{igrid}})$ is obtained by descending the AMR tree
from each sink position to its leaf cell.
The division by $\twotondim$ distributes the grid
cost evenly among its eight cells.

This cost function ensures that ranks hosting dense haloes (many
particles per cell) receive fewer cells, preventing memory exhaustion
on particle-heavy ranks.
Our Cosmo256 test (for details, see Section~\ref{sec:perf_config}, \rev{$256^3 \approx 16.78\,\mathrm{M}$} particles, 12 ranks) shows that
memory-weighted balancing reduces the peak-to-mean memory ratio from
2.5 to 1.3, and keeps the per-rank memory imbalance
$M_{\max}/M_{\min} \le 1.05$ across 2--64 ranks
(Section~\ref{sec:perf_membal}), without affecting physics results such as identical numbers of new stars, sinks,
$e_{\rm cons}$ (fractional total energy conservation error),
$e_{\rm pot}$ (gravitational potential energy), and
$e_{\rm kin}$ (kinetic energy)
to within the expected level after advancing ten coarse time steps.

In addition to pure memory-based balancing, \curamses\ supports a
\emph{hybrid} mode that incorporates computation-time feedback.  Each
rank measures the wall-clock time spent in the main physics loop
(hydrodynamics, cooling, feedback, and refinement flagging) at every
AMR level, and a global average cost-per-cell $\bar{t}$ is computed
via \code{MPI\_ALLREDUCE}.  A blending parameter $\alpha_t$
(\code{time\_balance\_alpha}, default 0) controls the correction:
\begin{equation}
C_{\rm hybrid} = C_{\rm cell} \times
\bigl[1 + \alpha_t\,(t_{\rm local}/\bar{t} - 1)\bigr],
\label{eq:hybrid_cost}
\end{equation}
where $t_{\rm local}$ is the local cost-per-cell and the correction
factor is clamped to $[0.5,\,2.0]$ to prevent runaway rebalancing.
When $\alpha_t = 0$ the scheme reduces to pure memory-based balancing;
setting $\alpha_t = 0.3$--$0.5$ allows the balancer to shift load away
from computationally expensive ranks (e.g.\ those hosting active AGN
or dense star-forming regions) while still respecting memory
constraints.
In extreme environments such as the central regions of massive galaxy
clusters, where thousands of sink particles and deep AMR hierarchies
concentrate on a few ranks, pure memory balancing can leave a
residual \emph{compute} imbalance because per-cell work varies by
orders of magnitude between quiescent and actively star-forming
or feedback-dominated cells.  The hybrid mode with
$\alpha_t \sim 0.3$--$0.5$ partially compensates for this effect
by up-weighting slow ranks during the next rebalancing cycle, though
the correction is limited to the EMA-smoothed cost ratio (clamped to
$[0.5,\,2]$).  Fully resolving such extreme imbalances may require
task-based or sub-rank work-stealing strategies that are beyond the
scope of the current implementation.

\section{AUTO-TUNING MPI COMMUNICATIONS}
\label{sec:ksec_exchange}

Ghost-zone (virtual boundary) exchange is the most
communication-intensive operation in \ramses, invoked multiple times
per fine time step for hydro variables, gravitational potential,
particle flags, and density contributions. Each exchange involves
the same algorithmic pattern (pack data from local grids, transfer to
remote ranks, and unpack into ghost zones), but the underlying \mpi\
transport can be realised in fundamentally different ways.
\curamses\ implements two additional methods and adaptively selects one of the three communication backends (MPI\_ALLTOALL/P2P/K-Section) at runtime.

\subsection{Exchange Methods}
\label{sec:exchange_backends}

\subsubsection{MPI\_ALLTOALLV}
\label{sec:backend_alltoall}

The \code{MPI\_ALLTOALLV} collective exchanges variable-length
messages among all ranks in a single call. This backend packs all
emission data (data to be sent to other ranks) into a contiguous buffer with displacements computed
from per-rank send counts, invokes \code{MPI\_ALLTOALLV}, and unpacks
the received data. Although it involves all $\nrank$ ranks (with zero
counts for non-neighbours), ``high-quality'' \mpi\ implementations exploit
the sparsity internally and can outperform explicit P2P on certain
interconnects, particularly for small messages on fat-tree topologies.
However, the implicit global synchronisation and $\Olog{\nrank}$ memory
overhead make it less attractive at very large rank counts.

\subsubsection{Point-to-point (P2P)}
\label{sec:backend_p2p}

The P2P backend follows the original \ramses\ pattern of non-blocking
sends/receives (\code{MPI\_ISEND}/\code{MPI\_IRECV}) with
\code{MPI\_WAITALL}.  The number of communication partners is set by
the domain geometry (typically 6--8), independent of $\nrank$, making
P2P the best choice when neighbour sets are sparse.

\subsubsection{Hierarchical K-Section Exchange}
\label{sec:backend_ksection}

The \ksec\ backend replaces the global communication pattern with a
hierarchical tree walk using the domain decomposition (\dd) tree described
in Section~\ref{sec:ksec_tree}.  The algorithm walks the tree from
root to leaf, where each node represents a contiguous group of \mpi\
ranks.  At each level~$\ell$, a node is
partitioned into $k_\ell$ child nodes and each rank communicates with at
most $k_\ell - 1$ correspondents (one per sibling subtree).  This spreads the communication load evenly across the
sibling rather than funnelling all traffic to a single rank.

\begin{figure}
\centering
\includegraphics[width=\columnwidth]{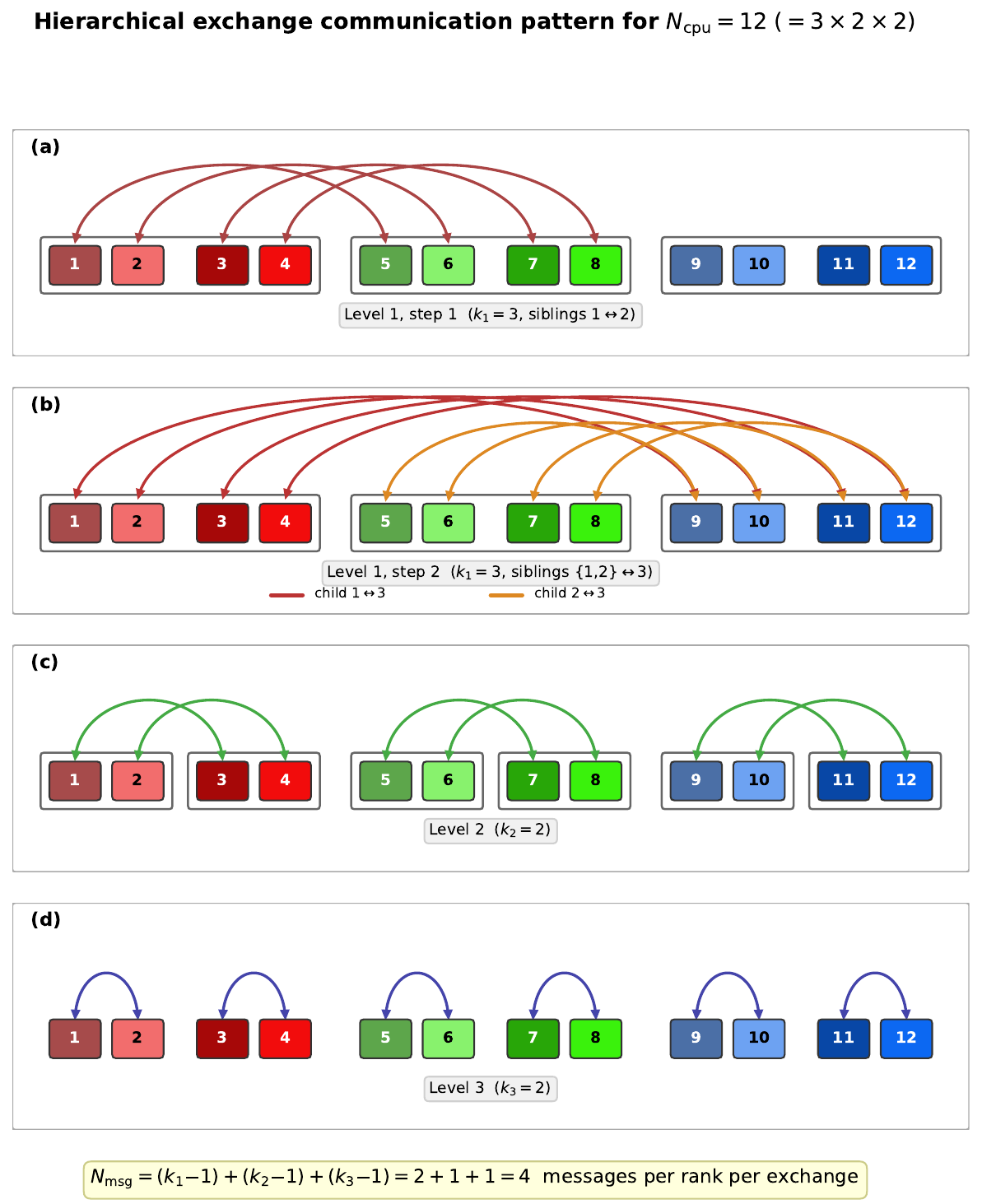}
\caption{Hierarchical exchange communication pattern for $\nrank = 12
\,(= 3 \times 2 \times 2)$. Coloured rectangles represent \mpi\ ranks
numbered 1--12, using the same colour scheme as
Fig.~\ref{fig:ksec_progressive}. Arcs denote bidirectional
point-to-point exchanges.  At level~1 ($k_1 = 3$), two steps connect
each rank with correspondents in the two sibling subtrees, and the
two colours in step~2 distinguish the children~1$\leftrightarrow$3
(dark red) and children~2$\leftrightarrow$3 (orange) pairings that
proceed concurrently.  At levels~2 and~3 ($k_2 = k_3 = 2$), the
communication range contracts to within each group and then to
adjacent pairs.  The total communication count per rank is
$N_{\rm com} = (3{-}1) + (2{-}1) + (2{-}1) = 4$.}
\label{fig:hier_exchange}
\end{figure}

The total number of communications per rank per exchange is
\begin{equation}
N_{\rm com} = \sum_{\ell=1}^{L} (k_\ell - 1) = \sum_i m_i (p_i - 1),
\label{eq:msg_count}
\end{equation}
which for $\nrank = 1024 = 2^{10}$ gives $N_{\rm com} = 10$, which is two
orders of magnitude fewer than the ${\sim}1024$ communications in an
all-to-all pattern. Even in the best case, where the \mpi\ library
internally optimises \code{MPI\_ALLTOALLV} to skip zero-length
communications, a rank with six neighbours must still perform at least six
pairwise transfers (and often more on non-ideal interconnects), so the
\ksec\ communication count of $\Olog{\sum_\ell k_\ell}$ remains competitive.
Crucially, $N_{\rm com}$ depends only on the prime factorization of
$\nrank$ and is completely independent of the degree of particle
clustering or the AMR refinement level, so the communication topology remains fixed regardless of how complex the simulation becomes. Fig.~\ref{fig:hier_exchange} illustrates the communication pattern
for $\nrank = 12$.  Each rank sends $N_{\rm com} = (3{-}1) + (2{-}1) +
(2{-}1) = 4$ messages per exchange, independent of $\nrank$.

The pairing structure at each tree level follows directly from the
branching factor. When a node has $p$ children, data from every child
must reach every other child, accomplished in $p - 1$ sequential
steps. In step~$s$ ($s = 1, \ldots, p-1$), child~$s+1$ exchanges
with each of the children $1, \ldots, s$, yielding $s$ concurrent
point-to-point communications. After all $p - 1$ steps, every child
possesses the complete data set of the entire subtree.

A trade-off is that the total data volume per rank exceeds that of
direct P2P or \code{MPI\_ALLTOALLV} since items may be forwarded
through multiple tree levels. In the worst case, a rank relays its
entire buffer at every level, giving at most $L \times V_{\rm local}$, where $V_{\rm local}$ is the local data volume.
In practice, items are filtered by child index at each level and
successive levels operate on progressively smaller subsets. The key
advantage, namely reducing the number of communication partners from
$\Olog{\nrank}$ to $\Olog{\sum_\ell k_\ell}$, more than compensates for the
modest volume increase particularly at large $\nrank$, where the startup latency in the network hand-shaking dominates.

Table~\ref{tab:complexity} compares the communication complexity of
the three backends. The \ksec\ exchange also extends to
ghost-zone operations (forward, reverse, integer, and bulk variants),
communication structure construction (replacing the original
\code{MPI\_ALLTOALL}-based build\_comm), and feedback collectives,
eliminating all global communication patterns from the AMR
infrastructure.

\begin{table}
\centering
\caption{Communication complexity per exchange operation.
$N_{\rm nb}$ is the number of neighbour ranks,
$N_{\rm gh}$ the number of ghost grids per rank,
$k_l$ the branching factor at tree level $l$, and
$k_{\rm max} = \max(k_l)$.}
\label{tab:complexity}
\begin{tabular}{lccc}
\toprule
& P2P & \code{ALLTOALLV} & \ksec \\
\midrule
Msgs   & $\Olog{N_{\rm nb}}$          & $\Olog{\nrank}$                        & $\Olog{\sum_\ell k_\ell}$ \\
Buffer & $\Olog{N_{\rm nb} N_{\rm gh}}$ & $\Olog{\nrank N_{\rm gh}}$           & $\Olog{k_{\rm max} N_{\rm gh}}$ \\
Sync.  & none                    & barrier                      & none \\
\bottomrule
\end{tabular}
\end{table}

\subsection{Adaptive Method Selection}
\label{sec:autotune}

No single backend dominates in performance across all exchange types, rank counts,
and simulation stages. During the initial mesh setup and I/O phases,
communication is sparse and P2P excels. After load rebalancing, the
neighbour topology changes and a previously inferior method may become
optimal. As the simulation evolves and AMR levels deepen, ghost-zone
volumes grow and the relative cost of each backend shifts.
\curamses\ therefore implements a continuous per-component auto-tuning
framework that adapts to the evolving communication pattern throughout
the simulation.

\subsubsection{Identification of exchange components}
\label{sec:autotune_components}

Seven independent exchange components are identified, each with its
own data size and call frequency.
(i) \code{fine\_dp}, forward double-precision ghost exchange (hydro,
potential).
(ii) \code{fine\_int}, forward integer ghost exchange (\code{cpu\_map},
flags).
(iii) \code{reverse\_dp}, reverse accumulation (density).
(iv) \code{reverse\_int}, reverse integer accumulation.
(v) \code{pair\_int}, paired integer exchange.
(vi) \code{bulk\_dp}, bulk forward exchange ($N_{\rm var}$ variables
in one call).
(vii) \code{bulk\_rev\_dp}, bulk reverse accumulation.
Each component maintains its own auto-tune state, so a method that is
optimal for small integer exchanges need not be the same as for large
bulk hydro transfers.

\subsubsection{Auto-tune protocol}
\label{sec:autotune_protocol}

Each component independently progresses through four phases. (0)~trial
P2P, (1)~trial \ksec, (2)~production with the faster backend, and
(3)~periodic probing of the alternative.  Phase~2 tracks performance
via an exponential moving average (EMA, $\alpha = 0.05$) as
\begin{equation}
	\bar{t}_{n+1} = \alpha t_n + (1-\alpha) \bar{t}_n,
\end{equation} 
where $\alpha$ is the update rate, $t_n$ is the elapsed time of the current call and $\bar{t}_n$ is
the running EMA from the previous step.
When the
probe in Phase~3 finds the alternative faster by $>$20\,per\,cent, the
component switches.

\section{MORTON KEY HASH TABLE AND MEMORY OPTIMIZATIONS}
\label{sec:morton}

\subsection{The Neighbour Array (\texttt{nbor}) Problem}
\label{sec:morton_problem}

\ramses\ stores the octree connectivity in several arrays, the largest
of which is \code{nbor(1:ngridmax, 1:\rev{$2\,n_{\rm dim}$})}, a
six-column integer array (in 3D) that records, for each grid, the
cell index of its neighbour in each of the six Cartesian directions
($\pm x$, $\pm y$, $\pm z$). Each entry is a default-kind 32-bit integer
occupying \rev{4}\,bytes, so this array consumes \rev{$24\,\ngridmax$
bytes (120\,MB for $\ngridmax = 5\,\mathrm{M}$)}.
\rev{On a per-cell basis (each oct contains $2^{n_{\rm dim}}\!=\!8$ cells in three dimensions), this corresponds to 3 bytes per cell, representing approximately $2.9\,\%$ of the primary hydrodynamics and gravity state payload of roughly 104 bytes per cell ($\mathrm{NVAR}\!\times\!8 + 5\!\times\!8$ bytes per cell with $\mathrm{NVAR}=8$ hydrodynamic variables stored as double precision in \texttt{uold}, plus the five gravity variables $\rho$, $\phi$, $f_x$, $f_y$, $f_z$). A comprehensive per-array profiling which includes the Hilbert key and the load balancing scratch arrays is provided in Appendix ~\ref{app:memory} (Table~\ref{tab:mem_breakdown}). All values in that table were verified by direct \texttt{c\_sizeof} measurement under the production compile flags.}
Moreover, the \code{nbor} array must be maintained during grid creation,
deletion, defragmentation, and inter-rank migration, which is a significant
source of code complexity and potential bugs.

\subsection{Morton Key Encoding and Neighbour Arithmetic}
\label{sec:morton_encoding}

A Morton key (also known as a Z-order key) \citep{Morton1966,Samet2006} is an integer formed
by interleaving the bits of the three-dimensional integer coordinates
$(i, j, k)$ of a grid at its AMR level as
\begin{equation}
M(i, j, k) = \sum_{b=0}^{B-1}
\Bigl[
  \text{bit}_b(i) \cdot 2^{3b} +
  \text{bit}_b(j) \cdot 2^{3b+1} +
  \text{bit}_b(k) \cdot 2^{3b+2}
\Bigr],
\label{eq:morton_encode}
\end{equation}
where $B$ is the number of bits per coordinate and $\text{bit}_b(n)$
extracts bit $b$ of integer $n$.
Two key widths are supported via a compile-time flag:
a 64-bit key with $B = 21$ (default, compatible with the Intel \texttt{ifx}
compiler), and a 128-bit key with $B = 42$.
At AMR level $\ell$, the integer coordinate range is $\texttt{[}0,\, 
2^{\ell-1}n_x\texttt{)}$, where $n_x$ is the number of root-level cells per dimension.
With $B = 21$ the maximum allowed level is 22 ($n_x = 1$) or 20 ($n_x = 4$) while
$B = 42$ extends these to 43 and 41, respectively.

The integer coordinates of a grid at level $\ell$ are computed from its
floating-point centre position $\mathbf{r}_g$ as
\begin{equation}
i_d = \texttt{iFloor}( 2^{\ell-1}r_{g,d} ), \quad d \in \{x, y, z\},
\label{eq:grid_to_int}
\end{equation}
where \texttt{iFloor} is the integer floor function and coordinates are in units of the coarse grid spacing.
The integer coordinates use zero-based (C-style) indexing,
starting from $i_d = 0$ even though Fortran arrays are
conventionally one-based.
Note that AMR does not populate all possible grid positions at a given
level since only regions that satisfy the refinement criteria contain grids.
The Morton key therefore serves as a unique spatial address for each
\emph{existing} grid. The hash table (Appendix~\ref{app:hashtable})
stores only the grids that are actually allocated making the
look-up cost independent of the total number of potential grid
positions at that level.

The neighbour key in direction $j$ is obtained by decoding, shifting
the appropriate coordinate with periodic wrapping, and re-encoding
(Appendix~\ref{app:morton}).  Parent and child keys follow from
3-bit shifts as $M_{\rm parent} = M \text{\texttt{>>}} 3$, $M_{\rm child} =
(M \text{\texttt{<<}} 3) \,|\, i_{\rm child}$.

The Morton keys are stored in a per-level open-addressing hash table
that maps keys to grid indices, providing $\Olog{1}$ expected
neighbour look-up (Appendix~\ref{app:hashtable}).
Each hash table entry stores a 16-byte Morton key and a 4-byte grid
index (20 bytes total).  With a maximum filling factor of 1.4 (i.e.\ 70\,per\,cent occupancy), the
aggregate memory footprint across all levels is approximately
$1.4 \times 20 \times N_{\rm grids} = 28\,N_{\rm grids}$ bytes,
where $N_{\rm grids}$ is the number of allocated grids.
For a typical run with $N_{\rm grids} \approx 3\,\mathrm{M}$, this
amounts to roughly 83\,MB, comparable to the original
\code{nbor} array \rev{($24\,\ngridmax \approx 120$\,MB for
$\ngridmax = 5\,\mathrm{M}$)} but with the additional
$\Olog{1}$ neighbour look-up benefit.

\rev{Combined with the on-demand allocation of the Hilbert key and the
load-balancing scratch arrays (Appendix~\ref{app:memory}), the total
peak memory saved at $\ngridmax = 5\,\mathrm{M}$ approaches 1\,GB per
rank under \code{QUADHILBERT}.}

\section{POISSON SOLVER OPTIMIZATIONS}
\label{sec:multigrid}

Solving the Poisson equation $\nabla^2\phi = 4\pi G\rho$ for
self-gravity is one of the most expensive operations in cosmological
AMR simulations.  In baseline \ramses, performance profiling reveals that the
Poisson solver accounts for approximately half of the total
wallclock time per coarse step. And this fraction grows further as the
AMR structure develops and deeper levels are populated.  We describe two complementary
strategies, optimizing the iterative multigrid solver that operates on
all AMR levels (Section~\ref{sec:mg_method}), and replacing it with a
direct FFT solve on the uniform base level
(Section~\ref{sec:fftw3}).  Together these reduce the Poisson
fraction to approximately 40\,per\,cent.

\subsection{Multi-Grid Method}
\label{sec:mg_method}

\ramses\ employs a multigrid V-cycle
\citep{Brandt1977,Briggs2000,Trottenberg2001} with red-black
Gauss--Seidel (RBGS) smoothing \citep{Wesseling1992,Adams2001}
at each AMR level.

\subsubsection{V-Cycle Algorithm and Exchange Optimization}
\label{sec:mg_vcycle}

\begin{figure}
\centering
\includegraphics[width=\columnwidth]{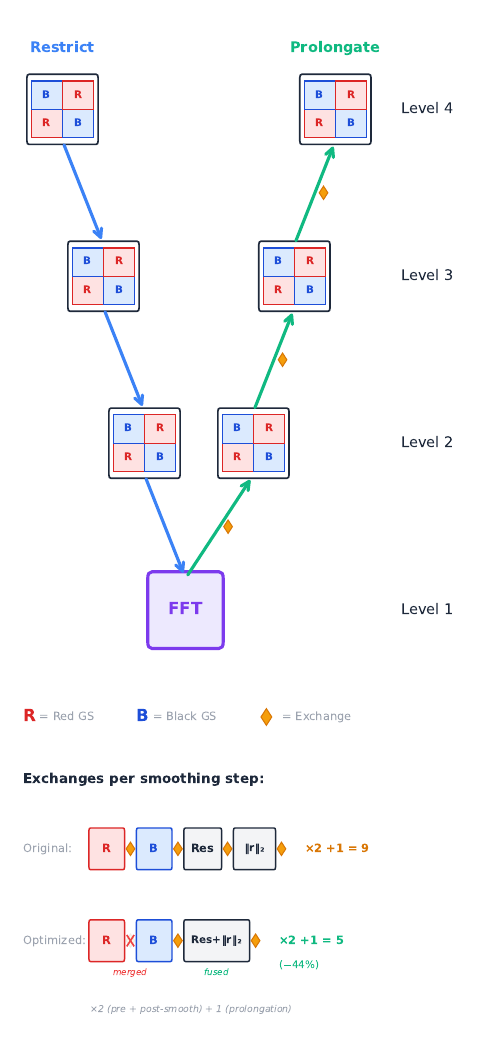}
\caption{Schematic of the multigrid V-cycle for four AMR levels.
  \emph{Upper:} The left (descending) leg restricts the residual from
  fine to coarse, while the right (ascending) leg prolongates the
  correction.  Orange diamonds on the prolongation arrows mark
  ghost-zone exchanges.
  \emph{Lower:} Exchange positions within one smoothing step.
  The original code requires 4 exchanges (after red, after black,
  residual, norm). However, the optimized code merges the red-black sweep and
  fuses the residual and norm computation, reducing the count to 2.
  With two smoothing steps (pre + post) plus one prolongation
  exchange, the total number decreases from~9 to~5 per level per iteration.}
\label{fig:mg_vcycle}
\end{figure}

Starting at the finest AMR level, the algorithm applies RBGS
pre-smoothing and then \emph{restricts} the residual to the next
coarser level.  Restriction requires no ghost exchange because each
coarse cell is the average of its eight fine children, all of which
reside on the same rank by the oct-tree construction.  This
restriction--smooth sequence descends to the coarsest level, where a
direct solve (or additional smoothing) is applied.  The correction is
then \emph{prolongated} (interpolated) back up through each level,
with post-smoothing at each step in order to damp high-frequency errors
introduced by the interpolation.  Unlike restriction, prolongation
\emph{does} require a ghost exchange. Specifically the trilinear (3D) interpolation stencil reads neighbouring coarse cells which may belong to a different rank.  Fig.~\ref{fig:mg_vcycle} illustrates the V-cycle structure and the exchange positions.

In a distributed-memory implementation each smoothing step requires
ghost-zone exchanges between \mpi\ ranks to update boundary values.
In the original code, every red sweep, black sweep, residual
computation, and norm reduction each triggers a separate communication
(ghost exchange or \code{MPI\_ALLREDUCE}),
totalling 9 communications per smoothing cycle per level (4 for
pre-smoothing, 4 for post-smoothing, plus 1 for prolongation).

We apply several targeted optimizations as follows.
\begin{enumerate}
\item \textbf{Precomputed neighbour cache.}  Neighbour grid indices
  are precomputed into a contiguous array before the V-cycle,
  reducing the cost of hash table lookups over all iterations
  (see Appendix~\ref{app:mg_details}).
\item \textbf{Merged red-black sweep.}  The inter-sweep ghost
  exchange between the red and black passes is removed. The black
  sweep proceeds using boundary values that have not yet been refreshed
  from neighbouring ranks.  Because the multigrid smoother is used only
  as a preconditioner (not as a standalone solver), these slightly
  \emph{outdated} boundary values do not affect convergence since any small error
  they introduce will be corrected by the next iteration. This means that we need not be bothered by the fresh communication of data from neighbours. This saves a number of ghost communications accordingly.
\item \textbf{Fused residual and norm.}  The residual
  $r_i \equiv f_i - (\nabla^2_h\phi)_i$, where $f_i \equiv 4\pi G\rho_i$
  is the source term and $\nabla^2_h$ is the discrete Laplacian,
  and its squared norm $\sum_i r_i^2$ are computed in a single loop,
  eliminating one exchange per iteration.
\end{enumerate}
These reduce the exchange count from 9 to 5 per iteration,
as shown in the lower panel of
Fig.~\ref{fig:mg_vcycle}.  The merged red-black technique is a
standard relaxed-synchronisation strategy for distributed multigrid
solvers \citep{Adams2001,Briggs2000}.  Implementation details are given in
Appendix~\ref{app:mg_details}.

\subsubsection{Performance Impact}
\label{sec:mg_performance}

Combining all optimizations, the MG Poisson solver's share of total
runtime is reduced from approximately 50\,per\,cent to 40\,per\,cent in a
Cosmo256 test (\rev{$256^3 \approx 16.78\,\mathrm{M}$} particles, 12 ranks, 10 coarse
steps). The iteration counts are unchanged (Level~8, 5 iterations and
Level~9, 4 iterations), confirming that the merged red-black exchange
does not degrade convergence.

\subsection{FFTW3 CPU Direct Poisson Solver}
\label{sec:fftw3}

As a complement to the iterative optimizations described above,
we replace the MG V-cycle
on fully uniform AMR levels with a single direct FFT solve.  At the
base level (\code{levelmin}), all $N = N_x  N_y  N_z$
cells exist and the grid is periodic making the Poisson equation
$\nabla^2\phi = 4\pi G\rho$ solvable exactly via spectral methods.
A further advantage of the Fourier-space representation is that
modified gravity models (e.g.\ massive neutrinos, coupled dark energy
perturbations) can be incorporated by multiplying the Green's function
with a scale- and time-dependent transfer function without altering
the real-space solver infrastructure.

The solver uses \textsc{fftw3} \citep{Frigo2005} and operates in two
regimes depending on the grid size.

\subsubsection{Small grid ($N \le 256^3$)}
Each rank gathers its local right-hand side into a local array and an
\code{MPI\_ALLREDUCE} sums the contributions.  A local in-place R2C
FFT (multi-threaded via \textsc{fftw3}+OpenMP) solves the system.

\subsubsection{Large grid ($N > 256^3$)}
The global grid is distributed across ranks using FFTW's native slab
decomposition along the $x$-axis.  Data redistribution between
RAMSES's spatial \dd\ and FFTW's slab layout requires
an all-to-all exchange in both forward and reverse directions.

\subsubsection{Sparse point-to-point exchange}
We replace \code{MPI\_ALLTOALLV} with sparse
\code{MPI\_ISEND}/\code{MPI\_IRECV} targeting only actual
communication partners, since each rank's domain overlaps only a
small number of FFTW slabs.  Partner lists are precomputed from the
spatial bounding boxes and cached until the \dd\
changes, reducing the communication count from $O(\nrank)$ to
$O(n_{\text{partners}})$.

\subsubsection{Performance}
\label{sec:fftw_perf}
The FFTW3 solver reduces the base-level Poisson time from 241\,s
(MG V-cycle) to 28.9\,s (\textbf{8.3$\times$ speedup}) on 12~ranks
with a $512^3$ base grid, matching the cuFFT GPU result (28\,s)
without requiring a GPU.
The solver is enabled via \code{use\_fftw=.true.} in
\code{\&RUN\_PARAMS} and requires compilation with
\code{make USE\_FFTW=1}.

Partner lists are computed from bounding boxes via
\code{MPI\_ALLGATHER}, so the solver works identically with both
Hilbert and \ksec\ orderings.

\subsubsection{cuFFT profiling}
\label{sec:cufft_breakdown}

\rev{The cuFFT GPU direct solver (28\,seconds) does not outperform the
CPU FFTW3 implementation (28.9\,seconds) even though it runs on a
high end accelerator. To investigate the bottleneck, we set CUDA timers in the cuFFT solver  to measure each phase separately.
Table~\ref{tab:cufft_phases} lists the per-call profiling results for a
$512^3$ uniform grid on an NVIDIA A100 SXM4 (80\,GB GPU memory
connected by a PCIe~Gen4 slot). The arithmetic operations themselves---comprising the forward FFT, Green's function multiplication, and inverse FFT---collectively account for only
$2.4\,$per\,cent of the wall-clock time of a single solve. In contrast, uploading the density to the GPU and downloading the potential back to the host consume the remaining 
$97.6\,$per\,cent of time.  Consequently, the cuFFT direct solver is limited by the PCIe bandwidth rather than 
the GPU's computational power. This heavy transfer overhead explains why it provides 
no advantage on PCIe-attached accelerators over a well-tuned multi-threaded CPU 
FFTW3 implementation that avoids the round trip transfer entirely. The implication 
for tightly coupled architectures such as the NVIDIA Grace Hopper (GH200), which 
eliminates the PCIe bottleneck, is clear
(see Section~\ref{sec:hybrid_perf}).}

\begin{table}
\centering
\caption{\rev{Breakdown of the wall-clock time for a single execution of the cuFFT GPU direct solver on a $512^3$ uniform grid (NVIDIA A100 SXM4, connected via a PCIe Gen4 slot), measured using internal CUDA timers. The arithmetic phases (forward FFT, Green's function multiplication, and inverse FFT) collectively account for only $2.4\,$per\,cent of the total time. In contrast, data transfer over the PCIe bus dominates the wall-clock time at $97.6\,$per\,cent.}}
\label{tab:cufft_phases}
\footnotesize
\begin{tabular}{@{}lrr@{}}
\toprule
Phase                       & Time (s) & Fraction \\
\midrule
H2D copy of $\rho$          & 0.278    & 46.9\,\% \\
Forward FFT (D2Z)           & 0.004    &  0.7\,\% \\
Green's function multiply   & 0.006    &  1.0\,\% \\
Inverse FFT (Z2D)           & 0.004    &  0.7\,\% \\
D2H copy of $\phi$          & 0.301    & 50.7\,\% \\
\midrule
Total per solve             & 0.593    & 100.0\,\% \\
\midrule
Compute (fwd $+$ G $+$ inv) & 0.014    &  2.4\,\% \\
Transfer (H2D $+$ D2H)      & 0.579    & 97.6\,\% \\
\bottomrule
\end{tabular}
\end{table}

\section{FEEDBACK SPATIAL BINNING}
\label{sec:feedback}
Feedback routines must identify all cells within a finite blast
radius of each event.  We introduce spatial hash binning to
accelerate this search, reducing it from an
$\Olog{N_{\rm cells} \times N}$ pairwise scan to an
$\Olog{N_{\rm cells}}$ local lookup.  With \ksec\ ordering, the
binning grid covers only the local domain bounding box (plus an
$r_{\rm max}$ halo), avoiding global memory allocation and
naturally restricting neighbouring interactions to a compact
region.

\subsection{The Brute-Force Bottleneck}
\label{sec:fb_problem}

The Type~II supernova (SNII) feedback implementation in \ramses\
involves two computationally expensive routines.

\begin{itemize}
\item \code{average\_SN}: averages hydrodynamic quantities within the
  blast radius of each SN event, accumulating volume, momentum, kinetic
  energy, mass loading, and metal loading. The original implementation
  loops over all cells with respect to all SNe, yielding
  $\Olog{N_{\rm cells} \times N_{\rm SN}}$ complexity.

\item \code{Sedov\_blast}: injects the blast energy and ejecta into
  cells within the blast radius. Same $\Olog{N_{\rm cells} \times
  N_{\rm SN}}$ complexity.
\end{itemize}

In production simulations with $\sim$2000 simultaneous SN events,
these routines consume 66\,s and 11\,s per call respectively,
dominating the feedback time step.

\subsection{Spatial Hash Binning}
\label{sec:fb_binning}

We partition the simulation domain into a uniform grid of
$n_{\rm bin}^3$ bins, where
\begin{equation}
n_{\rm bin} = \texttt{max}\bigl[1,\, \texttt{min}\bigl(128,\,
\texttt{iFloor}(L_{\rm box} \,/\, r_{\rm max})\bigr)\bigr]	
\end{equation}
and $r_{\rm max}$ is the maximum SN blast radius (the larger of
\code{rcell} $\times$ \code{dx\_min} and \code{rbubble}). Each SN
event is assigned to a bin based on its position, and a linked list
threads the events within each bin.

\begin{figure}
\centering
\includegraphics[width=\columnwidth]{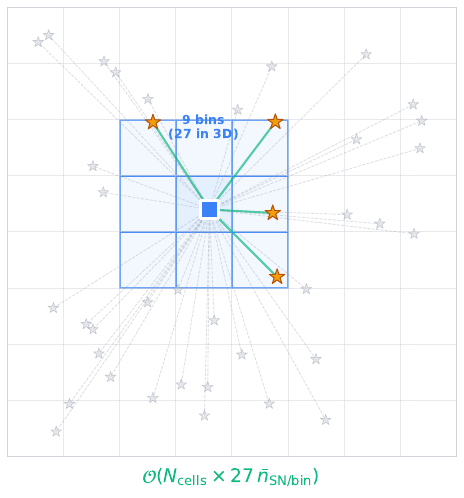}
\caption{Gathering scheme on the spatial hash binning for feedback (shown in 2D for clarity).
  The domain is partitioned into a uniform grid of bins.  The target
  cell (blue square) checks only the 9 neighbouring bins (27 in 3D),
  shown as shaded regions.  Solid green lines connect the cell to SN
  events (orange stars) within the neighbourhood.  Grey dashed lines
  indicate distant SN events that are skipped, reducing the complexity
  from $\Olog{N_{\rm cells} \times N_{\rm SN}}$ to
  $\Olog{N_{\rm cells} \times 27\,\bar{n}_{\rm SN/bin}}$.}
\label{fig:spatial_binning}
\end{figure}

For each cell, we compute its bin index and check only the 27
neighbouring bins in three dimensions. Since $r_{\rm max}$ is at most the bin size by
construction, this 27-bin neighbourhood is guaranteed to contain all
SNe that could influence the cell. The complexity becomes
$\Olog{N_{\rm cells} \times \bar{n}_{\rm SN/bin} \times 27}$,
where $\bar{n}_{\rm SN/bin} = N_{\rm SN} / n_{\rm bin}^3$ is the
average number of SNe per bin. Fig.~\ref{fig:spatial_binning}
illustrates this approach.

\subsection{Performance Results}
\label{sec:fb_performance}

Both routines are parallelised with OpenMP over grids using a
\emph{gathering} scheme in which the outer loop iterates over cells
(distributed across threads by grid ownership), and each cell looks
up nearby SN events from the spatial bins.  This gathering approach
is well-suited to OpenMP because each thread processes its own set
of grids.  The alternative \emph{scattering} approach---iterating
over SN events and modifying all affected cells---would incur race
conditions whenever multiple events overlap, requiring expensive
locking.  In the gathering scheme, \code{Sedov\_blast} writes only
to locally owned cells (one thread per grid) and needs no
synchronisation, while \code{average\_SN} accumulates contributions
into shared per-SN arrays using lightweight \code{!\$omp atomic}
updates.

Measured on the Cosmo512 test (Section~\ref{sec:perf_config};
12~\mpi\ ranks, \ksec\ ordering) at the restart epoch with
approximately 2000 simultaneous SN events, the combined gains from
spatial binning and OpenMP gathering are as follows.
\begin{itemize}
\item \code{average\_SN}: 66\,s $\rightarrow$ 0.25\,s
  ($\sim$260$\times$ speedup)
\item \code{Sedov\_blast}: 11\,s $\rightarrow$ 0.07\,s
  ($\sim$157$\times$ speedup)
\end{itemize}

The same spatial binning technique is applied to the AGN feedback
routines (\code{average\_AGN} and \code{AGN\_blast}), which suffer
from the same $\Olog{N_{\rm cells} \times N_{\rm AGN}}$ brute-force
scaling. In production simulations with tens of thousands of active
AGN sink particles, these routines dominate the sink-particle time
step.

The AGN feedback involves three distinct interaction modes (saved
energy injection, jet feedback, and thermal feedback), each with a
different geometric distance criterion. The spatial binning is
agnostic to these distinctions and simply reduces the candidate AGN set from
the full population to only those in the 27 neighbouring bins, while
preserving all distance-check logic and physical calculations
unchanged. The linked-list construction and 27-bin traversal follow
the same pattern as the SNII implementation
(\S\ref{sec:fb_binning}), with \code{bin\_head} and \code{agn\_next}
arrays replacing the SN-specific versions.

With approximately 32\,000 active AGN particles, the binned
\code{average\_AGN} achieves a $30\times$ speedup and \code{AGN\_blast}
a $14\times$ speedup, reducing the total AGN feedback time by a factor
of $\sim$4. Verification confirms bit-identical conservation
diagnostics compared to the original brute-force implementation.

\subsection{Sink Particle Merger: Oct-Tree FoF with Auto-Tuning}
\label{sec:fof}

Sink particles in \ramses\ represent unresolved compact objects (black
holes, proto-stars) and must be merged when they approach within a
linking length $r_{\rm link} = r_{\rm merge}\,\Delta x_{\rm min}$.
The standard merge routine uses a Friends-of-Friends (FoF) algorithm
with $\Olog{N_{\rm sink}^2}$ brute-force pair enumeration, in which for each
connected component (group) it swaps particles into a contiguous block
and iterates until no new links are found.  Although $N_{\rm sink}$ is
typically small (${\lesssim}10^3$), production simulations of galaxy
clusters can reach $N_{\rm sink} \sim 10^5$
\citep[e.g.\ Horizon Run~5;][]{Lee2021}, at which point the
$\Olog{N^2}$ scaling becomes prohibitive.

We provide three FoF backends with automatic selection.

\paragraph{Sequential brute-force.}
The original $\Olog{N^2}$ algorithm, preserved as a correctness
reference. But note that the order of sink array changes by swapping a pair of sinks during the FoF link.

\paragraph{Serial oct-tree.}
An oct-tree is built over the sink positions.  For each sink particle,
tree nodes whose bounding sphere does not overlap the linking sphere
are pruned, reducing the average neighbour search to
$\Olog{N\log N}$.  Linked particles are erased from the tree
so subsequent searches skip them.  This is
equivalent to the brute-force result but dramatically faster for
$N_{\rm sink} \gtrsim 500$.

\paragraph{OpenMP oct-tree.}
Every \mpi\ rank builds the same oct-tree from the full (replicated)
sink particle list, so no \mpi\ data distribution is involved that the
parallelism is purely intra-rank via OpenMP.  The particle index range
is divided among OpenMP threads, and all threads share the same
read-only tree but maintain private inclusion and group-ID arrays.
A thread-safe variant of \code{FoF\_link}
(\code{Omp\_FoF\_link}) replaces the sequential tree-erasure step
with a thread-local \code{included} array, enabling concurrent
traversal without locks.

Cross-boundary groups are rolled back to the owning thread, ensuring
deterministic results at negligible cost ($< 0.1$\,per\,cent of
particles form multi-member groups).  Thread-local results are
consolidated after the parallel phase and the final result is identical on all ranks.

\paragraph{Auto-tuning.}
An \code{Auto\_Do\_FoF} wrapper probes all three backends during the
first call, measures their wall-clock time, and selects the fastest
via \code{MPI\_BCAST} consensus.  Subsequent calls bypass the probe
and invoke the winner directly.

\begin{figure}
\centering
\includegraphics[width=\columnwidth]{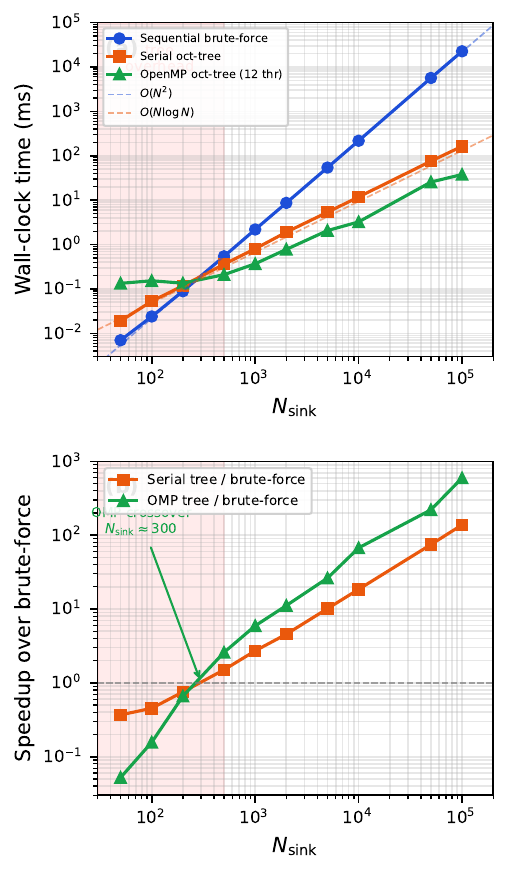}
\caption{FoF merger wallclock time as a function of the number of
  sink particles for the three backends. We randomly distribute the test sink particles. The sequential brute-force
  (blue) scales as $\Olog{N^2}$, the serial oct-tree (orange) scales
  as $\Olog{N\log N}$, and the OpenMP oct-tree (green, 12~threads)
  achieves a further parallel speedup of about 4.3.
  The test uses a massively populated, highly clustered particle
  distribution in which nearly every particle has close neighbours,
  representing a worst-case scenario for the FoF algorithm.
  The dashed line shows the $\Olog{N^2}$ scaling for reference.
  At $N_{\rm sink}=10^5$, the parallel tree is
  about 600 times faster than brute-force.}
\label{fig:fof_scaling}
\end{figure}

All three methods produce identical group partitions for every test
case ($N_{\rm sink}$ from 1 to $10^5$), verified by an $\Olog{N}$
bidirectional group-mapping check.  At $N_{\rm sink} = 10^5$ the
OpenMP oct-tree is about 600 times faster than brute-force
(Fig.~\ref{fig:fof_scaling}).

\section{VARIABLE-NCPU RESTART}
\label{sec:varcpu}

\subsection{HDF5 Parallel I/O and Restart}
\label{sec:hdf5}

Standard \ramses\ writes one binary file per \mpi\ rank per output. Therefore, restarting with a different number of ranks is not directly supported requiring an intermediate step of reading with the original rank count,
redistributing, and re-writing.

\begin{figure}
\centering
\includegraphics[width=0.75\columnwidth]{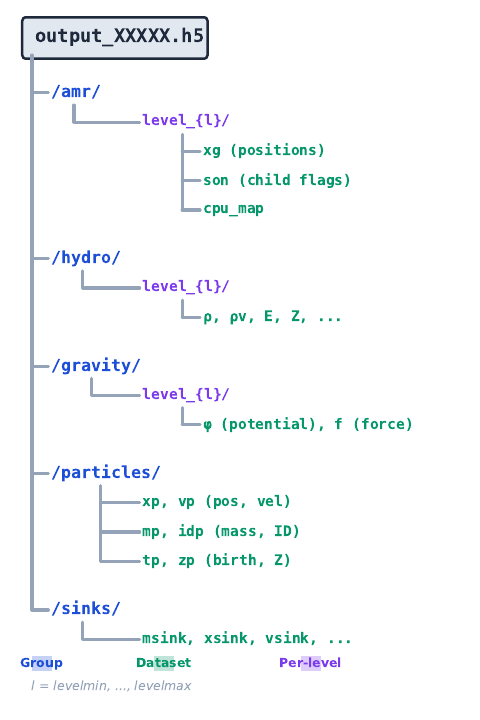}
\caption{HDF5 output file hierarchy.  AMR, hydro, and gravity data are
  stored per level while particles and sinks are stored as flat arrays.
  All ranks write to a single shared file via \mpi\-IO.}
\label{fig:hdf5_schema}
\end{figure}

We implement HDF5 parallel I/O using the HDF5 library's \mpiio\ 
backend. All ranks write to and read from a single HDF5 file,
with datasets organized hierarchically. Fig.~\ref{fig:hdf5_schema} illustrates this hierarchy.

When the number of ranks in the output file ($\nfile$) 
differs from the current run (when $\nfile \ne \nrank$), the following procedure
executes during restart.
\begin{enumerate}
\item Before reading in data files, build a uniform \ksec\ tree for the new $\nrank$ (homogeneous equal-volume partitioning, without load-balance adjustment).
\item Read all grids from the HDF5 file. Since the file is a single
  shared file, all ranks can access all data.
\item For each grid, compute the CPU ownership from the father cell's
  position.
\item Each rank retains only the grids assigned to it, building the
  local AMR tree incrementally.
\item Hydro, gravity, and particle data are read and scattered to
  locally owned grids using a precomputed file-index-to-local-grid
  mapping.
\item On the first coarse step after restart, a forced load-balance
  operation redistributes grids for optimal balance under the new rank
  configuration.
\end{enumerate}

This approach requires that all ranks temporarily hold the full grid
metadata (positions and \code{son} flags) during the reconstruction phase.
For typical production outputs (${\sim}10\,\mathrm{M}$ total grids), this
temporary overhead is a few hundred MB, well within the memory
budget freed by the optimizations of Section~\ref{sec:morton}.

\subsection{Restarting Issues}
\label{sec:varcpu_binary}

We implement a distributed I/O strategy that
enables variable-$\nrank$ restart even from the per-rank binary files written by standard \ramses.  The original binary format stores one file per
\mpi\ rank (\code{amr\_XXXXX.outYYYYY},
\code{hydro\_XXXXX.outYYYYY}, etc.) so that the number of CPUs in the previous run equals the number of files
($\nfile$), which may differ from the current number of ranks, $\nrank$. This limitation arises when transitioning between computing environments or when dynamically scaling resources for extended simulation runs.

The restart proceeds in three stages.
First, files are distributed among ranks by round-robin assignment.
Second, each rank reads its assigned AMR files, computes the
destination rank for each grid from its father cell position,
and a hierarchical \ksec\ exchange (Section~\ref{sec:ksection})
routes grids to their new owners with $O(\log_k \nrank)$ memory
per rank, replacing the original \code{MPI\_ALLTOALLV} which
required $O(\nrank)$ send/receive buffers.
On the receiving side, each grid is inserted into the local
Morton-key hash table so that subsequent data stages can
locate grids by position in $O(1)$ time without storing
per-rank metadata.
Third, hydro and gravity data are redistributed using the
same \ksec\ exchange and Morton-hash lookup.
This stage supports an optional \emph{chunked} mode
controlled by the \code{varcpu\_chunk\_nfile} parameter ($N_\textrm{chunk}$) as
when set to~$N_\textrm{chunk}>0$, only $N_\textrm{chunk}$ files are read and exchanged
at a time, reducing peak memory by a factor of
$[( \nfile + \nrank-1) / \nrank ] / N_\textrm{chunk}$.
The default ($N_\textrm{chunk}=0$) processes all local files at once,
recovering the original single-pass behaviour.
Particle files are filtered by position into the local
\ksec\ domain.

\subsection{Format Compatibility}
\label{sec:varcpu_compat}

Table~\ref{tab:restart_compat} summarises the supported
input--output combinations for variable-$\nrank$ restart.
Both the \ksec\ and Hilbert \dd\ can restore from
any input format and rank count as \ksec\ rebuilds the spatial
tree from scratch, while Hilbert recomputes a uniform
\code{bound\_key} partition for the new $\nrank$.
When the same-$\nrank$ path detects that the file's \dd\ ordering differs from the run ordering, it automatically
redirects to the variable-$\nrank$ path, which rebuilds the
decomposition from scratch.  This allows cross-ordering restarts
with any number of \mpi\ ranks for both original binary and HDF5 formats.
\begin{table}
\centering
\caption{Supported restart format combinations.
All combinations of format, file decomposition, and run decomposition
are allowed with an arbitrary number of \mpi\ ranks.}
\label{tab:restart_compat}
\small
\begin{tabular}{@{}lllc@{}}
\toprule
Format & File DD & Run DD & Allowed $\nrank$ \\
\midrule
Binary & Hilbert & Hilbert & Any \\
Binary & Hilbert & \ksec   & Any \\
Binary & \ksec   & \ksec   & Any \\
Binary & \ksec   & Hilbert & Any \\
HDF5   & Hilbert & Hilbert & Any \\
HDF5   & Hilbert & \ksec   & Any \\
HDF5   & \ksec   & \ksec   & Any \\
HDF5   & \ksec   & Hilbert & Any \\
\bottomrule
\end{tabular}
\end{table}

\section{PERFORMANCE RESULTS}
\label{sec:performance}

\subsection{Test Configuration}
\label{sec:perf_config}

We use three test configurations of increasing size, referred to
throughout as \textbf{Cosmo256}, \textbf{Cosmo512}, and
\textbf{Cosmo1024}.  All three share the same $\Lambda$CDM cosmology
($\Omega_m = 0.3111$, $\Omega_\Lambda = 0.6889$, $\Omega_b = 0.047$,
$h = 0.6766$) in a periodic box of side 256\,$h^{-1}$\,Mpc, with
initial conditions generated by \textsc{music} \citep{Hahn2011} at
$z = 29.5$.  All runs include radiative cooling \citep[Haardt \&
Madau background;][]{HaardtMadau2012}, star formation, Type~II
supernova (SNII) and AGN feedback, Bondi black hole accretion, and BH spin with magnetically arrested disc (MAD) jets as included in \cite{Lee2021}.

The \textbf{Cosmo256} test is designed for code consistency
verification.  \rev{It has $256^3 \approx 16.78\,\mathrm{M}$ dark
matter particles on a base grid of $256^3$ (\code{levelmin}=8) with
adaptive refinement up to \code{levelmax}=10, i.e.\ the standard
$N_{\rm part} = N_{\rm cell}$ uniform configuration produced directly
by \textsc{music}.}  The simulation is restarted from an HDF5
output at $z \approx 22$ and evolved for 5 coarse steps.
This small problem fits comfortably in a single node and serves as
the reference for regression testing every optimisation against a
standard Hilbert-ordering run.

The \textbf{Cosmo512} test, used for intra-node strong scaling
(Section~\ref{sec:perf_timing}), is a cosmological zoom-in
simulation sized to fit within a single compute node.  It contains
$135.7\,\mathrm{M}$ dark matter particles with a base grid of
$512^3$ (\code{levelmin}=9) and \code{levelmax}=16.  At the
restart epoch ($z \approx 4.3$, $a = 0.19$) levels~9--14 are active,
comprising about $54\,\mathrm{M}$ AMR grids, ${\sim}291$k stellar
particles, and ${\sim}32$k sink (black hole) particles.

The \textbf{Cosmo1024} test, used for multi-node scaling
(Section~\ref{sec:perf_multinode}), is a larger cosmological
zoom-in simulation designed to evaluate parallel efficiency across
multiple compute nodes.
It contains $1024^3$ (${\approx}\,1.07\,\mathrm{B}$) dark matter
particles with a base grid of $1024^3$ (\code{levelmin}=10) and
\code{levelmax}=18.  At the restart epoch ($z \approx 5.2$,
$a = 0.16$) levels~10--15 are active, comprising
${\sim}\,362\,\mathrm{M}$ AMR grids, ${\sim}941$k stellar
particles, and ${\sim}35$k sink particles.  The total particle
count including dark matter exceeds $1.15 \times 10^9$.
The run is restarted from an HDF5 output using the
variable-$\nrank$ restart feature.

\rev{All three benchmarks utilize the default 64-bit versions of the Hilbert
and Morton keys. The maximum refinement depth reached (for example, \code{levelmax}=18
in Cosmo1024) is well within the 21-level range addressable by a 64-bit key
in three dimensions. Therefore, the upstream extended-precision (\texttt{QUADHILBERT})
Hilbert path is not required here. The corresponding 128-bit (\texttt{MORTON128})
path for the new Morton-key octree described in Section~\ref{sec:morton}
has been verified by a dedicated unit test suite that exercises every dual-\texttt{int8}
arithmetic branch (encode and decode round-trip, parent and child traversal,
periodic neighbor walking, hash insert and lookup, and bit shifts spanning
the 64-bit boundary) at coordinates that map to key bits 64--125, covering
refinement levels up to the maximum addressable depth of 43. The same
\texttt{MORTON128} binary additionally completes a specially-designed test run
(grafic IC, \code{levelmax}=23) past the first coarse step with the hash
self-consistency check returning zero mismatches. A production-scale
\texttt{MORTON128} benchmark at \code{levelmax}~$\sim 25$ may be reserved for an
upcoming large-scale simulations.}

The Cosmo256 and Cosmo512 tests are run on a dual-socket AMD EPYC
7543 node (64 physical cores) with 512 GB of DDR4
memory.  The Cosmo1024 test is run on the \textit{Grammar}
cluster\footnote{\url{https://cac.kias.re.kr/systems/clusters/grammar}}
(up to 32~nodes, dual-socket AMD EPYC 7543 per node, 64~cores/node,
512\,GB RAM per node, 100\,Gbps InfiniBand interconnect).

\subsection{Strong Scaling}
\label{sec:perf_timing}

We separate the scaling analysis into two complementary parts.  The
full-physics application benchmark in this section tests intra-node
strong scaling with all solver components active, providing precise
per-component attribution.  The multi-node communication scaling
behaviour, governed by the \ksec\ \dd\ and sparse P2P exchange
patterns, is characterised independently through dedicated
microbenchmarks on a 16-node cluster
(Section~\ref{sec:bench_ghost}).

We measure strong scaling by restarting the Cosmo512 simulation
from an HDF5 output at $z \approx 4.3$.  The variable-$\nrank$ restart feature (\S\ref{sec:varcpu}) allows the
output to be read with any number of \mpi\ ranks. A forced
\code{load\_balance} on the first coarse step ensures optimal grid
distribution before timing begins. We then measure two additional
coarse steps. All runs set \code{OMP\_NUM\_THREADS=1}.

The wall-clock time and per-component timer averages for $\nrank = 1$--64
are measured.  The multigrid Poisson solver dominates the total time
(approximately 37\,per\,cent at 1~rank) and scales from 4034\,s to
106\,s at 64~ranks (speedup of 38.2), reflecting the effectiveness of
the Morton hash-based neighbour lookup and precomputed cache arrays.
The Godunov solver achieves a super-linear speedup of 58.4 (2536\,s to
43\,s) because smaller per-rank working sets fit in the L3 cache.
Particle operations show comparable scaling (538\,s to 14\,s, speedup
of 38.4).  Sink particle operations scale well up to 24~ranks (1346\,s
to 69\,s) but flatten beyond 32~ranks due to global \code{ALLREDUCE}
operations in AGN feedback.  Load-balancing overhead converges to
approximately 16\,s beyond 32~ranks and is further reduced by the
\code{nremap} interval (every 5 coarse steps in production).  The
overall speedup reaches 33.9 at 64~ranks (parallel efficiency of
53\,per\,cent).  Although individual components such as the MG solver
and particle routines scale nearly ideally, the aggregate efficiency is
limited by synchronisation barriers and serial fractions that become
relatively more significant at high rank counts on a single node.

\begin{figure*}
\centering
\includegraphics[width=\textwidth]{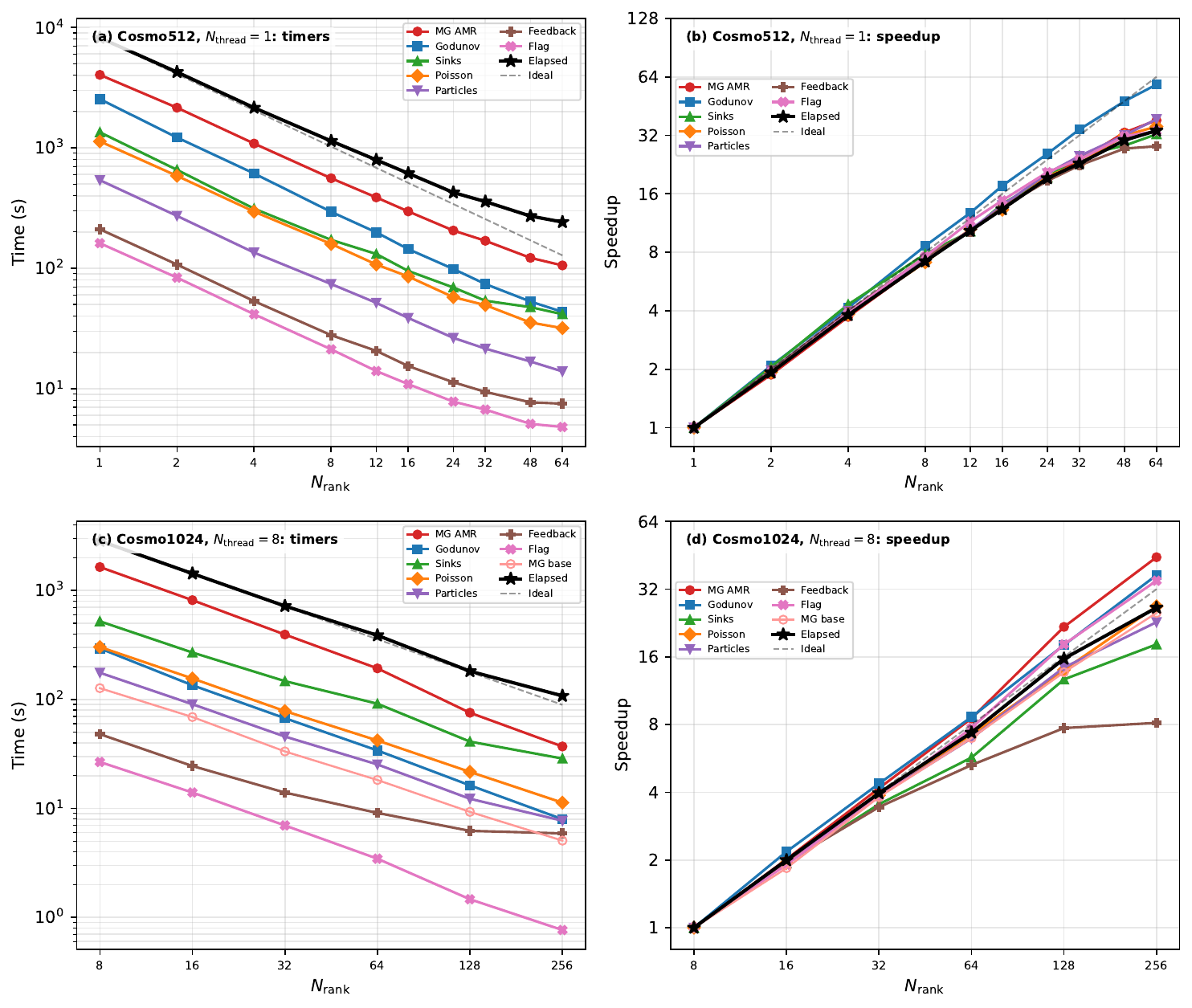}
\caption{Strong scaling of \curamses.
  \textbf{Top row:} Cosmo512 on a single dual-socket AMD EPYC 7543
  node (64~cores), $\nthread=1$.
  (a)~Elapsed and per-component times versus $\nrank$.
  (b)~Speedup relative to 1~rank, where the overall speedup reaches
  $33.9\times$ at 64~ranks (53\,per\,cent efficiency).
  \textbf{Bottom row:} Cosmo1024 on the Grammar cluster
  (1--32 nodes, 64~cores/node), $\nthread=8$.
  (c)~Per-component times versus $\nrank$.
  (d)~Speedup relative to $\nrank=8$ (1~node), where the overall speedup
  reaches $26.5\times$ at 32~nodes (83\,per\,cent efficiency).
  In both panels the dashed line shows ideal scaling.  Some
  individual components exceed ideal scaling (efficiency
  ${>}\,100$\,per\,cent) because reducing the per-rank working set improves
  cache utilisation and lowers memory-bandwidth pressure, producing
  a super-linear speedup that the elapsed time (which includes
  communication and synchronisation overhead) does not fully realise.}
\label{fig:scaling}
\end{figure*}

Fig.~\ref{fig:scaling} visualises these trends.  Panel~(a) shows the
elapsed time and per-component timer values as a function of $\nrank$
on a log--log scale.  All major components follow the ideal scaling
line closely up to $\sim$16 ranks, with gradual deviation at higher
rank counts due to communication overhead.
Panel~(b) shows the speedup relative to the single-rank baseline as
particle operations and the flag routine achieve near-ideal scaling
while the elapsed time reaches $33.9\times$ at 64 ranks.

\subsection{OpenMP Thread Scaling}
\label{sec:perf_omp}

To evaluate intra-rank parallelism we fix $\nrank = 4$ and vary \code{OMP\_NUM\_THREADS} from 1 to 30, using the same Cosmo512 problem as Section~\ref{sec:perf_timing}.  The total core count ranges from 4 to 120, and the physical core limit of the dual-socket node is 64.

Several trends distinguish the OpenMP scaling from the \mpi-only results as \emph{Multigrid Poisson solver} benefits the most from threading, improving from $1085 \to 121$\,s at 16 threads ($8.9\times$) with continued gains beyond 16 threads reaching $10.5\times$ at 30 threads.
 The precomputed neighbour arrays and fused residual loops (Section~\ref{sec:multigrid}) expose substantial loop-level parallelism. In contrast, the \emph{Godunov solver} scales $11.4\times$ from 1 to 16 threads
  ($610 \to 54$\,s).  Beyond 16 threads, performance degrades slightly due to memory bandwidth saturation and \texttt{NUMA} effects (mismatch between CPU cores and memory banks).
On the other hand, \emph{Sinks and Poisson} exhibit modest scaling (3.0 times speedups for each at 16 threads) because sink operations involve global
  reductions and sequential sections which limit parallelism.
\emph{Overall speedup} plateaus at 5.8 times speedup with 16 threads (64 cores) mainly limited by the sequential parts of \mpi\ communication and
  sink-particle operations.  Beyond the physical core count,
  oversubscription yields no further improvement.

\begin{figure*}
\centering
\includegraphics[width=\textwidth]{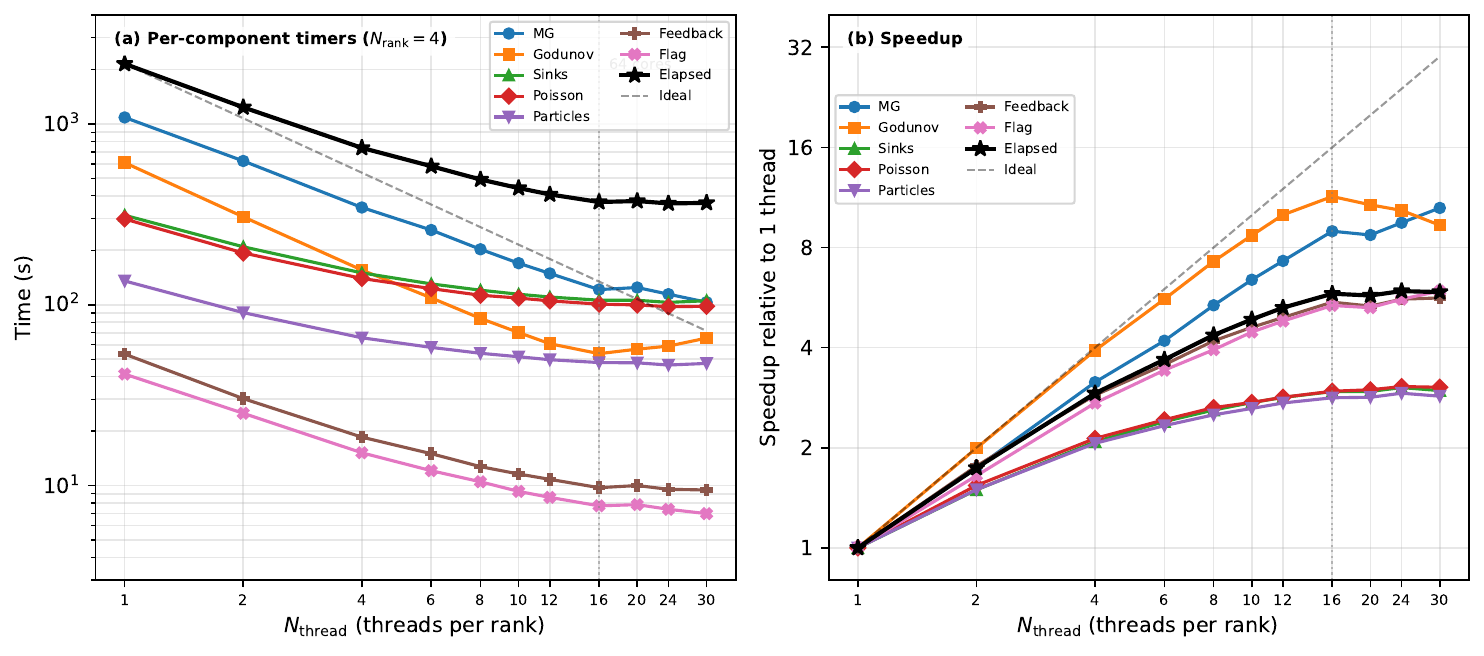}
\caption{OpenMP thread scaling of \curamses\ with $\nrank = 4$ \mpi\
ranks on a dual-socket AMD EPYC 7543 node (64 cores).
(a)~Per-component wall-clock times versus $\nthread$. The dashed
line shows ideal scaling from the single-thread baseline.
The vertical dotted line marks the physical core limit (16 threads
$\times$ 4 ranks $=$ 64 cores).
(b)~Speedup relative to 1 thread.  The MG solver achieves $10.5\times$
speedup, while the overall elapsed time plateaus at ${\sim}5.8\times$.}
\label{fig:omp_scaling}
\end{figure*}

Fig.~\ref{fig:omp_scaling} compares per-component timers and speedup
curves as a function of $\nthread$.  The vertical dotted line marks
the physical core limit (64 cores).  Panel~(a) shows that the MG solver
and Godunov components track the ideal scaling relation closely, while sinks and Poisson saturate early.  Panel~(b) confirms that the overall
speedup reaches a ceiling near $6\times$, indicating that further
performance gains require additional \mpi\ ranks rather than more threads
per rank.

\subsection{Memory-Weighted Load Balancing}
\label{sec:perf_membal}

The memory-weighted cost function, equation~(\ref{eq:cost_function}),
assigns $w_{\rm grid}$ automatically from \code{nvar} (2256\,Bytes for
\code{nvar}=14 in our test) and $w_{\rm part} = 12$ bytes per particle.
Table~\ref{tab:membal} quantifies the load balance achieved by this
scheme across the strong-scaling runs described in the preceding section.
\begin{table}
\centering
\caption{Memory load balance for the strong-scaling test
  ($54\,\mathrm{M}$ grids, $135.7\,\mathrm{M}$ particles).
  $M_{\min}$ and $M_{\max}$ are the minimum and maximum per-rank
  resident memory after the final load-balance remap (step 243).
  The level-10 grid counts illustrate per-rank work imbalance
  at the most populated level ($19.5\,\mathrm{M}$ grids total).}
\label{tab:membal}
\small
\begin{tabular}{rrrrrrr}
\hline
$\nrank$ & $M_{\min}$ & $M_{\max}$ & $M$ ratio
  & \multicolumn{3}{c}{Lv-10 grids/rank ($\times 10^3$)} \\
  &  (GB) & (GB) &
  & min & max & ratio \\
\hline
  1 & 147.2 & 147.2 & 1.000  &\multicolumn{3}{c}{---} \\
  2 &  87.0 &  87.4 & 1.005  & 9120 & 10338 & 1.13 \\
  4 &  50.1 &  50.5 & 1.008  & 4546 &  5210 & 1.15 \\
  8 &  28.0 &  28.6 & 1.021  & 2200 &  2675 & 1.22 \\
 12 &  20.9 &  21.4 & 1.024  & 1370 &  1828 & 1.33 \\
 16 &  16.7 &  17.1 & 1.024  & 1021 &  1376 & 1.35 \\
 24 &  12.5 &  13.1 & 1.048  &  671 &   937 & 1.40 \\
 32 &  10.7 &  11.1 & 1.037  &  521 &   711 & 1.36 \\
 48 &   8.2 &   8.6 & 1.049  &  306 &   479 & 1.56 \\
 64 &   8.2 &   8.6 & 1.049  &  225 &   362 & 1.61 \\
\hline
\end{tabular}
\end{table}

The memory-weighted balancer keeps the per-rank memory imbalance remarkably low, with $M_{\max}/M_{\min} \le 1.05$ for all rank counts tested from 2 to 64.  This is achieved despite substantial grid-count imbalance
at the most populated refined level (level~10, containing
$19.5\,\mathrm{M}$ grids), where the per-rank max/min ratio grows from
1.13 at 2~ranks to 1.61 at 64~ranks because the \ksec\ sub-domains
become smaller relative to the AMR clustering scale.

The low memory imbalance is key for production runs because each rank
must pre-allocate arrays sized to its local $\ngridmax$. A high imbalance
forces over-allocation on lightly loaded ranks.  With memory-weighted
balancing, the 64-rank run requires only 8.6\,GB per rank at peak,
compared with the 147.2\,GB needed in a single-rank run, a factor of
17.1 reduction, close to the ideal 64$\times$ scaling modulo the
$\sim$4\,GB fixed per-rank overhead (\mpi\ buffers, hash tables, coarse
grid).  The fixed overhead explains why per-rank memory saturates at
$\sim$8\,GB for 48 and 64 ranks.

\begin{figure*}
\centering
\includegraphics[width=\textwidth]{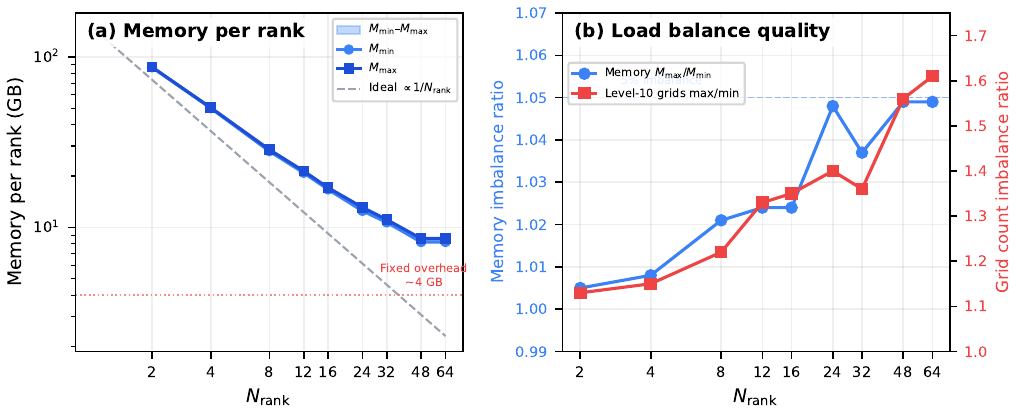}
\caption{Memory-weighted load balance across the strong-scaling test
  suite.
  (a)~Per-rank memory: $M_{\min}$ and $M_{\max}$ follow the ideal
  $1/\nrank$ scaling (dashed grey) until a fixed per-rank overhead of
  $\sim$4\,GB dominates at high rank counts.
  (b)~Load balance quality, showing the memory imbalance ratio
  $M_{\max}/M_{\min}$ (blue, left axis) remains below 5\,per\,cent
  for all rank counts tested ($2$--$64$), while the level-10 grid
  count imbalance (red, right axis) grows to 1.61 at 64 ranks,
  demonstrating that the memory-weighted cost function successfully
  decouples memory balance from grid-count balance.}
\label{fig:memory_balance}
\end{figure*}

The load-balance remap itself costs 16--26\,s for $\nrank \ge 8$,
growing from 2.3 to 5.1\,per\,cent of the
total runtime as the computation shrinks under strong scaling, a
modest price for near-perfect memory balance.
Fig.~\ref{fig:memory_balance} visualizes these trends.

An optional runtime \emph{key} parameter \code{time\_balance\_alpha} ($\alpha_t$,
default~0) blends measured per-level wall-clock cost into the memory-weighted
cost function.  For each level $\ell$, the code accumulates the local time
$T_\ell$ spent in the domain-size-dependent section (Godunov, cooling,
particle moves, star formation, and flagging) between successive remap
calls.  A per-level blend factor is defined as
\begin{equation}
  f_\ell = 1 + \alpha_t \bigl(\bar{c}_\ell^{\rm loc}/\bar{c}_\ell^{\rm avg} - 1\bigr),
  \quad f_\ell \in [0.5,\,2.0],
\end{equation}
where $\bar{c}_\ell = T_\ell / N_{\rm cells,\ell}$ is the cost per cell,
multiplies each cell's memory-weighted cost.  With $\alpha_t = 0$ (the
default), the balancer reduces to the pure memory-weighted scheme described
above.  Setting $\alpha_t \sim 0.3$--$0.5$ is recommended for
feedback-heavy or particle-dense simulations where runtime asymmetry
exceeds memory asymmetry.  The Poisson multigrid V-cycle, which imposes
a global barrier, is deliberately excluded from the timing measurement
because its cost is independent of the local domain size.

\subsection{Multi-Node Scaling (Cosmo1024)}
\label{sec:perf_multinode}

The preceding single-node tests demonstrate intra-node scaling
efficiency but cannot probe the inter-node communication overhead
that dominates production runs.  We therefore perform a multi-node
strong-scaling study using the Cosmo1024 configuration.

\subsubsection{MPI rank scaling}
\label{sec:perf_mpi_scaling}

We restart the Cosmo1024 simulation from its HDF5 output at
$z \approx 5.2$ and run 5 additional coarse steps, discarding
the first step (which includes the \code{load\_balance} remap) as
warm-up.  The number of \mpi\ ranks is varied from
$\nrank = 8$ to 256, with \code{OMP\_NUM\_THREADS}=8 and
$\nrank/8$ ranks per node, so that the total core count ranges
from 64 (1~node) to 2048 (32~nodes).  The variable-$\nrank$
restart feature allows each rank count to read the same checkpoint.

Table~\ref{tab:mpi_scaling} and Fig.~\ref{fig:scaling}(c,d)
summarise the results.  The steady-state wall-clock time per coarse
step decreases from 2858\,s for 8 ranks on 1~node to 108\,s for 256 ranks on 32 nodes, yielding a $26.5\times$ speedup.
Up to 16~nodes the scaling is essentially ideal, for example, $2.0\times$
at 2~nodes (99.8\,per\,cent efficiency), $4.0\times$ at 4~nodes
(99.0\,per\,cent), and $15.7\times$ at 16~nodes (98.4\,per\,cent).  Even at the
largest configuration of 32~nodes (256~ranks, 2048~cores), the
parallel efficiency remains 82.8\,per\,cent, which is a strong result for an AMR
code with full subgrid physics.

The \texttt{poisson-mg AMR} multigrid V-cycle dominates the runtime
at all rank counts (43--46\,per\,cent), followed by sink operations
(15--21\,per\,cent) and the combined Poisson base-level solver
(7--12\,per\,cent).  The hierarchical \ksec\ exchange
(Section~\ref{sec:ksec_exchange}) keeps the number of
communication partners bounded at $O(\log P)$, preventing the
all-to-all scaling bottleneck that would otherwise appear at high
rank counts with Hilbert ordering.  The efficiency dip at 8~nodes
(92.1\,per\,cent) followed by recovery at 16~nodes (98.4\,per\,cent) reflects
the interplay between inter-node latency and improved per-rank cache
utilisation as the working set shrinks.  At 32~nodes, the per-rank
grid count drops to ${\sim}1.4\,\mathrm{M}$, and the communication
overhead in the multigrid V-cycle begins to dominate because it requires global convergence
checks and boundary exchanges at every relaxation step.

\begin{table}
\centering
\caption{Multi-node MPI strong scaling (Cosmo1024,
  ${\sim}362\,\mathrm{M}$ grids, $1.15 \times 10^9$ particles,
  OMP\,=\,8, 8~ranks/node).  Timings are averaged over 4
  steady-state coarse steps (steps 220--223), excluding the first
  step which includes the \code{load\_balance} remap.  The
  $\mu$s/pt column gives the wall-clock time normalised by the
  total number of cell-updates across all AMR sub-steps.}
\label{tab:mpi_scaling}
\begin{tabular}{@{\,}r@{\;\;}r@{\;\;}r@{\;\;}r@{\;\;}r@{\;\;}r@{\;\;}r@{\,}}
\hline
$\nrank$ & Nodes & Cores & Time\,(s) & $\mu$s/pt & Speedup & Eff.\,(\%) \\
\hline
   8 &  1 &   64 & 2858 & 7.66 &  1.00 & 100.0 \\
  16 &  2 &  128 & 1432 & 7.67 &  2.00 &  99.8 \\
  32 &  4 &  256 &  721 & 7.73 &  3.96 &  99.0 \\
  64 &  8 &  512 &  388 & 8.30 &  7.36 &  92.1 \\
 128 & 16 & 1024 &  182 & 7.78 & 15.74 &  98.4 \\
 256 & 32 & 2048 &  108 & 9.24 & 26.51 &  82.8 \\
\hline
\end{tabular}
\end{table}

\subsubsection{MPI/OpenMP hybrid scaling}
\label{sec:perf_multinode_omp}

To identify the optimal \mpi/OpenMP balance, we fix the total
resource at 8~nodes (512~cores) and vary the \mpi/OMP split with
$\nrank \times \nthread = 512$, where $\nthread$ ranges from
1 (pure \mpi, 512 ranks) to 64 (8~ranks, 64~threads each).  Each
configuration allocates $\nrank/8$ ranks per node, so the per-node
memory footprint varies inversely with $\nrank$ while the total
compute budget remains constant.

\begin{figure*}
\centering
\includegraphics[width=\textwidth]{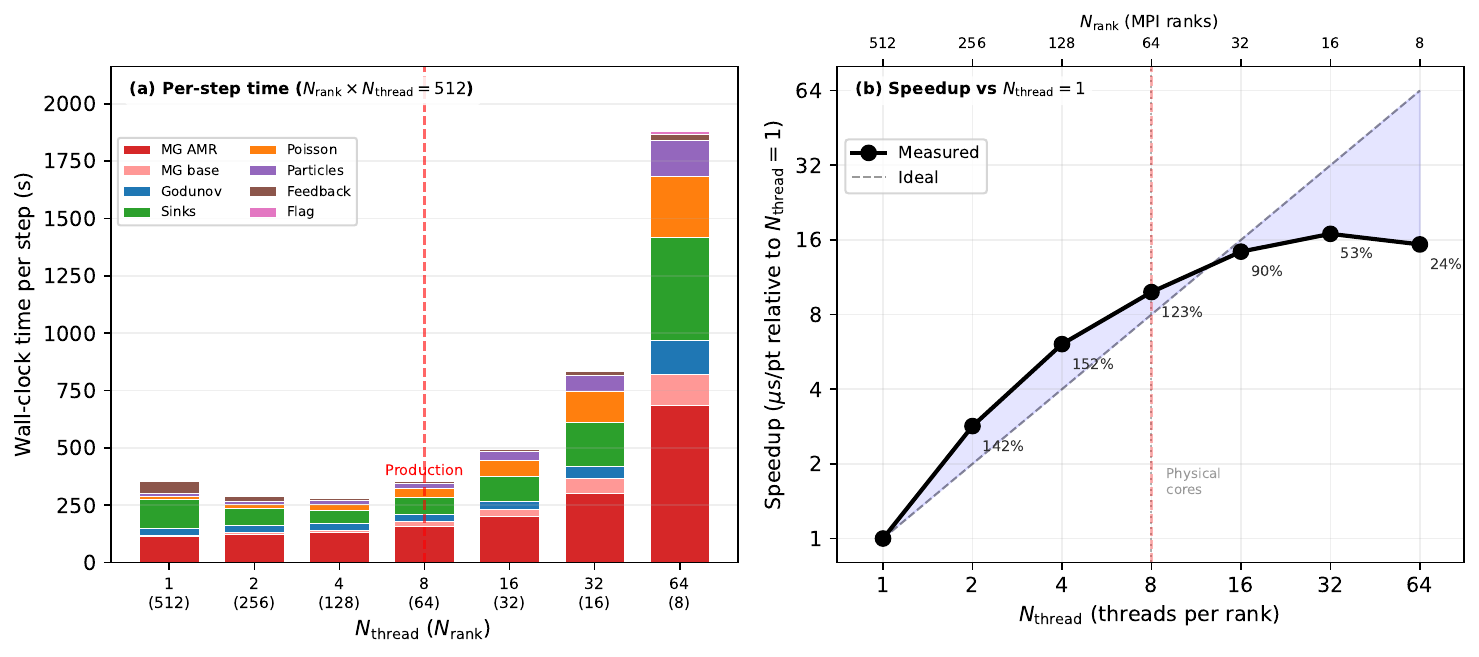}
\caption{Hybrid \mpi/OpenMP scaling of \curamses\ (Cosmo1024)
  on 8~nodes (512~cores total), varying the \mpi/OMP split
  ($\nrank \times \nthread = 512$).
  (a)~Per-step wall-clock time decomposed into solver components;
  the vertical dashed line marks the production configuration
  ($\nrank = 64$, $\nthread = 8$).
  (b)~Speedup relative to the pure-\mpi\ baseline
  ($\nthread = 1$, 512~ranks), and percentages indicate parallel
  efficiency at each point.  Super-linear speedup at
  $\nthread = 2$--$8$ reflects the communication savings from
  fewer \mpi\ ranks.  The vertical grey line marks the physical
  core limit (8~threads/rank), and beyond it hyperthreading yields
  diminishing returns.}
\label{fig:cosmo1024_omp}
\end{figure*}

Fig.~\ref{fig:cosmo1024_omp} shows the per-step wall-clock
time and relative speedup as a function of $\nthread$.
The production configuration ($\nrank = 64$, $\nthread = 8$)
achieves $9.9\times$ speedup over pure \mpi, a super-linear
efficiency of 123\,per\,cent that reflects the large communication
overhead of 512 ranks.  The minimum per-step time occurs near
$\nthread = 32$ ($4.17\,\mu$s/pt) but with only 53\,per\,cent
thread efficiency. We therefore adopt $\nthread = 8$ as the
production setting, balancing communication savings against
hyperthreading overhead and memory consolidation.
Beyond $\nthread = 8$ (the physical core limit), additional
threads share execution units and the MG~AMR solver scales poorly due to its inherently sequential level-by-level structure, which dominates the total time.

\subsubsection{Weak scaling}
\label{sec:perf_weak}

The strong-scaling tests above keep the problem size fixed while
increasing resources. Complementarily, we assess \emph{weak scaling}
by growing the problem in proportion to the core count so that each
rank handles a constant workload of ${\sim}8.4\times10^{6}$
cells/rank.  Three configurations are used: Cosmo256 ($256^{3}$,
2~ranks, 1~node, 16~cores), Cosmo512 ($512^{3}$, 16~ranks, 2~nodes,
128~cores), and Cosmo1024 ($1024^{3}$, 128~ranks, 16~nodes,
1024~cores).  All runs use identical cosmology
(Section~\ref{sec:perf_config}), $\ell_{\max}=\ell_{\min}+5$, and
$\nthread=8$.

\begin{figure*}
\centering
\includegraphics[width=\textwidth]{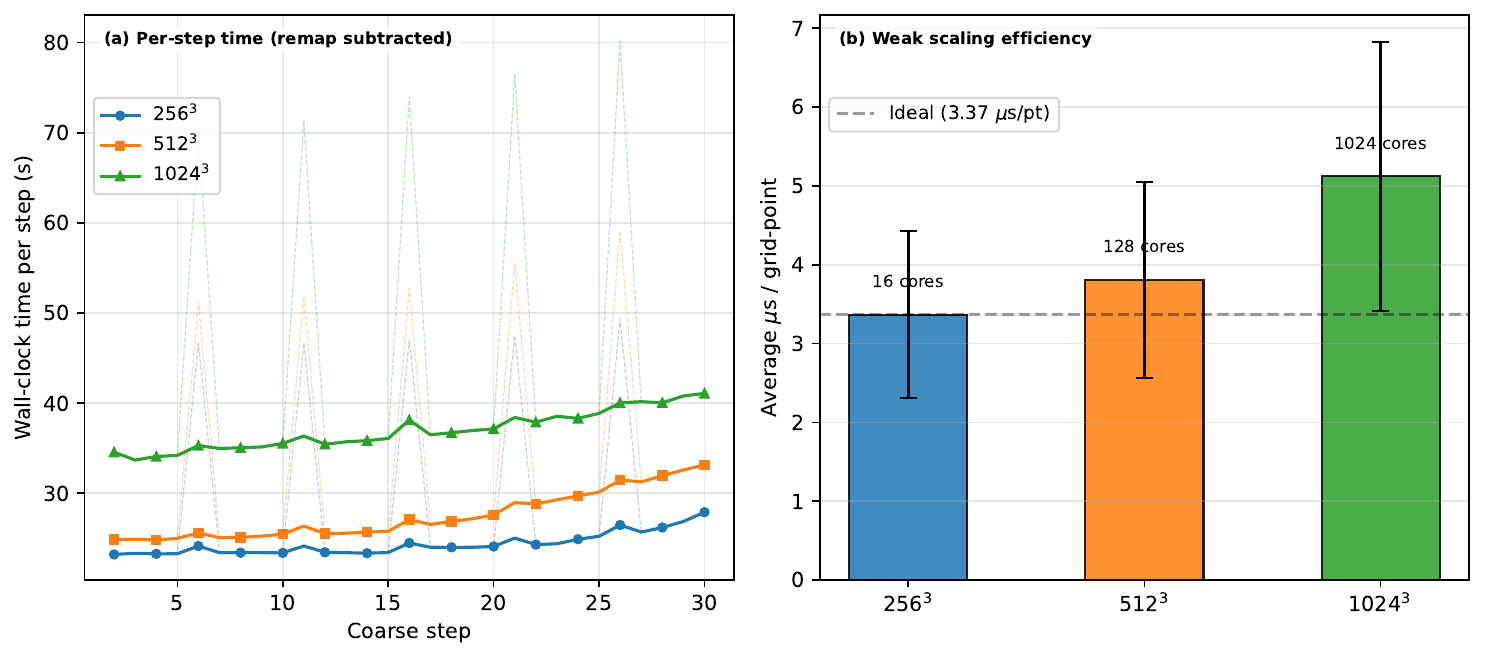}
\caption{Weak scaling of \curamses\ on the Grammar cluster
  ($\nthread=8$, same cosmology, $\ell_{\max}=\ell_{\min}+5$).
  (a)~Wall-clock time per coarse step for the three problem sizes.
  (b)~Average $\mu$s per grid-point, where perfect weak scaling corresponds
  to a constant value (dashed line).  The efficiency is 88.5\,per\,cent at
  128~cores and 65.7\,per\,cent at 1024~cores relative to the 16-core
  baseline.}
\label{fig:weak_scaling}
\end{figure*}

Fig.~\ref{fig:weak_scaling} shows the per-step wall-clock time and
the average microseconds per grid-point ($\mu$s\,/\,pt), the standard
weak-scaling efficiency metric.  From Cosmo256 to Cosmo512 the
efficiency is 88.5\,per\,cent, and scaling further to Cosmo1024 yields 65.7\,per\,cent.
A per-component breakdown reveals where this efficiency loss originates.
Of the ${\sim}\,505$\,s additional wall time from Cosmo256 to
Cosmo1024, sink-particle processing accounts for 53\,per\,cent
($+267$\,s, $2.5\times$ growth), driven by the $\Olog{\nrank}$
all-reduce synchronisation in accretion and merging diagnostics and
the increasing number of sinks at later evolutionary stages.
Load balancing contributes a further 20\,per\,cent ($+100$\,s,
$1.6\times$ growth) as the remap communication volume grows with
the number of inter-node domain boundaries.  By contrast, the
hydrodynamic solver scales nearly perfectly ($+3$\,s, $1.02\times$),
confirming that the \ksec\ ghost-zone exchange maintains constant
communication cost per rank.  The Poisson solvers (MG base + AMR)
grow by $1.2\times$, consistent with the additional AMR levels in
the larger problem.

\subsection{Communication Microbenchmarks}
\label{sec:perf_bench}

To validate the communication advantages of the \ksec\ framework at
scale, we design and execute a suite of dedicated microbenchmarks on the
\textit{Grammar} cluster (16~nodes, dual-socket AMD EPYC 7543,
64~cores/node, 512\,GB RAM/node, 100\,Gbps InfiniBand, Intel~\mpi\ 2021.17).
All benchmarks are self-contained single-source Fortran programs using
plane-by-plane ownership computation [$O(N^2)$ memory per rank] so that
grid sizes up to $N=2048$ are feasible.

\subsubsection{Ghost zone exchange}
\label{sec:bench_ghost}

This benchmark constructs a uniform $N^3$ periodic grid, assigns cell
ownership via either Hilbert or \ksec\ \dd, and
measures ghost zone exchange time for three communication backends:
\textit{P2P} (non-blocking \texttt{MPI\_ISEND/IRECV}),
\texttt{MPI\_ALLTOALLV}, and the hierarchical \ksec\ exchange described
in Section~\ref{sec:ksec_exchange}.  Each cell carries $n_{\rm var}=5$
double-precision variables (matching \ramses\ hydro), and timings are
averaged over 3~repeats.

Results for $N=1024$ ($10^9$~cells)
on 64--1024~\mpi\ ranks show the following trends:

\begin{itemize}
\item The ghost list \emph{build time} is dominated by the Hilbert curve
  computation (${\sim}453$--$534$\,s, rising at high rank counts) versus
  the \ksec\ tree traversal which scales from 200\,s (64~ranks) to
  354\,s (1024~ranks).  The \ksec\ build is $1.5$--$2.3\times$ faster
  across all rank counts.
\item The \emph{exchange time} decreases with rank count as the
  per-rank ghost volume shrinks.  P2P and \texttt{ALLTOALLV} are
  effectively tied (7--42\,ms at 64--1024 ranks), while the hierarchical
  \ksec\ exchange is ${\sim}5$--$10\times$ slower (52--368\,ms) due to
  its $O(\log P)$ serialised stages.  The gap narrows at high rank
  counts, and at 1024~ranks the \ksec\ exchange drops to 52\,ms versus
  7\,ms for P2P.
  The \ksec\ exchange is designed for variable-size AMR communications where
  all-to-all collectives become prohibitive. On this uniform-grid test
  its overhead is expected.
\item The average number of communication partners is constant at 6.0
  for \ksec\ across all rank counts, compared to 6.0--9.1 for Hilbert.
  Non-power-of-two rank counts (96, 192, 320, 448) increase the Hilbert
  neighbour count to 8.4--9.1, while \ksec\ remains at exactly 6,
  reflecting the compact cuboid sub-domains produced by recursive
  bisection.
\item The surface-to-volume ratio is consistently lower for \ksec\
  than for Hilbert at non-power-of-two ranks (e.g.\ 0.027 vs.\ 0.033
  at 96~ranks), confirming that the cuboid decomposition minimises
  ghost volume.
\end{itemize}

\subsubsection{MPI collective performance}
\label{sec:bench_mpi}

A second benchmark profiles the raw performance of five \mpi\ collectives,
namely \texttt{MPI\_ALLREDUCE}, a manual binary-tree allreduce (reduce to
root then broadcast, $O(\log_2 P)$ stages), \texttt{MPI\_BCAST},
neighbour-to-neighbour \texttt{MPI\_SENDRECV} (1D ring), and
\texttt{MPI\_ALLTOALL}, over message sizes from 1 to $10^7$ doubles
(8\,B to 80\,MB).  Each test executes a warmup round followed by 100
timed repeats, and the per-repeat maximum across all ranks is recorded.
These measurements provide a baseline for interpreting the dominant
communication costs in \curamses.

\subsubsection{Load balance quality}
\label{sec:bench_lb}

To isolate the impact of \dd\ geometry on load balance,
a third benchmark assigns non-uniform cell weights drawn from 20
Gaussian clusters (amplitude 10--100, width $\sigma=0.02$--0.10)
superposed on a unit background.  For each DD method (Hilbert vs.\
\ksec) the benchmark reports (i)~the weight imbalance ratio
$W_{\max}/W_{\mathrm{mean}}$, (ii)~the ghost cell ratio (fraction of
owned cells with at least one face neighbour on a different rank), and
(iii)~the number of \mpi\ communication partners per rank.

\subsubsection{Poisson solver}
\label{sec:bench_poisson}

The fourth benchmark compares three iterative solvers (Jacobi,
red-black Gauss-Seidel, and unpreconditioned conjugate gradient)
on a distributed $N^3$ periodic grid with slab decomposition in~$z$.
The source term is $f = -12\pi^2 \sin(2\pi x)\sin(2\pi y)\sin(2\pi z)$,
giving an exact solution $\phi = \sin(2\pi x)\sin(2\pi y)\sin(2\pi z)$.
Each solver separates computation (stencil update) from communication
(ghost plane exchange via \texttt{MPI\_SENDRECV}), enabling direct
measurement of the compute/communication ratio.

\rev{This microbenchmark measures the per-iteration cost of the three reference solvers 
in terms of arithmetic operations and communication overhead. It is not intended 
to predict the number of iterations required for convergence. For a three-dimensional 
Poisson problem with $N$ unknowns on a uniform grid, the Jacobi and red-black 
Gauss-Seidel methods converge in $\mathcal{O}(N^{2/3})$ iterations resulting in a 
total computational cost of $\mathcal{O}(N^{5/3})$. An unpreconditioned conjugate 
gradient (CG) solver converges in $\mathcal{O}(N^{1/3})$ iterations with a total 
work of $\mathcal{O}(N^{4/3})$, whereas only the geometric multigrid method 
achieves the optimal $\mathcal{O}(N)$ total complexity \citep{Trottenberg2001,Briggs2000}. 
The smoother employed within the \curamses\ multigrid V-cycle is red-black 
Gauss-Seidel. In this benchmark, the conjugate gradient, Jacobi, and red-black 
Gauss-Seidel solvers serve strictly as baseline references to measure the 
per-iteration arithmetic-to-communication ratio.}

\section{GPU OFFLOADING: APPROACH AND ASSESSMENT}
\label{sec:hybrid}

\rev{We begin this section by stating the fundamental principle that governs every design decision discussed below: \emph{the ultimate design of GPU utilisations  must rely on the simulation data being fully resident in GPU memory.} On current hardware, the bandwidth asymmetry between on-package HBM and the host--device interconnect (quantified in the next paragraph) is so pronounced that any offloading scheme requiring the simulation state to be streamed across the bus on every kernel invocation becomes uncompetitive. Consequently, the GPU chip is starved of work long before its compute units saturate. The remainder of Section~\ref{sec:hybrid} is organised around this principle. The dispatch and batching machinery (Sections~\ref{sec:hybrid_dispatch}--\ref{sec:hybrid_superbatch}) exists solely to make this residency model practical within an existing CPU-side AMR code, whereas the speedups reported in Section~\ref{sec:hybrid_perf} stem directly from the residency strategies described in Sections~\ref{sec:hybrid_gather} and \ref{sec:hybrid_mg}. Finally, the projected GH200 expectations are precisely what one expects when the host--device bottleneck is eliminated altogether.}

Modern GPU-equipped compute nodes offer substantial floating-point
throughput and a natural question for simulation practitioners is
whether enabling GPU offloading can meaningfully accelerate their
production runs.  The answer turns out to depend strongly on the
CPU--GPU interconnect. On current PCIe-connected hardware the net
speedup is a modest ${\sim}20$\,per\,cent (but still worthwhile for
multi-month campaigns) whereas tightly coupled architectures such as
the NVIDIA GH200 promise ${\sim}2\times$.  In this section we
describe the hybrid dispatch design, benchmark it on three GPU
platforms (H100, A100, A40), and derive an analytical performance
model that enables users to predict the benefit for their own hardware.

Certain compute-intensive routines, namely the Godunov solver, gravity force
computation, hydrodynamic synchronisation, CFL timestep, prolongation,
and radiative cooling, are amenable to GPU acceleration.  Rather than
running entire time steps on the GPU, \curamses\ adopts a
\emph{hybrid dispatch} model where OpenMP threads dynamically choose
between CPU and GPU execution at runtime.

\subsection{Dynamic Dispatch Model}
\label{sec:hybrid_dispatch}

At the start of each parallel region, each OMP thread attempts to acquire a GPU stream slot via an atomic counter.  Threads that acquire a slot accumulate grid data into a \textbf{batched grid buffer} of configurable size (by default, 4096 grids) and launch GPU kernels asynchronously when the buffer becomes full.  Threads that do not acquire an available slot execute the standard Fortran solver on the CPU side.  The \code{schedule(dynamic)} clause ensures load balancing so that while a GPU thread is waiting for kernel completion, remaining loop iterations are picked up by CPU threads.

This design requires no code duplication since the CPU branch is the
original Fortran subroutine, and the GPU branch is an alternative
within the same \texttt{OpenMP do} loop.

\subsection{Batched Grid Buffering}
\label{sec:hybrid_superbatch}

Launching a GPU kernel incurs a fixed overhead of 10--50\,$\mu$s, which can dominate the computation time when each kernel processes only a single grid at a time.  To mitigate this overhead, the GPU thread in the
\code{OpenMP do} loop does not launch a kernel immediately.  Instead,
it collects the integer \emph{index list} (the
cell and neighbour identifiers which define each grid's stencil) into
a host-side buffer.  Once the buffer reaches 4096 grids, a
\emph{flush} is triggered. Then, the index list is uploaded to the device (or GPU) in a single transfer, and consecutive GPU kernels execute on the entire batch.

A key optimisation is the \textbf{on-device compactification}.  In a na\"{i}ve
implementation, the intermediate flux data for every cell in the batch
would be copied back to the host (${\sim}98$\,MB per flush), creating
a severe PCIe bandwidth bottleneck.
Instead, the scatter-reduce kernel completes the conservative update
directly on the device, writing only the final per-grid result to
\code{unew}.  The resulting device-to-host transfer is
${\sim}5$\,MB per flush, a $20\times$ reduction in PCIe traffic.

\subsection{GPU-Resident Mesh Data}
\label{sec:hybrid_gather}
The initial GPU implementation gathers the stencil data for each batch
of grids on the CPU, copies it to the device, and copies the results
back after the kernel finishes.  Profiling reveals that this
host-to-device (H2D) transfer accounts for the majority of the total
GPU time, completely negating the kernel speedup.

We therefore adopt a \emph{GPU-resident mesh} strategy in which the full AMR
mesh arrays (\code{uold}, \code{unew}, \code{phi}, \code{f},
\code{son}) are uploaded to the device once at the beginning of each
level update.  The gather phase, which formerly assembled the stencil
on the host, is moved to a GPU kernel that reads directly from
device-resident arrays.  After the Godunov sweep the scatter kernel
writes updated fluxes back to \code{unew} on the device, and only the
modified portion is copied back to the host.

Furthermore, GPU device arrays and pinned host memory registrations
are kept \emph{persistent} across level updates and fine timesteps.
Rather than allocating, pinning, freeing, and unpinning the mesh
buffers at every level (approximately 95 cycles per three coarse
steps), the code allocates once and reuses the buffers whenever the
mesh size remains unchanged.  Reallocation occurs only when the cell
count changes (e.g.\ after load balancing).  Final cleanup is deferred
to program termination.  This eliminates the dominant
\code{cudaHostRegister}/\code{cudaHostUnregister} overhead (measured at
${\sim}3$--$4$\,s per cycle for 23\,GB of host memory), reducing the
per-level GPU overhead to the \code{cudaMemcpy} cost alone.

This reorganisation reduces the host-to-device transfer to less than
10\,per\,cent of wall-clock time.  The gather time drops from 70.7\,s to
2.1\,s ($34\times$) and the data transfer from 6.6\,s to 0.24\,s
($27\times$).  The approach requires one \mpi\ rank per GPU because
the mesh upload consumes 4--9\,GB of device memory per rank.  When the
device memory is insufficient the code automatically runs on the CPU
instead.

With the Godunov solver running efficiently on the GPU, we next turn
to the other major compute kernel, the multigrid Poisson solver, which
dominates the remaining runtime.

\subsection{GPU Multigrid Solver}
\label{sec:hybrid_mg}

We extend the GPU acceleration to the multigrid (MG) Poisson solver.
All four V-cycle kernels such as red-black Gauss--Seidel smoothing,
residual computation, restriction, and interpolation, are implemented
as CUDA kernels with the GPU-side data (\code{phi}, \code{f},
neighbour arrays) totalling approximately 6.4\,GB.
A \emph{ghost-zone-only} exchange transferring only the boundary cells
between host and device instead of the full \code{phi} array reduces
per-sweep PCIe volume from ${\sim}2.4$\,GB to ${\sim}6$\,MB.
The remaining bottleneck is the MPI halo exchange after each kernel,
which requires device-to-host and host-to-device transfers of
boundary data at every smoothing sweep and level transition.
The complete GPU-accelerated V-cycle algorithm is detailed in
Appendix~\ref{app:vcycle}.

\subsection{GPU Performance Assessment}
\label{sec:hybrid_perf}

Table~\ref{tab:gpu_perf} summarises the GPU performance on the
\textit{Syntax} cluster\footnote{\url{https://cac.kias.re.kr/systems/clusters/syntax}}, which hosts NVIDIA H100~NVL GPUs
(93.1\,GB HBM3 per device, PCIe Gen5~$\times 16$), using 4~\mpi\ ranks
$\times$ 8~OMP threads, running 3 coarse steps of the Cosmo1024
test (Section~\ref{sec:perf_config}).  Each \mpi\ rank is bound to
one H100~NVL device (4~GPUs per node).

\begin{table*}
\centering
\caption{GPU performance comparison on H100~NVL (4~ranks $\times$
8~threads, 3 coarse steps, Cosmo1024).  The GPU run uses persistent
mesh buffers and ghost-zone-only MG exchange.
\rev{The GPU row is decomposed into a compute (kernel) and a transfer
(H2D + D2H over PCIe) subrow for the two Poisson columns: the FFTW3
column uses the per-call cuFFT phase split of
Table~\ref{tab:cufft_phases} (2.4\,per\,cent compute, 97.6\,per\,cent
transfer), and the MG AMR column uses Eq.~(\ref{eq:gpu_model})
with $T_{\rm CPU}^{\rm fixed}+T_{\rm GPU}^{\rm kernel}=383+178$\,s
and $V_{\rm PCIe}/B=1820/22$\,s. The Godunov column is reported as a
single number because the bottleneck is the
${\sim}23\,$GB-per-level full-mesh \code{cudaMemcpy} that was not
separately CUDA-event-timed.}}
\label{tab:gpu_perf}
\footnotesize
\begin{tabular}{@{}lrrrrr@{}}
\toprule
Mode & Total & MG AMR & FFTW3  & Godunov & Other \\
     & (s)   & (s)    & (s)        & (s)     & (s)   \\
\midrule
CPU-only                          & 1852 & 1094 & 104 & 181 & 473 \\
GPU                               & 1534 &  643 & 120 & 263 & 508 \\
\rev{\quad compute (kernel)}      &      & \rev{561} & \rev{3}   &     &     \\
\rev{\quad transfer (H2D$+$D2H)}  &      & \rev{82}  & \rev{117} &     &     \\
\midrule
Speedup    & $1.21\times$ & $1.70\times$ & --- & $0.69\times$ & --- \\
\bottomrule
\end{tabular}
\end{table*}

The multigrid Poisson solver on AMR levels is the dominant
beneficiary of GPU offloading since the MG~AMR component drops from
1094\,s to 643\,s, a substantial speedup of 1.70 times that accounts for the
entire net improvement.  However, the Godunov hydrodynamics solver, by
contrast, is \emph{slower} on the GPU (181\,s~$\to$~263\,s).
The stencil gather and scatter kernels execute efficiently on the
device, but the per-level \code{cudaMemcpy} of the full mesh
(${\sim}23$\,GB per upload) is costly enough to negate the kernel
speedup.  It is important to note that the GPU mode dedicates one of
the $r$ OMP threads to GPU management (kernel launch, stream
synchronisation, PCIe transfer), so the comparison at $r = 8$ is
effectively \emph{7 CPU threads + 1~GPU} versus \emph{8 CPU threads}
in the CPU-only run.

\subsubsection{Bandwidth-limited performance model}

To quantify the GPU efficiency as a function of the CPU--GPU
interconnect bandwidth, we decompose the MG~AMR time into three
components as
\begin{equation}
T_{\mathrm{MG}}(B) \;=\; T_{\mathrm{CPU}}^{\mathrm{fixed}}
  \;+\; T_{\mathrm{GPU}}^{\mathrm{kernel}}
  \;+\; {V_{\mathrm{PCIe}}}{B^{-1}},
\label{eq:gpu_model}
\end{equation}
where $T_{\mathrm{CPU}}^{\mathrm{fixed}}$ is the time spent in
MPI halo exchanges and coarse-level solves that cannot be offloaded
(${\approx}\,383$\,s, ${\sim}\,35$\,per\,cent of the CPU-only time), $T_{\mathrm{GPU}}^{\mathrm{kernel}}$ is the
GPU stencil compute time (${\approx}\,178$\,s, corresponding to a
${\sim}\,4\times$ kernel speedup over CPU), $V_{\mathrm{PCIe}}$ is
the cumulative PCIe transfer volume (${\approx}\,1820$\,GB over
the full run), and $B$ is the effective CPU--GPU bandwidth.

\begin{figure*}
\centering
\includegraphics[width=\textwidth]{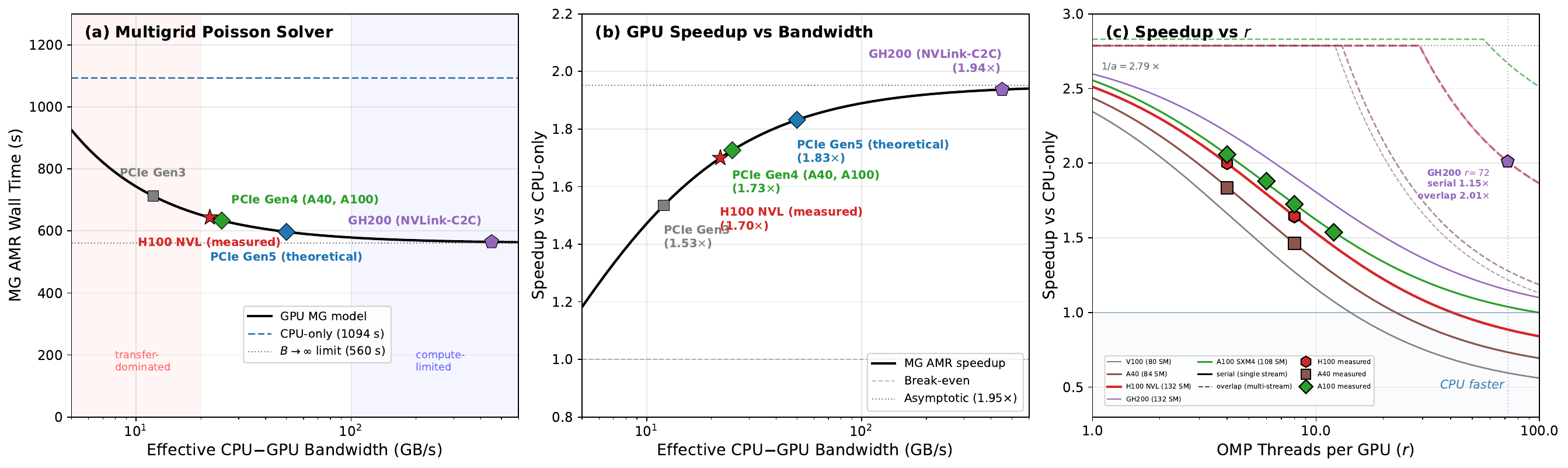}
\caption{Multigrid Poisson GPU performance model.
\textit{(a)}: MG AMR wall-clock time as a function of effective
CPU--GPU bandwidth $B$, equation~(\ref{eq:gpu_model}), compared with
the CPU-only baseline (dashed) and the infinite-bandwidth limit
(dotted).  The red star marks the measured H100~NVL data point.
\textit{(b)}: corresponding speedup over CPU-only.  The asymptotic
limit of $1.95\times$ is set by the Amdahl fraction of
CPU-bound work.
\textit{(c)}: speedup as a function of $r$ (number of OMP threads per GPU)
(see equation~\ref{eq:gpu_model_r}).  Solid and dashed curves show the
serial (single-stream) and overlap (multi-stream) models,
respectively.  Filled markers show measured MG~AMR speedups:
hexagons for H100~NVL, squares for A40 (both on the same node),
and diamonds for A100~SXM4 (separate node with independently
calibrated $T_{\mathrm{CPU}}(r)$, see text).  The A100 curve
is fitted from measurements at $r = 4$ \& $12$.}
\label{fig:gpu_mg_model}
\end{figure*}

Fig.~\ref{fig:gpu_mg_model} shows the model predictions.  On the
H100~NVL (measured $B \approx 22$\,GB/s with pinned memory),
the model reproduces the observed $1.70\times$ speedup.  The
curve reveals three regimes. First, in the \textit{Transfer-dominated} ($B \lesssim 20$\,GB/s), the
      PCIe bottleneck limits the speedup to ${\lesssim}\,1.5\times$.
      Even legacy PCIe Gen3 ($B \approx 12$\,GB/s) still yields a
      net speedup of $1.53\times$ because the MG stencil computation
      is expensive enough to offset the transfer cost.
Second, in the \textit{Current hardware} ($20 \lesssim B \lesssim 60$\,GB/s),
      PCIe Gen4--Gen5 accelerators achieve $1.7$--$1.8\times$ speedup.
      The H100~NVL falls in this regime.
And the last, in the \textit{Compute-limited} ($B \gtrsim 100$\,GB/s), the speedup
      saturates at an asymptotic limit of ${\sim}\,1.95\times$, set
      by the Amdahl fraction of CPU-only work.  Even the NVIDIA
      GH200 with its NVLink-C2C coherent interconnect
      ($B \approx 450$\,GB/s) reaches only $1.94\times$, indicating
      that further bandwidth improvements yield diminishing returns.

However, equation~(\ref{eq:gpu_model}) was evaluated at a fixed thread count $r = 8$.  Since GPU mode dedicates one thread to device management, the effective comparison is $(r{-}1)$ CPU threads plus one GPU versus $r$ CPU threads.  Defining the total serial MG~AMR work as $C_{\mathrm{total}} = r \, T_{\mathrm{CPU}}(r)$ and the serial CPU-fixed work as $C_{\mathrm{fixed}} = (r{-}1)\, T_{\mathrm{CPU}}^{\mathrm{fixed}}$, the generalised speedup ($S(r,B)\equiv T_\mathrm{CPU}/T_\mathrm{MG}(B)$) becomes
\begin{equation} 
S(r, B) = \frac{C_{\mathrm{total}} / r}
  {C_{\mathrm{fixed}} / (r{-}1) \;+\; T_{\mathrm{GPU}}^{\mathrm{kernel}}
   \;+\; V_{\mathrm{PCIe}} \, B^{-1}},
\label{eq:gpu_model_r}
\end{equation}
where $r$ is the number of OMP threads per \mpi\ rank (each rank bound to one GPU).  From the Cosmo1024 benchmark at $r = 8$ and $B = 22$\,GB/s we obtain $C_{\mathrm{total}} = 8752$\,s$\cdot$thread, $C_{\mathrm{fixed}} = 2681$\,s$\cdot$thread, $T_{\mathrm{GPU}}^{\mathrm{kernel}} = 178$\,s, and $V_{\mathrm{PCIe}} = 1824$\,GB.

Panel (c) of Fig.~\ref{fig:gpu_mg_model} shows $S(r)$ for several
GPU cards.  For a given card the speedup peaks at $r \approx 4$
and decreases for both smaller and larger $r$. At $r = 2$ the loss
of half the CPU threads outweighs the GPU contribution, while at
large $r$ the GPU overhead exceeds the kernel saving.
The card-dependent quantity is $T_{\mathrm{GPU}}^{\mathrm{kernel}}$.
Because the AMR multigrid stencil accesses octree data through
irregular, pointer-chasing patterns, it does not achieve peak HBM
streaming bandwidth. Instead, the effective throughput scales with
the number of streaming multiprocessors (SMs), which determines
the degree of memory-level parallelism available to hide access
latency.  Accordingly, we set
$T_{\mathrm{GPU}}^{\mathrm{kernel}} \propto 1 / N_{\mathrm{SM}}$,
calibrated from the H100~NVL ($N_{\mathrm{SM}} = 132$,
$T_{\mathrm{GPU}}^{\mathrm{kernel}} = 178$\,s).
Separate benchmarks on A40~GPUs ($N_{\mathrm{SM}} = 84$) at $r = 4$
and $r = 8$ confirm this scaling within $5\,\mathrm{per\,cent}$ for the
4-rank configuration (check filled squares in Panel (c) of Fig.~\ref{fig:gpu_mg_model}).

To validate the model on a different hardware platform, we repeat
the $r$-sweep benchmark on A100~SXM4 GPUs ($N_{\mathrm{SM}} = 108$,
80\,GB HBM2e, PCIe Gen4).  Because the A100 node has a faster host CPU, 
$T_{\mathrm{CPU}}(r)$ is recalibrated independently.  Fitting
equations~(\ref{eq:gpu_model})--(\ref{eq:gpu_model_r}) to the A100
data yields $a = 0.354$ (within 1.5\,per\,cent of the H100 value
$a = 0.359$), confirming that the CPU-fixed fraction is a
hardware-independent property of the MG algorithm.  The A100 data
span $r = 4$ and 12 and are reproduced by the fitted curve within
1\,per\,cent (green diamonds in Panel (c) of Fig.~\ref{fig:gpu_mg_model}).
Notably, the A100 MG~AMR speedup reaches $2.06\times$ at $r = 4$,
exceeding the H100 value of $2.00\times$, because the A100's
HBM2e memory subsystem (2\,TB/s) provides higher effective per-SM
throughput for the memory-intensive AMR stencil than the simple
$1/N_{\mathrm{SM}}$ scaling from H100 predicts so that the fitted kernel
time $T_{\mathrm{GPU}}^{\mathrm{kernel}}$ is 37\,per\,cent faster
than the SM-count extrapolation.

Table~\ref{tab:a100_rsweep} summarises the A100 $r$-sweep results.
The GPU speedup peaks at $r = 4$ ($1.49\times$ overall, $2.06\times$
MG~AMR) and decreases monotonically through $r = 12$ ($1.09\times$
overall).  At $r = 16$ ($4 \times 16 = 64$ threads), CPU
oversubscription degrades both modes, breaking the trend.

\begin{table}
\centering
\caption{GPU $r$-sweep benchmark on A100~SXM4 (4~ranks, 3~coarse
steps, Cosmo1024).  Speedup is GPU vs CPU-only at each~$r$.}
\label{tab:a100_rsweep}
\footnotesize
\begin{tabular}{@{}rrrrrr@{}}
\toprule
$r$ & CPU & GPU & Speedup & MG CPU & MG GPU \\
    & (s) & (s) &         & (s)    & (s)    \\
\midrule
 4 & 2424 & 1624 & $1.49\times$ & 1663 &  808 \\
 6 & 1924 & 1442 & $1.33\times$ & 1237 &  658 \\
 8 & 1623 & 1341 & $1.21\times$ &  974 &  565 \\
12 & 1353 & 1246 & $1.09\times$ &  739 &  481 \\
16 & 1803 & 1562 & $1.15\times$ &  962 &  573 \\
\bottomrule
\end{tabular}
\end{table}

Our benchmark configuration $r = 8$ yields $S = 1.70$ on H100 and
$S = 1.73$ on A100, both sitting on the slowly declining branch past
the optimum, indicating that a system with a higher GPU-to-CPU ratio
(fewer threads per GPU) would benefit more from GPU offloading.

All V-cycle arithmetic such as smoothing, residual, restriction, and
interpolation, already executes on the GPU
(Appendix~\ref{app:vcycle}).  The irreducible CPU-bound fraction
(${\sim}\,35$\,per\,cent) is dominated by MPI halo exchanges
({\sc isend}/{\sc irecv} + {\sc waitall}, ${\sim}\,10$ calls per
V-cycle iteration) and by the coarse-level solve that falls below
the GPU-dispatch threshold.  Nevertheless, the current hybrid
approach already provides a worthwhile $1.7\times$ MG speedup on
both H100 and A100 GPUs at $r = 8$, with the A100 reaching
$2.1\times$ at $r = 4$, reducing the total simulation time by
17--33\,per\,cent depending on the GPU-to-CPU ratio.

\subsubsection{GH200 extrapolation}

The NVIDIA GH200 Grace Hopper superchip pairs an H100 GPU
($N_{\mathrm{SM}} = 132$) with a 72-core Grace ARM CPU via
NVLink-C2C ($B \approx 450$\,GB/s), placing it firmly in the
compute-limited regime of panels~(a) and (b) of Fig.~\ref{fig:gpu_mg_model}.
Panels~(a) and (b) predict that the GH200 achieves
$1.94\times$ MG speedup at $r = 8$ (nearly the asymptotic limit
of $1.95\times$) because the NVLink-C2C bandwidth virtually
eliminates the PCIe transfer term
$V_{\mathrm{PCIe}}/B$ in equation~(\ref{eq:gpu_model}).

However, the GH200's default operating set is $r = 72$ (one GPU per 72~CPU cores), which lies deep on the declining branch of the $S(r)$ curve.  At such large $r$, the per-thread CPU workload $T_{\mathrm{CPU}}(r)$ is small, yet the GPU kernel time $T_{\mathrm{GPU}}^{\mathrm{kernel}}$ remains so fixed that the GPU contribution becomes a bottleneck rather than an accelerator.  The serial (single-stream) model predicts only $S(72) = 1.15\times$ for the GH200 (the purple marker in Panel (c) of Fig.~\ref{fig:gpu_mg_model}), a marginal improvement.

However, multi-stream pipelining recovers much of this loss.  When GPU
kernel execution overlaps with the CPU-fixed work, the effective
GPU time reduces from
$a \, T_{\mathrm{CPU}} + T_{\mathrm{GPU}}^{\mathrm{kernel}}$
to $\max(a \, T_{\mathrm{CPU}},\;
T_{\mathrm{GPU}}^{\mathrm{kernel}})$, yielding an overlap
speedup of $S_{\mathrm{overlap}}(72) = 2.01\times$ for the
GH200 (dashed curve).  This illustrates that on high-$r$
platforms such as the GH200, multi-stream overlap is not merely
an optimization but a prerequisite for the GPU offloading to
remain beneficial.

\subsubsection{Practical recommendations for users}

Based on the performance model and benchmarks across three GPU
platforms, we offer the following practical guidance:
\begin{enumerate}
\item \textbf{Enable only the MG Poisson GPU solver} (\code{gpu\_poisson=.true.})
      while keeping the Godunov solver on the CPU
      (\code{gpu\_hydro=.false.}).  The Godunov solver is PCIe-limited
      and currently \emph{slower} on the GPU ($0.69\times$).
\item \textbf{Use $r = 4$--$6$ threads per GPU} for maximum MG~AMR
      speedup (${\sim}2\times$).  At higher $r$, the GPU contribution
      diminishes as the per-thread CPU workload shrinks.
\item \textbf{Set \code{gpu\_auto\_tune=.true.}} (the default).  The
      auto-tuner benchmarks both paths during the first few steps and
      selects the faster option per level, avoiding the need for manual
      tuning.
\item \textbf{A 20\,per\,cent net improvement} on PCIe-connected
      hardware (H100, A100) translates to considerable savings in
      multi-month production campaigns, even though it falls short of
      the order-of-magnitude speedups sometimes associated with GPU
      computing.
\end{enumerate}

\subsubsection{Pathway to higher GPU speedup}

The performance model, equation~(\ref{eq:gpu_model}), shows that
$T_{\mathrm{CPU}}^{\mathrm{fixed}}$ sets the ultimate speedup
ceiling at $1/a \approx 2.8\times$.  Because
$T_{\mathrm{CPU}}^{\mathrm{fixed}}$ is dominated by MPI halo
exchanges that traverse the host CPU (first, device-to-host copy, \mpi\
send/recv over InfiniBand, and finally host-to-device copy) a net speedup
beyond ${\sim}\,2\times$ requires reducing or bypassing this
CPU-mediated communication path.

GPU-Direct RDMA (GDR) and NVLink-based GPU-to-GPU interconnects
can eliminate the host-memory staging entirely by allowing
\mpi\ libraries (e.g.\ NCCL, NVSHMEM, or CUDA-aware \mpi) to
transfer halo data directly between GPU memory spaces.  On
NVLink-connected multi-GPU nodes (e.g.\ DGX~A100, GH200), the
inter-GPU bandwidth reaches 600--900\,GB/s, two orders of magnitude
faster than the ${\sim}\,25$\,GB/s PCIe path that currently
dominates $T_{\mathrm{CPU}}^{\mathrm{fixed}}$.  In this regime the
halo exchange would overlap almost completely with the GPU stencil
computation, effectively reducing $a$ toward the coarse-level solve
fraction alone (${\lesssim}\,10$\,per\,cent).

Implementing GPU-direct halo exchange requires (i)~allocating the
AMR grid arrays in device memory as the primary copy, (ii)~packing
ghost-zone buffers on the GPU, and (iii)~invoking CUDA-aware \mpi\
or NCCL collective calls that operate directly on device pointers.
While these changes are non-trivial for an octree AMR code with
irregular communication patterns, they represent the most promising
path toward achieving the ${\gtrsim}\,5\times$ MG speedups that the
compute-limited regime of Fig.~\ref{fig:gpu_mg_model}(a) suggests
is physically attainable.

\section{DISCUSSIONS}
\label{sec:discussions}

\rev{The \curamses\ code base derives from the publicly distributed \ramses\ code
at commit \texttt{d92cd7d9} of the master branch of
\href{https://bitbucket.org/rteyssie/ramses}{bitbucket.org/rteyssie/ramses}
(8 November 2016, corresponding to \ramses\ version 3.10). Since the fork, the
descendant code base has been continuously developed and maintained in house
and was used to produce the Horizon Run~5 simulation \citep{Lee2021}. The
Horizon Run~5 patch reimplements the sub grid physics modules first
introduced for the Horizon-AGN simulation
\citep{Dubois2014} (black hole seeding, Bondi Hoyle accretion, dual mode AGN
feedback, and Type~II supernova feedback) on top of the post-2016 \ramses\
base. The original Horizon-AGN simulation itself predates this fork point
and was carried out with an earlier \ramses\ version so it should be
regarded as an ancestor of the sub grid \emph{physics} rather than of the
code base used here. The optimisations described in the present paper are
layered on top of that in house lineage of the software development.}

\rev{For the same reason, it is worth stating explicitly which of the optimizations 
described in this paper have public upstream counterparts and which do not. 
We benchmarked \curamses\ against the upstream master branch (commit \texttt{d7139a0}, 
1 November 2024) available at \href{https://bitbucket.org/rteyssie/ramses}{bitbucket.org/rteyssie/ramses}. 
A recursive bisection routine for domain decomposition is already implemented 
upstream in \texttt{amr/bisection.f90} as a binary tree that splits the domain 
along alternating coordinate directions. Our recursive \ksec\ scheme extends this 
approach from a binary ($K\!=\!2$) to a general $K$-way split at each tree level, 
with a hierarchical communication tree constructed on top of the resulting partition 
(Section~\ref{sec:ksection}). }

\rev{Additionally, an HDF5 I/O module (\texttt{pario/hdf5.f90}) is already distributed 
upstream\footnote{The upstream \texttt{pario/hdf5.f90} carries the header 
``Copyright (c) 2007--2011 IDRIS/CNRS. Author: Philippe Wautelet (IDRIS/CNRS). 
Distributed under the CeCILL 2.0 licence.'' The parallel I/O layer is documented 
in \citet{WauteletKestener2012}, a PRACE white paper which uses \ramses\ as one 
of its parallel I/O test cases on the CURIE supercomputer.}, meaning that HDF5 output 
capability itself is not a novel contribution. Instead, the key novelty in \curamses\ 
is the variable-$\nrank$ restart mechanism described in Section~\ref{sec:varcpu}. 
This framework allows a snapshot written by a specific rank count to be replayed by 
an arbitrary number of ranks via parallel hyperslab selection and recursive \ksec\ 
redistribution. Neither of these two core features---the general $K$-way recursive 
\ksec\ exchange tree and the variable-$\nrank$ HDF5 restart---is present in the 
upstream master branch at the time of writing.}

\rev{It is reasonable to question whether the hierarchical \ksec\ exchange 
introduced above offers advantages that a well-tuned MPI library would 
not already provide automatically. MPI offers a collective routine, 
\texttt{MPI\_ALLTOALLV}, designed precisely to enable every process to 
send varying amounts of data to all other processes. The library is free to 
select from several internal communication patterns depending on message sizes 
and process counts. However, the application retains no direct control over this 
selection, which is fundamentally based on the total volume of data exchanged 
rather than \emph{which} specific processes share data. }

\rev{In practice, our scheme behaves differently due to three key factors. First,
AMR ghost-zone exchanges are spatially sparse. For instance, in a typical
Cosmo1024 step across $\nrank = 128$ processes, an average process transmits
real data to only about six geometric neighbors. Conversely, a generic library
routine must treat the exchange as potentially dense, often falling back to
direct pairwise sends. Second, our \ksec\ exchange guarantees a predictable
number of messages (of order $K \log_K \nrank$) and a nearly constant data
volume per stage, regardless of the rank count or load imbalance while a generic
library routine cannot provide such guarantees for such irregular patterns.
Third, and more subtly, the speedup reported in Section~\ref{sec:bench_ghost}
originates not from the data transfer itself, but from \emph{constructing the
communication topology} (i.e., identifying neighbor pairs). Performing this
topology construction on the \ksec\ tree, using the Hilbert key as a spatial
sort, dominates the execution time. We are not aware of an equivalent
spatially-aware sorting step within any of the publicly documented collective
algorithms that library-level tuning could select.}

\rev{We emphasise, however, that the preceding comparison is restricted to the
\emph{publicly available} MPI implementations we were able to inspect
directly (MPICH, OpenMPI, UCC, and the published Intel MPI tuning interface).
It is entirely possible that one or more vendor-proprietary MPI stacks
shipped on specific supercomputers---for example the Cray MPICH variant on
HPE/Cray EX systems, the Tofu-aware MPI on Fugaku, or IBM Spectrum MPI on the
former Summit/Sierra platforms---already implement an analogous hierarchical,
sparsity-aware exchange internally. Such implementations are closed source
and not visible to the application layer and, therefore, we cannot confirm or
rule out their existence. The \ksec\ exchange is accordingly presented as a
portable, application-side realisation of this idea that is available wherever
a standard MPI is available, rather than as a categorically new algorithm.}

\rev{For full transparency we also note that the abstract communication
graph of the recursive \ksec\ tree is a $K$-ary butterfly, the same
abstract topology as the radix-$K$ Bruck schedule already shipped in
several MPI implementations (e.g.\ MPICH and Intel MPI). At $K\!=\!2$
the two schedules coincide up to a reversal of stage order, with
\ksec\ partners at level $\ell$ given by $r\oplus 2^{L-1-\ell}$ rather
than the standard Bruck $r\oplus 2^{\ell}$ (with $L=\log_{2}\nrank$).
The novelty of the present scheme is therefore not the abstract
butterfly itself, but four concrete design choices applied on top
of it. First, the stages are traversed from the most significant
digit downward, which is the natural direction for a recursive
spatial partition. Second, the rank permutation that determines
which physical process occupies which butterfly node is fixed by a
spatial sort key (either the Hilbert index or the \ksec\ index,
both of which preserve geometric locality) rather than by rank
arithmetic. Third, each pair of partners performs a \emph{mutual
symmetric} exchange in which a single peer is both the send and the
receive endpoint at every stage, in contrast to the one-way
rotation pattern of Bruck in which rank $r$ sends to
$r + d \cdot K^{j}$ and receives from $r - d \cdot K^{j}$ (mod
$\nrank$) so that the send and receive peers are distinct. Fourth,
the payload is sparsity-aware, transmitting only the data that
physically crosses each partition boundary instead of the dense
block rotation assumed by Bruck. The first two choices align the
butterfly with the geometry of the simulation, the third turns the
schedule into a sequence of symmetric peer-to-peer exchanges that
maps cleanly onto \texttt{MPI\_Sendrecv} primitives, and the fourth
converts the dense $O(\nrank^{2})$ communication pattern of a
generic all-to-all into the $O(K\log_{K}\nrank)$ messages with
roughly six geometric neighbours per rank observed in
Section~\ref{sec:bench_ghost}.}

\rev{The merged red-black optimization has one consequence worth noting 
explicitly. Because the black sweep reads boundary values from the 
previous iteration—whose contents inherently depend on the domain decomposition 
among processes—the residual at any given iteration is no longer bit-for-bit 
identical when running the same problem on different numbers of \mpi\ ranks. 
Nevertheless, the V-cycle reliably converges to the same numerical solution 
because the smoother acts as a preconditioner and the prolongation step 
refreshes the boundaries. Consequently, the converged solutions vary across 
rank counts only at a level comparable to the underlying discretization error 
of the Laplacian. }

\rev{For users who require strict bit-for-bit reproducibility across various rank numbers 
(e.g., when validating a new compiler build or a new architecture), we have 
introduced the namelist switch \code{mg\_merged\_rb} within \code{\&RUN\_PARAMS}. 
Setting \code{mg\_merged\_rb=.false.} reinserts the boundary exchange between 
the red and black sweeps, thereby eliminating the residual's dependence on the 
rank count. This replication comes at the cost of one extra boundary exchange per 
smoothing step (increasing the overhead to seven exchanges per level per iteration 
instead of five). }

\rev{A quantitative comparison of the merged and strict $\phi$ fields as a function of 
$\nrank$ calibrated against an FFTW3 reference solver is provided in 
Appendix~\ref{app:c4_phi}. The differences between the merged and strict variants 
as well as the rank-induced variances are three to four orders of magnitude 
smaller than the inherent truncation error of the multigrid scheme itself. 
Thus, the merged variant remains the highly recommended default for scientific production 
 runs except for the bit-for-bit comparison during further code development.}

\rev{The $8.3\times$ speedup of the FFTW3 solver reported in
Section~\ref{sec:bench_poisson} applies specifically to the regime $N=1024^3$ with a
few hundred ranks. Extrapolating to larger problems requires care for two reasons.
First, the FFTW3 slab decomposition performs a global all-to-all transpose per
Poisson solve, whose per-rank communication volume scales as $N^{3}/\nrank^{2/3}$
in the bandwidth-bisection limit \citep{Czechowski2012} and is therefore
interconnect-bound at large rank numbers. Second, the multigrid V-cycle exchanges
only with face-neighboring ranks and, therefore, its compute-bound cost scales as
$N^{3}/\nrank$, which is intrinsically more favorable in strong scaling.}

\rev{To quantify these competing effects, we benchmarked both solvers on a
$(N,\nrank)$ grid spanning $N\in\{512,1024,2048\}$ and
$\nrank\in\{64,128,256,512,1024,2048\}$ and fit standard cost models to the
measured per-step Poisson times (details are presented in
Appendix~\ref{app:fftw_mg_crossover}). The FFTW3 solver is faster than multigrid
by a factor of $8\!-\!12\times$ across every configuration we surveyed up to
$N=2048^{3}$ and $\nrank=2048$. Multigrid retains a clear advantage only for
deeply nested AMR zoom-in simulations, where the effective base grid is small
while many refinement levels are active. The V-cycle's surface-term cost, then,
depends only on the local patch resolution whereas FFTW3 still pays the global
slab transpose. The \code{use\_fftw} runtime switch selects between the two
solvers on a per-run basis.}

\rev{The feedback routine optimized in this section is the legacy thermal-plus-kinetic 
Type~II supernova implementation inherited from the Horizon-AGN production line 
\citep{Dubois2014}, whose core design traces back to the cosmological supernova 
feedback prescription of \citet{Dubois2008}. Several alternative feedback formulations 
coexist within the \ramses\ ecosystem. For instance, \citet{Kimm2014} introduced a 
mechanical supernova feedback model that explicitly queries the neighboring cells 
affected by each event while \citet{Agertz2013} developed a multi-channel prescription 
integrating radiation, stellar winds, and Type~II supernova ejecta. }

\rev{Because these newer routines natively operate using local neighbor lists or octree searches, 
they do not suffer from the global cells-times-supernovae scan that motivates the 
spatial binning optimization introduced here. Consequently, the optimization described 
in this work is specific to the legacy thermal Type~II routine and is not intended to 
supersede these alternative formulations. Instead, it restores competitive computational 
performance to the Horizon-AGN and HR5 lineages without altering the underlying physical model.}

\section{CONCLUSIONS}
\label{sec:conclusions}

We have presented \curamses, a set of algorithmic and implementation
improvements to the \ramses\ cosmological AMR code that collectively
address the key scaling bottlenecks, namely communication overhead, memory
consumption, solver efficiency, and hardware utilisation, encountered
in large-scale cosmological simulations. While previous efforts such as
OMP-RAMSES \citep{Lee2021} and RAMSES-yOMP \citep{Han2026} introduced
\mpi+OpenMP hybrid parallelism, \curamses\ extends this to a three-level
\mpi+OpenMP+CUDA paradigm targeting the heterogeneous architectures of
current petascale and upcoming exascale supercomputers, although we
stress that the GPU benefit is currently bandwidth-limited on
PCIe-connected platforms (Section~\ref{sec:hybrid}). The main
contributions are:
\begin{enumerate}
\item \textbf{Recursive \ksec\ domain decomposition}
  (Section~\ref{sec:ksection}): hierarchical \mpi\ exchange with
  $\Olog{\sum_\ell k_\ell}$ communications substituting all \code{MPI\_ALLTOALL}
  calls with memory-weighted load balancing reducing peak-to-mean
  imbalance to ${<}\,5$\,per\,cent.

\item \textbf{Morton key hash table} (Section~\ref{sec:morton}):
  $\Olog{1}$ neighbour lookup replacing the \code{nbor} array, with the
  hash table footprint comparable to the eliminated \code{nbor} array.

\item \textbf{Memory optimizations}: on-demand allocation saves
  more than 1\,GB per rank (Appendix~\ref{app:memory}).

\item \textbf{Multigrid solver optimizations}
  (Section~\ref{sec:multigrid}): merged red-black sweeps and fused
  residual-norm reduce MG communication by 44\,per\,cent, lowering
  the Poisson share from 50 to 40\,per\,cent.

\item \textbf{FFTW3 direct Poisson solver} (Section~\ref{sec:fftw3}):
  $8.3\times$ speedup at the base level with sparse P2P exchange.

\item \textbf{Feedback spatial binning} (Section~\ref{sec:feedback}):
  ${\sim}260\times$ SNII and $30\times$ AGN speedups.

\item \textbf{Variable-$\nrank$ restart} (Section~\ref{sec:varcpu}):
  HDF5 and binary I/O with arbitrary rank counts.

\item \textbf{GPU acceleration} (Section~\ref{sec:hybrid}):
  a hybrid CPU+GPU dispatch model with GPU-resident mesh data
  achieves a $1.70\times$ MG Poisson speedup on H100~NVL GPUs,
  but the Godunov solver is $0.69\times$ (slower) due to per-level
  23\,GB \code{cudaMemcpy} overhead.  The net improvement is a
  modest $1.21\times$ overall, indicating that PCIe-connected GPUs
  provide limited benefit for AMR codes with frequent inter-level
  data movement.
\end{enumerate}

All optimizations preserve conservation diagnostics.  Every global diagnostic ---
$e_{\rm cons}$, $e_{\rm pot}$, $e_{\rm kin}$, $e_{\rm int}$,
$m_{\rm tot}$, $n_\star$, $n_{\rm sink}$, and the total number of AMR
grids --- is verified to match the Hilbert-ordering reference run to
within roundoff.  The sole exception is the merged red-black sweep
(Section~\ref{sec:multigrid}), which intentionally uses stale boundary
values at inter-sweep ghost zones, introducing a 0.5\,per\,cent shift
in $e_{\rm cons}$ relative to the standard sequential sweep.  This shift
is comparable to the effect of changing the domain decomposition itself
and does not grow with integration time. The merged sweep is algebraically
equivalent to one full Gauss--Seidel iteration with a slightly different
update ordering, and the V-cycle converges to the same residual tolerance
regardless.  We have used a deterministic, cell-index-seeded random
number generator, which ensures that star formation stochasticity is
independent of the \dd\ and thread scheduling.

The techniques presented here --- Morton hash table, \ksec\
decomposition, spatial binning, and auto-tuning \mpi\ backends --- are
general and applicable to other octree AMR codes beyond \ramses.
While our multi-node benchmarks reach 2048~cores (32~nodes), the
\ksec\ exchange is designed for much larger concurrencies because each rank
communicates with at most $\sum_\ell k_\ell$ neighbours independent
of $\nrank$, and the hierarchical tree depth grows only as
$\Olog{\log \nrank}$.  For a typical binary tree ($k = 2$) with
$\nrank = 10^5$ ranks, this amounts to ${\sim}\,17$ exchange levels
with ${\sim}\,34$ partners, compared to the
$\Olog{\nrank^2}$ all-to-all traffic of the original Hilbert
implementation.  Verifying this scaling at ${\gtrsim}\,10^4$ ranks is
a priority for future work.
The FFTW3 base-level solver opens the door to non-standard
cosmologies (e.g.\ massive neutrinos, coupled dark energy) via
Fourier-space transfer functions without altering the real-space AMR
infrastructure.

The GPU acceleration experiment (Section~\ref{sec:hybrid}) provides
both a practical speedup and a cautionary lesson.  On current
\emph{PCIe-connected} hardware, the MG Poisson solver benefits from
GPU offloading ($1.70\times$ on H100~NVL, $2.06\times$ on A100~SXM4
at $r = 4$), but the Godunov solver is actually \emph{slower} on the
GPU ($0.69\times$) because per-level \code{cudaMemcpy} of the full
23\,GB mesh negates the kernel speedup.  The net overall improvement
of $1.21\times$ is modest but still translates to meaningful wall-clock
savings in production campaigns spanning months.  A generalised
performance model, equation~(\ref{eq:gpu_model_r}), shows that the
speedup is sensitive to the CPU-to-GPU thread ratio $r$, and the
current benchmark ($r = 8$) sits past the optimum of $r \approx 4$.
Notably, the AMR stencil kernel throughput scales with the number
of SMs rather than peak HBM bandwidth because the irregular octree
access pattern is latency-bound, a scaling confirmed by independent
A40 and A100 benchmarks within 5\,per\,cent.
Looking ahead, \emph{tightly coupled} architectures such as the
NVIDIA GH200 (NVLink-C2C, 450\,GB/s) are predicted to reach
${\sim}2\times$ MG speedup with multi-stream pipelining, and
GPU-direct halo exchange could push the ceiling beyond $5\times$.
Achieving efficient GPU utilisation for the full AMR V-cycle remains
an open challenge requiring device-residency of the restriction,
prolongation, and coarse-level solve operations.

Strong scaling on a single 64-core node reaches $33.9\times$ speedup
(53\,per\,cent parallel efficiency) while OpenMP thread scaling
achieves $5.8\times$ with 16 threads per rank.  Multi-node strong
scaling on the Cosmo1024 test ($3.6 \times 10^8$ grids,
$1.15 \times 10^9$ particles) achieves $26.5\times$ speedup from 1
to 32~nodes (83\,per\,cent parallel efficiency) with near-ideal scaling up
to 16~nodes (98\,per\,cent).  The variable-$\nrank$
restart capability supporting both HDF5 and binary formats, provides
operational flexibility for production campaigns where hardware
availability changes between runs.

\curamses\ is in production use for the next-generation cosmological simulation project and will be made publicly available upon completion of the benchmark campaign.

\section*{Acknowledgements}

We thank the anonymous referee for a careful and constructive report whose suggestions substantially improved the clarity and scope of this paper. 
This work was supported by the Korea Institute for Advanced Study. Computational resources were provided by the KIAS Center for Advanced Computation.  JK is supported by KIAS Individual Grants (KG039603) at the Korea Institute for Advanced Study. JK acknowledges the support of the National Research Foundation of Korea (NRF) grant funded by the Korea government (MSIT)(2022M3K3A1093827). The author thanks Romain Teyssier for the public release of the \ramses\ code and the \ramses\ developer community for continued maintenance and improvements. The author also thanks Yohan Dubois for sharing his Horizon-AGN version of the \ramses\ code, on which the present in-house lineage is based, and Taysun Kimm for the chemical enrichment model (alpha-element and stellar-wind yields) used in this work. The author further acknowledges Christophe Pichon, S\'ebastien Peirani, and Julien Devriendt for their contributions to design discussions on the Horizon-AGN code and simulation.

\section*{Data Availability}

The modified code is available at
\url{https://github.com/kjhan0606/cuRAMSES}. Test configurations
and analysis scripts will be shared upon reasonable request to the
author.

\bibliographystyle{mnras}

\appendix

\section{MORTON KEY ENCODING DETAILS}
\label{app:morton}

The Morton key interleaving for a single coordinate value $v$ with
$B$ bits per coordinate ($B = 21$ for 64-bit keys, $B = 42$ for 128-bit keys)
is computed by the following bit-manipulation loop:

\begin{algorithm}[H]
\caption{Morton key encoding of $(i_x, i_y, i_z)$}
\label{alg:morton_encode}
\begin{algorithmic}[1]
\REQUIRE Integer coordinates $(i_x, i_y, i_z)$
\ENSURE $3B$-bit Morton key $M$
\STATE $M \leftarrow 0$
\FOR{$b = 0$ \TO $B - 1$}
  \STATE $M \leftarrow M \,|\, (\text{bit}_b(i_x) \text{\texttt{<<}} 3b)$
  \STATE $M \leftarrow M \,|\, (\text{bit}_b(i_y) \text{\texttt{<<}} (3b+1))$
  \STATE $M \leftarrow M \,|\, (\text{bit}_b(i_z) \text{\texttt{<<}} (3b+2))$
\ENDFOR
\end{algorithmic}
\end{algorithm}

The neighbour key computation decodes, shifts the appropriate
coordinate, applies periodic wrapping, and re-encodes:

\begin{algorithm}[H]
\caption{Morton neighbour key in direction $j$}
\label{alg:morton_neighbor}
\begin{algorithmic}[1]
\REQUIRE Morton key $M$, direction $j$, grid counts $(n_x, n_y, n_z)$
\ENSURE Neighbour Morton key $M'$ (or $-1$ if out of bounds)
\STATE $(i_x, i_y, i_z) \leftarrow$ \textsc{Decode}($M$)
\STATE Adjust $i_d$ by $\pm 1$ according to direction $j$
\STATE Apply periodic wrapping: $i_d \leftarrow i_d \bmod n_d$
\STATE $M' \leftarrow$ \textsc{Encode}$(i_x, i_y, i_z)$
\end{algorithmic}
\end{algorithm}

\section{PER-LEVEL HASH TABLE AND REPLACEMENT FUNCTIONS}
\label{app:hashtable}

We maintain one open-addressing hash table (probing consecutive slots
on collision) per AMR level, mapping Morton keys $M$ to grid indices
\code{igrid}.

The hash function first reduces the key to 64 bits (for
128-bit keys, an XOR fold of the upper and lower halves, and for 64-bit
keys a direct cast), then applies multiplicative hashing
with an additional mixing step:
\begin{align}
t   &= \bigl[(h \times \phi_1) \text{\texttt{\textasciicircum}} (h \text{\texttt{>>}} 16)\bigr] \times \phi_2,
\nonumber \\
h(M) &= t \text{\texttt{\textasciicircum}} (t \text{\texttt{>>}} 13),
\label{eq:hash_func}
\end{align}
where $h$ is the 64-bit reduced key,
$\text{\texttt{\textasciicircum}}$ is bitwise XOR,
$\text{\texttt{>>}}$ is a logical right bit-shift,
$\phi_1 = 2654435761$ and $\phi_2 = \texttt{0x9E3779B97F4A7C15}$
are constants chosen for good bit mixing, and the table capacity is
always a power of two to allow bitmask modular arithmetic. Collisions
are resolved by linear probing, and the fill fraction is kept below 0.7 by
automatic rehashing (doubling capacity).

The hash table is maintained incrementally.
\begin{itemize}
\item \code{morton\_hash\_insert} is called during \code{make\_grid\_coarse} and \code{make\_grid\_fine}.
\item \code{morton\_hash\_delete} is called during \code{kill\_grid}.
\item A full rebuild is performed after defragmentation (\code{morton\_hash\_rebuild}).
\end{itemize}

A companion array \code{grid\_level(igrid)} stores the AMR level of
each grid, enabling Morton key computation from the grid index alone.

Two wrapper functions provide drop-in replacements for the original
\code{nbor}-based access patterns.

\begin{itemize}
\item \code{morton\_nbor\_grid(igrid, ilevel, j)} returns the grid
  index of the same-level neighbour in direction $j$, replacing the
  pattern \code{son(nbor(igrid, j))}. The implementation computes the
  Morton key, shifts by direction, and looks up the result in the hash table.

\item \code{morton\_nbor\_cell(igrid, ilevel, j)} returns the father
  cell index of the neighbour, replacing the pattern
  \code{nbor(igrid, j)}. For level 1 it returns the coarse cell index
  directly. For finer levels it computes the parent grid via the
  hash table at level $l-1$ and the octant index from the coordinate
  parity.
\end{itemize}

The \code{nbor} array is reduced to \code{allocate(nbor(1:1, 1:1))}
, effectively eliminated while maintaining compilation compatibility
with any remaining references.

\section{MEMORY SAVINGS DETAILS}
\label{app:memory}

\subsection{Hilbert Key and Histogram Array Elimination}
\label{app:mem_hilbert}

When using \ksec\ ordering, the Hilbert key array
\code{hilbert\_key(1:ncell)} is no longer needed for domain
decomposition. We replace it with \code{allocate(hilbert\_key(1:1))},
saving $16\,\ngridmax\,\twotondim$ bytes
under \code{QUADHILBERT} (128-bit keys stored as two 64-bit integers).
For $\ngridmax = 5\,\mathrm{M}$, this is approximately 640\,MB.

The defragmentation routine, which previously required Hilbert keys
for reordering, uses a local scratch array (\code{defrag\_dp})
allocated only during the defragmentation pass and immediately
deallocated.

Similarly, the arrays \code{bisec\_ind\_cell} and \code{cell\_level},
each of size $\ngridmax\,\twotondim$ default-kind integers
(\rev{4 bytes each}), are used exclusively during load balancing to
build the bisection histogram.
We allocate them on entry to \code{init\_bisection\_histogram} and
deallocate them after \code{cmp\_new\_cpu\_map} returns, saving
\rev{$8\,\ngridmax\,\twotondim \approx 320$\,MB}
for $\ngridmax = 5\,\mathrm{M}$ (160\,MB per array). Since load
balancing occurs only every \code{nremap} coarse steps, these arrays
are absent during the vast majority of the simulation.

\subsection{Memory Savings Summary}
\label{app:mem_summary}

\rev{The original \code{nbor} cost is $24\,\ngridmax$ bytes
(120\,MB at $\ngridmax = 5\,\mathrm{M}$). The replacement
\code{grid\_level} array adds $4\,\ngridmax$ bytes (20\,MB), and the
multi-level hash table adds $28\,N_{\rm grids}$ bytes under
\code{MORTON128} (17\,$N_{\rm grids}$ bytes for the 64-bit Morton
build). With $N_{\rm grids} \ll \ngridmax$ in practice (typical
occupancy 30--60\,per\,cent), the net neighbour-related saving is
about 30--60\,MB per rank, roughly a 25--50\,per\,cent reduction
relative to the original \code{nbor} cost. The dominant memory benefit
of the morton-octree refactor comes not from this neighbour-related
saving alone, but from the concurrent elimination of the
\code{hilbert\_key} and the load-balancing scratch arrays summarised
in Table~\ref{tab:mem_breakdown}.}

The computational cost of a hash lookup is $\Olog{1}$ expected time,
with worst-case linear probing bounded by the fill fraction. In
practice, the precomputed neighbour caches described in
Appendix~\ref{app:mg_details} alleviate any per-lookup overhead in the
performance-critical Poisson solver.

\rev{The combined peak memory savings for $\ngridmax = 5\,\mathrm{M}$
approach 1.1\,GB per rank under \code{QUADHILBERT}, including
\code{nbor} removal (120\,MB), \code{hilbert\_key} elimination
(640\,MB), on-demand \code{bisec\_ind\_cell} and \code{cell\_level}
(160\,MB each, 320\,MB combined at the load-balancing peak), and
defrag scratch (40\,MB), minus the hash table overhead
(${\sim}50$\,MB). For the standard 64-bit Morton build the
\code{hilbert\_key} saving is 320\,MB, lowering the cumulative figure
to ${\sim}0.75$\,GB per rank.}

\rev{For convenience of comparison with the main hydrodynamics and gravity
state payload of about 104 bytes per cell
($\mathrm{NVAR}=8$ hydrodynamic variables stored as double precision
in \code{uold}, plus $\rho$, $\phi$, $f_x$, $f_y$, $f_z$),
Table~\ref{tab:mem_breakdown} summarises the per array cost in
both absolute units and as a percentage of that state vector.
Eliminating \code{nbor} saves about $2.9\,\%$ of the per cell state
vector at all times. The 64 bit Hilbert key (default build) adds
roughly another $7.7\,\%$, while the extended precision
\code{QUADHILBERT} build doubles that to about $15\,\%$. The
load balancing scratch arrays \code{bisec\_ind\_cell} and
\code{cell\_level} add a transient ${\sim}7.7\,\%$ in aggregate that
is now absent for the vast majority of the simulation. The largest
single saving therefore comes from the \code{hilbert\_key} array
under \code{QUADHILBERT}; \code{nbor} elimination and the transient
scratch arrays together provide a comparable additional saving. The
cumulative benefit is what allows deeper AMR hierarchies, or a larger
$\ngridmax$, within the same hardware budget.}

\begin{table*}
\centering
\caption{\rev{Memory cost of individual arrays in \curamses\ before
the optimisations, expressed as bytes per cell (each oct holds
$2^{n_{\rm dim}}\!=\!8$ cells in three dimensions) and as a percentage
of the main hydrodynamics and gravity state vector
($\mathrm{NVAR}=8$ hydrodynamic plus 5 gravity variables, about
104 bytes per cell). Absolute memory is per rank for
$\ngridmax = 5\,\mathrm{M}$. Items marked ``transient'' are present
only during load balancing remap steps (once every \code{nremap}
coarse steps).}}
\label{tab:mem_breakdown}
{\color{black}
\begin{tabular}{lrrrr}
\hline
Array & B/oct & B/cell & \% & MB/rank \\
\hline
\code{nbor} ($2\,n_{\rm dim} \times 4$\,B) & 24 & 3 & 2.9 & 120 \\
\code{hilbert\_key} (64-bit) & 64 & 8 & 7.7 & 320 \\
\code{hilbert\_key} (\code{QUADHILBERT}) & 128 & 16 & 15.4 & 640 \\
\code{bisec\_ind\_cell} (transient) & 32 & 4 & 3.8 & 160$^{\dagger}$ \\
\code{cell\_level} (transient) & 32 & 4 & 3.8 & 160$^{\dagger}$ \\
\code{defrag\_dp} (scratch) & --- & --- & --- & 40 \\
hash-table overhead (post-opt.) & --- & --- & --- & $-50$ \\
\hline
\multicolumn{4}{l}{Cumulative savings (64-bit, peak)} & ${\sim}0.75$\,GB \\
\multicolumn{4}{l}{Cumulative savings (\code{QUADHILBERT}, peak)} & ${\sim}1.07$\,GB \\
\hline
\multicolumn{5}{l}{\footnotesize $^{\dagger}$Each of the two transient
arrays costs 160\,MB independently; cumulative} \\
\multicolumn{5}{l}{\footnotesize transient footprint is 320\,MB,
present only during \code{nremap} steps.} \\
\end{tabular}
}\end{table*}

The freed memory enables deeper AMR hierarchies or larger $\ngridmax$
within the same hardware budget.

A diagnostic routine \code{writemem\_minmax} reports
the minimum and maximum resident set size across all ranks at each
coarse step, providing runtime verification of the memory savings.

\section{MULTIGRID SOLVER IMPLEMENTATION DETAILS}
\label{app:mg_details}

\subsection{Neighbour Grid Precomputation}
\label{app:mg_nbor_precompute}

The Gauss--Seidel (GS) smoother and residual computation both require
access to the six Cartesian neighbours of each grid. In the Morton
hash table approach (Section~\ref{sec:morton}), each neighbour lookup
involves a hash table query. While individual lookups are $\Olog{1}$,
the GS kernel accesses 6 neighbours per grid, 8 cells per grid, and
typically 4--5 V-cycle iterations, resulting in hundreds of hash
lookups per grid per solve.

We precompute all neighbour grids into a contiguous array before
entering the V-cycle iteration loop,
\begin{equation}
\text{nbor\_grid\_fine}(j,\, i) = \text{morton\_nbor\_grid}(
\text{igrid\_amr}(i),\, l,\, j),
\label{eq:nbor_precompute}
\end{equation}
where $l$ is the AMR level being solved, $\text{igrid\_amr}(i)$ is
the AMR grid pointer for the $i$-th active grid at level $l$,
$j = 0, 1, \ldots, 6$ (where $j = 0$ stores the grid's own AMR
index), and $i = 1, \ldots, N_{\rm grid}$. This
array is allocated before the iteration loop and deallocated after,
so its memory overhead is transient.

\subsection{Merged Red-Black Exchange and Fused Kernels}
\label{app:mg_merged_rb}

We remove the inter-sweep ghost exchange between red and black passes
(Red $\to$ Black $\to$ Exchange instead of Red $\to$ Exchange $\to$
Black $\to$ Exchange), reducing the per-iteration exchange count from 9
to 5.  Boundary cells in the black sweep use values from the previous
iteration. Because the MG solve serves as a preconditioner for the CG
outer iteration, this relaxed synchronisation does not affect
convergence.

We also remove two unnecessary residual exchanges per iteration,
reducing the total from 9 to 5 exchange calls per iteration, a
44\,per\,cent reduction in MG communication volume.

The same optimization is applied to the coarse-level solver
(direct solve, pre-smoothing, post-smoothing), where the merged
red-black pattern similarly halves the exchange count.

The MG algorithm requires both the residual $r = f - \nabla^2_h\phi$ and its
$L^2$ norm $\|r\|_2^2$ at specific points in the V-cycle. In the
original code, these are computed in separate passes. We add an
optional \code{norm2} argument to \code{cmp\_residual\_mg\_fine}:
when present, the norm is accumulated during the same loop that
computes the residual, saving one full grid traversal.
Since the subroutine is \code{external} (not module-contained), callers
must include an \code{interface} block to enable the optional-argument
dispatch.

The GS fast-path computation involves a division by
$2 N_{\rm dim} = 6$:
\begin{equation}
\phi_{\rm new} = \frac{\sum_j \phi_j - h^2 f}{2 N_{\rm dim}}.
\label{eq:gs_update}
\end{equation}
We replace the division by $2N_{\rm dim}$ with a multiplication by
the precomputed reciprocal $1/(2N_{\rm dim})$, which is faster
on most architectures.

\section{\rev{ACCURACY OF THE MULTIGRID POTENTIAL UNDER THE MERGED
RED BLACK SMOOTHER AND DIFFERENT DOMAIN DECOMPOSITIONS}}
\label{app:c4_phi}

\rev{To measure how much the converged Poisson potential $\phi$ is
affected by skipping the boundary exchange between the red and the
black sweeps inside the merged smoother
(Appendix~\ref{app:mg_merged_rb}), we ran a controlled experiment in
which we varied three sets of choice independently. These are the number of \mpi\
ranks, the way the box is divided between those ranks, and which
smoother variant is used. We then compared the resulting $\phi$
fields against a reference computed with a direct FFTW3 solver.}

\rev{Starting from a single Cosmo1024 restart at $z\!\approx\!5.2$
($a\!=\!0.16$), we advanced one coarse step in nine production
configurations. We used three rank numbers ($N_\textrm{rank}=$ 4, 8, and 12), two ways of
dividing the box between ranks (the \ksec\ ordering and the standard
Hilbert ordering), and two smoother variants. These are the merged red black
sweep and the strict version that puts the boundary exchange back in
between (controlled by the namelist switch
\code{mg\_merged\_rb}\,$\in\{\code{.true.},\,\code{.false.}\}$). We
also ran a tenth configuration using the FFTW3 direct solver
(\code{use\_fftw=.true.}, Section~\ref{sec:fftw3}) at the base grid
$N\!=\!512^3$, which gives us a ground truth potential
$\phi_{\rm FFTW}$ that we can measure the multigrid solutions
against at the level $\ell_\textrm{base}=9$.  After the coarse step we wrote $\phi$ to disk via the
parallel HDF5 path (Section~\ref{sec:varcpu}) and computed, for every
cell, the residual $|\Delta\phi|/|\phi_{\rm FFTW}|_{\rm rms}$, where
$|\phi_{\rm FFTW}|_{\rm rms}$ is the root mean square of the FFTW3
reference potential at the base level.}

\begin{figure*}
\centering
\includegraphics[width=\textwidth]{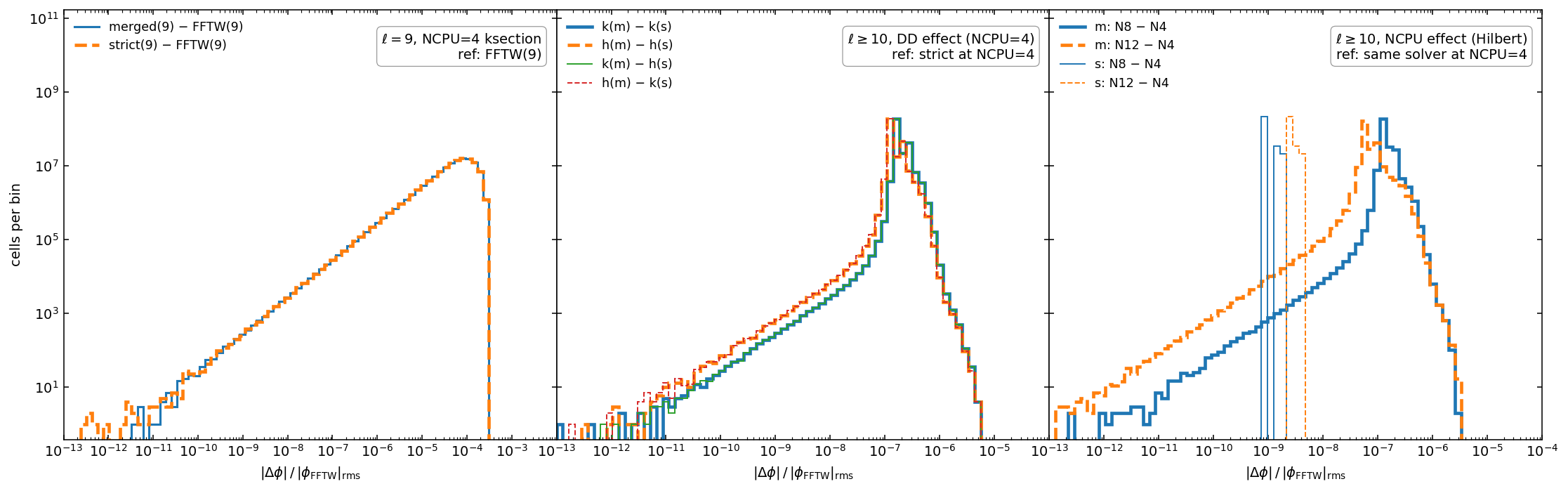}
\caption{\rev{Histograms of $|\Delta\phi|$, normalised by the
root mean square of the FFTW3 reference potential
$|\phi_{\rm FFTW}|_{\rm rms}$. \emph{Left}, the base level
($\ell_\textrm{base}\!=\!9$) multigrid potential compared to the FFTW ground truth
at $\nrank\!=\!4$ with the \ksec\ ordering. \emph{Middle}, the
merged minus strict differences at refined levels $\ell\!\geq\!10$
for all four combinations of ordering and smoother variant
(letters in the legend are $k$ for \ksec, $m$ for merged and $s$ for strict) at
$\nrank\!=\!4$. \emph{Right}, how each smoother varies with the
number of ranks at $\ell\!\geq\!10$, using the Hilbert ordering and
taking the $\nrank\!=\!4$ run as the self reference for each
smoother.}}
\label{fig:c4_phi_3panel}
\end{figure*}

\rev{Three key observations characterize the empirical reproducibility of the merged 
red-black optimization, all of which are clearly visible in Fig.~\ref{fig:c4_phi_3panel}.}

\rev{First, the reproduction error is a valid concern but remains small in amplitude. 
The relative difference between the merged and strict $\phi$ fields at refined levels 
(middle panel) peaks at $|\Delta\phi|/|\phi_{\rm FFTW}|_{\rm rms}\!\sim\!2\times10^{-7}$. 
Furthermore, all four combinations of loop ordering and smoother variants exhibit a single 
clustered distribution. The discrepancy introduced by reading neighboring boundary cells 
from the previous iteration is therefore a deterministic, systematic effect that correlates 
with the domain decomposition among ranks rather than random noise, as anticipated.}

\rev{Second, the strict mode achieves near bit-for-bit reproducibility across varying 
rank counts. With \code{mg\_merged\_rb=.false.} (thin curves in the right panel), 
increasing the rank count from 4 to 8 or 12 alters $\phi$ by at most 
$\sim\!3\times10^{-9}\,|\phi_{\rm FFTW}|_{\rm rms}$, which corresponds to the level 
expected from variations in floating-point summation order within \mpi\ reduction 
operations. In contrast, the same rank count survey using the merged smoother (thick
curves) produces differences that are four orders of magnitude larger, approaching 
$\sim\!10^{-7}\,|\phi_{\rm FFTW}|_{\rm rms}$. This result directly quantifies the 
residual's dependence on the rank count.}

\rev{The third observation is that the inherent truncation error of
the multigrid scheme dominates both of the systematic effects above.
The left panel compares the multigrid potential at the base level
$\ell\!=\!9$ against the FFTW direct reference. Both the merged and
the strict variants deviate from $\phi_{\rm FFTW}$ by
$|\Delta\phi|/|\phi_{\rm FFTW}|_{\rm rms}\!\sim\!10^{-4}$
(median $9.8\!\times\!10^{-5}$, $p_{99}\!\approx\!3\!\times\!10^{-4}$),
which means they overlap each other while both sit about three orders
of magnitude further from the FFTW solution than they are from each
other. This residual reflects the convergence of the V cycle at the
default tolerance $\epsilon_{\rm MG}\!=\!10^{-4}$ and is inherent to
the multigrid method itself, not a consequence of the merged
optimisation.}

\rev{Putting the three panels together, the merged red black
optimisation does break bit-for-bit reproducibility across rank
counts and the resulting error map is correlated with the
domain decomposition. But the amplitude of this systematic is three to four
orders of magnitude smaller than the inherent truncation error of
the same solver measured against the FFTW reference. For science
that is sensitive to absolute Poisson accuracy at the
$10^{-4}$ level, the appropriate remedy is to tighten the V cycle
tolerance $\epsilon_{\rm MG}$, or to enable the FFTW3 direct
base level solve via \code{use\_fftw=.true.}, rather than to disable
the merged optimisation. We therefore keep \code{mg\_merged\_rb=.true.}
as the default for production runs while
\code{mg\_merged\_rb=.false.} remains available for strict
bit reproducibility validation (for instance, regression testing
across compilers, \mpi\ libraries, or new hardware).}

\begin{figure*}
\centering
\includegraphics[width=\textwidth]{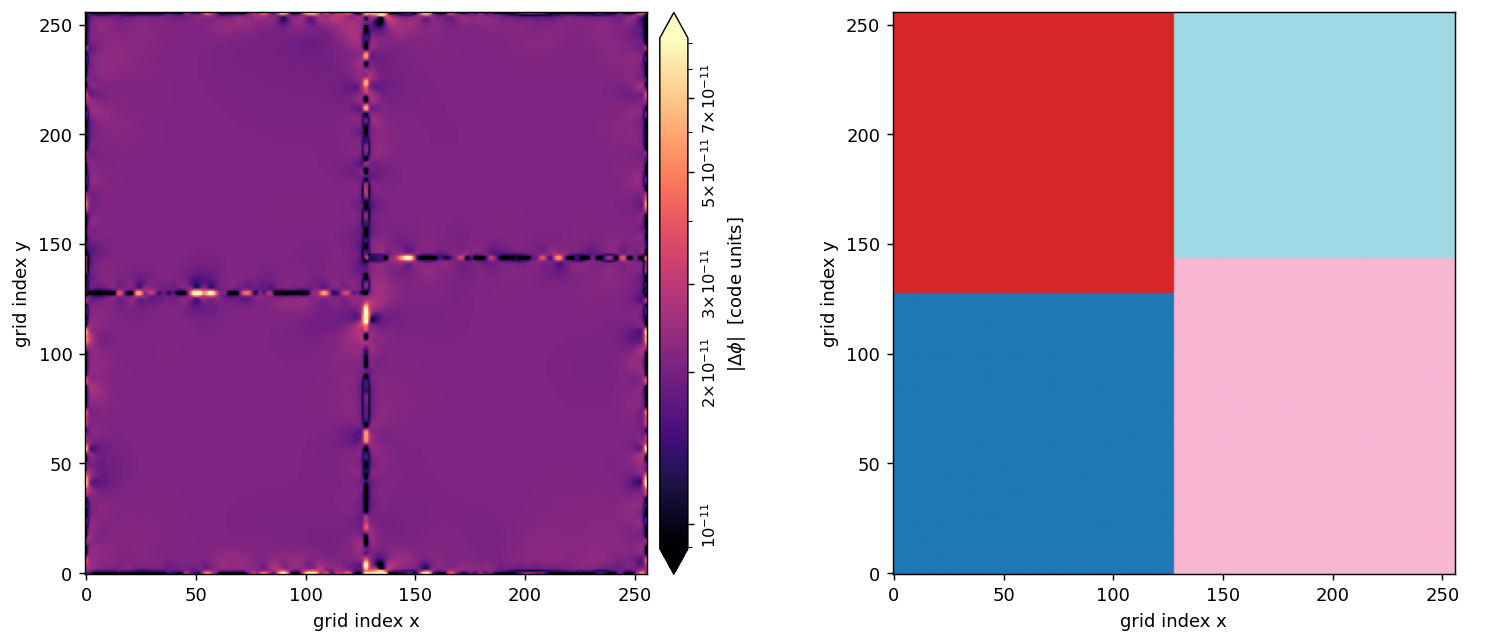}
\caption{\rev{Direct visualisation of how the merged smoother's
boundary lag error tracks the domain decomposition, for
$\nrank\!=\!4$ with the \ksec\ ordering. \emph{Left,}
$|\Delta\phi|\equiv|\phi_{\rm merged}-\phi_{\rm strict}|$ on the slice
$z\!=\!128/256$ at level $\ell_\textrm{base}\!=\!9$, on a logarithmic colour scale.
The interior sits at $\sim\!2\!\times\!10^{-11}$, while a cross
pattern of streaks along the rank boundaries reaches
$\sim\!4\!\times\!10^{-10}$. For reference, the FFTW solution on the
full level 9 box has $\phi_{\rm fftw,\,rms}\!\simeq\!1.18\!\times\!10^{-4}$
(code units), so the brightest streaks are $\sim\!3\!\times\!10^{-6}$
in relative amplitude and the bulk floor is $\sim\!2\!\times\!10^{-7}$.
\emph{Right,} the rank that owns each cell of the same slice (rank~1
in blue, rank~2 in red, rank~3 in pink, rank~4 in cyan). The
asymmetric $y\!=\!144$ split between ranks~3 and~4 reflects the
memory weighted load balancing of Section~\ref{sec:ksec_loadbal}.
The $|\Delta\phi|$ streaks line up with every rank boundary,
including the asymmetric one, confirming that the systematic
introduced by \code{mg\_merged\_rb=.true.} is deterministic and
correlated with the domain decomposition.}}
\label{fig:c4_stale_neighbor_slice}
\end{figure*}

\rev{To make the correlation between the systematic and the
domain decomposition (noted in the first of the three observations
above) directly visible,
Fig.~\ref{fig:c4_stale_neighbor_slice} shows a single $z$ slice of
$|\Delta\phi|\equiv|\phi_{\rm merged}-\phi_{\rm strict}|$ at level
$\ell_\textrm{base}\!=\!9$ ($z\!=\!128/256$) alongside the rank that owns each cell
of the same slice. Most of the slice sits at
$|\Delta\phi|\!\approx\!2\!\times\!10^{-11}$ (deep purple), but bright
streaks along $x\!=\!128$, $y\!=\!128$, and $y\!=\!144$ rise to
$\sim\!4\!\times\!10^{-10}$, a localised amplification of more than
an order of magnitude. To put these numbers on an absolute scale,
the FFTW reference has $\phi_{\rm fftw,\,rms}\!\simeq\!1.18\!\times\!10^{-4}$
over the full level 9 box.  Therefore, the bulk floor is
$\sim\!2\!\times\!10^{-7}$ and the brightest streaks are
$\sim\!3\!\times\!10^{-6}$ relative to the FFTW solution. The streaks
align exactly with the four \ksec\ rank boundaries shown in the
right panel including the asymmetric $y\!=\!144$ split between
ranks~3 and~4 that the load balancer chose so as to equalise the work
per rank. This is the geometric signature predicted by the
mechanism described above. Along every rank boundary the black sweep
of the merged smoother reads its red neighbours from the previous
iteration and, therefore, the discrete Laplacian effectively sees a one iteration
phase lag exactly at the boundary, and the extra residual accumulates
as a line in cell space that tracks the rank cut.}

\section{\rev{FFTW3 VERSUS MULTIGRID POISSON SOLVER: COST-MODEL FIT}}
\label{app:fftw_mg_crossover}

\rev{Section~\ref{sec:discussions} argues that the FFTW3 direct solver and the
multigrid V-cycle scale differently with rank count $\nrank$. The slab FFT is
limited by the global all-to-all transpose whereas the V-cycle exchanges only
with face neighbours and is therefore intrinsically more favourable in strong
scaling. In this appendix we quantify both costs with standard analytic models
fit to the measured data.}

\rev{We measured both solvers on a $(N,\nrank)$ grid covering
$N\in\{512,1024,2048\}$ and $\nrank\in\{64,128,256,512,1024,2048\}$, omitting
configurations with $\nrank>N^{2}/4$, where the slab decomposition becomes
degenerate (one rank carries fewer than four planes). Each configuration starts
from a Planck-cosmology initial condition and is integrated for two coarse time
steps. We report the per-step wall-clock time spent in the Poisson solver,
averaged across MPI ranks in Table~\ref{tab:fftw_mg_sweep}. The FFTW3 solver outperforms multigrid by a factor of
$8\!-\!12\times$ across the entire surveyed grid.}

\begin{table}
\centering
\caption{\rev{Per-step solver base time (seconds) for the FFTW3 direct Poisson solver
and the multigrid V-cycle, measured on the grammar cluster (Intel Xeon Sapphire
Rapids, EDR InfiniBand). A dash indicates a configuration omitted from the survey
($\nrank>N^{2}/4$).}}
\label{tab:fftw_mg_sweep}
\begin{tabular}{r r r r r}
\hline
$N$ & $\nrank$ & $T_{\rm FFTW3}$ & $T_{\rm MG}$ & $T_{\rm MG}/T_{\rm FFTW3}$ \\
\hline
 512 &   64 &   9.02 &  108.82 & 12.06 \\
 512 &  128 &   4.46 &   50.81 & 11.40 \\
 512 &  256 &   2.40 &   27.99 & 11.66 \\
 512 &  512 &   1.57 &   12.90 &  8.23 \\
1024 &  128 &  39.19 &  394.54 & 10.07 \\
1024 &  256 &  22.11 &  238.26 & 10.78 \\
1024 &  512 &  11.40 &  115.90 & 10.16 \\
1024 & 1024 &   6.67 &   56.10 &  8.41 \\
2048 &  512 &  88.55 & 1022.42 & 11.55 \\
2048 & 1024 &  47.77 &  528.95 & 11.07 \\
2048 & 2048 &  32.55 &  268.99 &  8.26 \\
\hline
\end{tabular}
\end{table}

\rev{To extrapolate to rank counts and resolutions outside the surveyed grid, we
fit standard cost models from the parallel-computing literature to each solver.
For a slab-decomposed 3-D FFT the total per-call time is the sum of (i) the
local 1-D FFT work, scaling as $N^{3}\log_{2}(N^{3}/\nrank)/\nrank$,
(ii) the global all-to-all transpose, which in the bisection-bandwidth limit
\citep{Czechowski2012} moves a volume scaling as $N^{3}/\nrank^{2/3}$ per rank,
and (iii) a constant latency floor associated with collective initialisation and
FFTW plan look-up \citep{Foster1995,Frigo2005}. We therefore adopt}
\begin{equation}
T_{\rm FFTW3}(N,\nrank) = \alpha_{\rm F}\,\frac{N^{3}}{\nrank^{2/3}}
        + \beta_{\rm F}
        + \gamma_{\rm F}\,\frac{N^{3}\log_{2}(N^{3}/\nrank)}{\nrank}.
\label{eq:fftw_cost}
\end{equation}

\rev{For the V-cycle, the standard textbook decomposition
\citep{Briggs2000,Trottenberg2001} writes the per-cycle cost as the sum of a
compute term $\propto N^{3}/\nrank$ (red-black Gauss-Seidel sweeps summed over
$\log_{2}N$ refinement levels), a surface-exchange term
$\propto N^{2}\log_{2}N/\nrank^{2/3}$, and a latency term
$\propto \log_{2}\nrank\,\log_{2}N$. In our measured regime the
$N^{2}\log_{2}N/\nrank^{2/3}$ and $\log_{2}\nrank\,\log_{2}N$ contributions are
two to three orders of magnitude smaller than the leading compute term, and a
joint fit returns coefficients consistent with zero. We therefore retain only
the leading-order, compute-bound expression as}
\begin{equation}
T_{\rm MG}(N,\nrank) = \alpha_{\rm M}\,\frac{N^{3}}{\nrank},
\label{eq:mg_cost}
\end{equation}
\rev{which captures the data with no loss of accuracy compared with the full
three-term form.}

\begin{figure*}
\centering
\includegraphics[width=\textwidth]{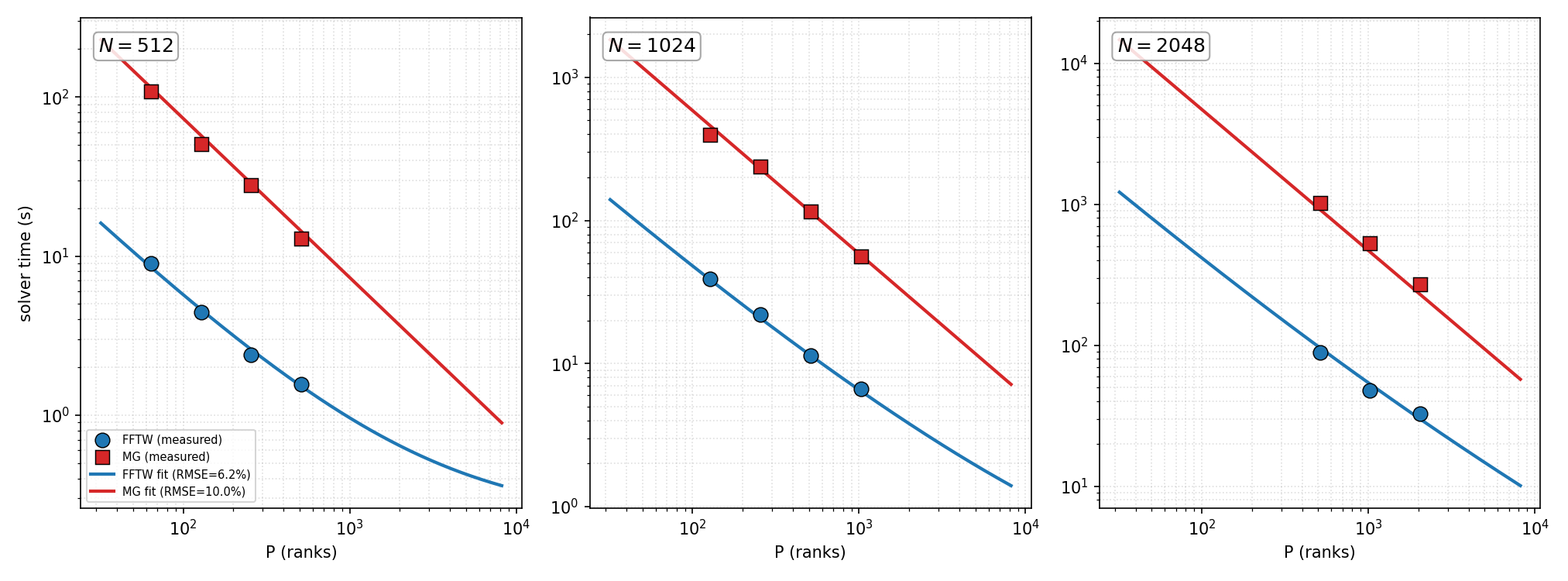}
\caption{\rev{Scaling relations for the
FFTW3 direct Poisson solver (blue circles) and the multigrid V-cycle (red
squares), shown for the three production resolutions $N=512^{3}$, $1024^{3}$,
$2048^{3}$. Solid curves are the best fits of Equations~(\ref{eq:fftw_cost})
and~(\ref{eq:mg_cost}). The fit RMS residuals are $6.2\%$ and $10.0\%$
respectively. The FFTW3 solver is uniformly $8\!-\!12\times$ faster than
multigrid across the surveyed regime.}}
\label{fig:c5_fftw_mg_fit}
\end{figure*}

\rev{We fit both models to the $N\geq 512^{3}$ entries of Table~\ref{tab:fftw_mg_sweep}
by minimising the sum of squared log-residuals
$\sum_{i}[\log T_{i}^{\rm pred}-\log T_{i}^{\rm meas}]^{2}$ with non-negative
coefficients. The best-fit values are measured as,}
\[
\begin{aligned}
& \alpha_{\rm F}=3.45\times 10^{-7}\;{\rm s},\;
  \beta_{\rm F}=0.22\;{\rm s},\;
  \gamma_{\rm F}=1.23\times 10^{-7}\;{\rm s}, \\
& \alpha_{\rm M}=5.46\times 10^{-5}\;{\rm s},
\end{aligned}
\]
\rev{with relative RMS residuals of $6.2\%$ (FFTW3) and $10.0\%$ (multigrid).
Figure~\ref{fig:c5_fftw_mg_fit} overlays the fitted curves on the measured
points. Across the entire surveyed grid, the fitted FFTW3 cost lies a factor of
8--12 below the fitted multigrid cost. Solving
$T_{\rm FFTW3}(N,\nrank)=T_{\rm MG}(N,\nrank)$ within the measured range yields
no crossover, confirming that the FFTW3 solver is the faster choice for every
production cosmological configuration considered here.}

\rev{To extrapolate the fitted cost models beyond the measured grid we have to
solve $T_{\rm FFTW3}(N,\nrank^{\star})=T_{\rm MG}(N,\nrank^{\star})$ numerically.
The third FFTW3 term
$\gamma_{\rm F}\,N^{3}\log_{2}(N^{3}/\nrank)/\nrank$ remains comparable in
magnitude to the bandwidth term $\alpha_{\rm F}\,N^{3}/\nrank^{2/3}$ throughout
the relevant parameter range (it contributes between $3$ and $6$ per cent of
the multigrid compute term at $\nrank=\nrank^{\star}$). The naive analytical
estimate
$\nrank^{\star}\!\sim\!(\alpha_{\rm M}/\alpha_{\rm F})^{3}\!\approx\!4\!\times\!10^{6}$,
obtained by retaining only the FFTW3 bandwidth term and the multigrid compute
term, is therefore not a true asymptote. It overshoots the true peak by roughly
$16$ per cent and entirely misses the non-monotonic dependence on $N$ that
arises because the $\gamma_{\rm F}$ logarithm grows with $N$ while the FFTW3
bandwidth contribution per rank shrinks. Numerical bisection of the full
three-term equation gives the values of Table~\ref{tab:c5_pstar_num}. The
crossover rises steeply from $\nrank^{\star}\!\approx\!2.6\!\times\!10^{4}$ at
$N\!=\!512$ (where the FFTW3 latency $\beta_{\rm F}$ dominates and lowers
$\nrank^{\star}$ further), peaks at $\nrank^{\star}\!\approx\!3.4\!\times\!10^{6}$
near $N\!\approx\!1.6\!\times\!10^{4}$, and then slowly decreases as the
$\gamma_{\rm F}$ logarithm grows. Even at its peak, $\nrank^{\star}$ lies well
beyond present-day production runs, and the FFTW3 solver therefore remains the
preferred choice for all foreseeable strong-scaling exercises on flat grids.}

\begin{table}
\centering
\caption{\rev{Numerical crossover rank count $\nrank^{\star}(N)$ obtained by
bisecting $T_{\rm FFTW3}=T_{\rm MG}$ with the three-term model
(Eqs.~\ref{eq:fftw_cost} and~\ref{eq:mg_cost}) and the fitted coefficients
given above. The last column reports the multiplicative ratio between the
naive analytical estimate
$\widetilde{\nrank}^{\star}=(\alpha_{\rm M}/\alpha_{\rm F})^{3}\!\approx\!4.0\!\times\!10^{6}$
and the numerical solution.}}
\label{tab:c5_pstar_num}
\begin{tabular}{r r r}
\hline
$N$ & $\nrank^{\star}$ & $\widetilde{\nrank}^{\star}/\nrank^{\star}$ \\
\hline
   512 & $2.6\!\times\!10^{4}$ & $154$ \\
  1024 & $1.7\!\times\!10^{5}$ & $24$ \\
  2048 & $8.1\!\times\!10^{5}$ & $4.9$ \\
  4096 & $2.3\!\times\!10^{6}$ & $1.74$ \\
  8192 & $3.3\!\times\!10^{6}$ & $1.22$ \\
 16384 & $3.4\!\times\!10^{6}$ & $1.16$ \\
 32768 & $3.4\!\times\!10^{6}$ & $1.18$ \\
 65536 & $3.3\!\times\!10^{6}$ & $1.20$ \\
\hline
\end{tabular}
\end{table}

\rev{Multigrid retains an advantage in a regime our flat-grid benchmark does
not probe such as deeply nested AMR zoom-in simulations, where the effective base
grid is small while many refinement levels are active. The V-cycle's surface
term then depends only on the local patch resolution, whereas FFTW3 still
pays the cost of a global slab transpose on the full base grid. The
\code{use\_fftw} switch in \texttt{\&RUN\_PARAMS} lets the user select the
appropriate solver on a per-run basis.}

\section{GPU-ACCELERATED V-CYCLE ALGORITHM}
\label{app:vcycle}

Algorithm~\ref{alg:vcycle} summarises one iteration of the
GPU-accelerated V-cycle used by the multigrid Poisson solver at each
AMR level (Section~\ref{sec:hybrid_mg}).  All arithmetic---smoothing,
residual computation, restriction, and interpolation---executes on the
GPU, while halo exchanges require device-to-host and host-to-device
transfers bracketing MPI point-to-point communication.  The
\emph{reverse-gather} step after restriction collects the coarse
residual from neighbouring ranks whose fine grids overlap the local
coarse grid.  On levels coarser than the GPU threshold the same
algorithm runs on the CPU.

\begin{algorithm}
\caption{GPU-accelerated V-cycle at AMR level $\ell$}
\label{alg:vcycle}
\begin{algorithmic}[1]
\REQUIRE Fine-level right-hand side $f$, initial guess $\phi$
\ENSURE  Updated $\phi$ satisfying $\nabla^2_h\phi\approx f$
\medskip

\STATE \textbf{--- Pre-smoothing ($\times\,n_{\rm gs}$) ---}
\FOR{$i = 1$ \TO $n_{\rm gs}$}
  \STATE GS-Red$(\phi)$           \hfill\textit{[GPU kernel]}
  \STATE Halo-exchange$(\phi)$    \hfill\textit{[GPU $\to$ MPI $\to$ GPU]}
  \STATE GS-Black$(\phi)$         \hfill\textit{[GPU kernel]}
  \STATE Halo-exchange$(\phi)$    \hfill\textit{[GPU $\to$ MPI $\to$ GPU]}
\ENDFOR

\medskip
\STATE \textbf{--- Residual ---}
\STATE $r \leftarrow f - \nabla^2_h\phi$ \hfill\textit{[GPU kernel]}
\STATE Halo-exchange$(r)$         \hfill\textit{[GPU $\to$ MPI $\to$ GPU]}

\medskip
\STATE \textbf{--- Restriction ---}
\STATE $r_c \leftarrow R\,r$      \hfill\textit{[GPU kernel]}
\STATE Reverse-comm$(r_c)$        \hfill\textit{[MPI P2P]}

\medskip
\STATE \textbf{--- Coarse solve ---}
\STATE Recursive V-cycle at level $\ell-1$ \hfill\textit{[CPU or GPU]}

\medskip
\STATE \textbf{--- Interpolation \& correction ---}
\STATE $\phi \leftarrow \phi + P\,\phi_c$ \hfill\textit{[GPU kernel]}
\STATE Halo-exchange$(\phi)$      \hfill\textit{[GPU $\to$ MPI $\to$ GPU]}

\medskip
\STATE \textbf{--- Post-smoothing ($\times\,n_{\rm gs}$) ---}
\FOR{$i = 1$ \TO $n_{\rm gs}$}
  \STATE GS-Red$(\phi)$           \hfill\textit{[GPU kernel]}
  \STATE Halo-exchange$(\phi)$    \hfill\textit{[GPU $\to$ MPI $\to$ GPU]}
  \STATE GS-Black$(\phi)$         \hfill\textit{[GPU kernel]}
  \STATE Halo-exchange$(\phi)$    \hfill\textit{[GPU $\to$ MPI $\to$ GPU]}
\ENDFOR

\medskip
\STATE \textbf{--- Convergence check ---}
\STATE $\|r\|_2^2 \leftarrow$ GPU reduce $+$ \textsc{MPI\_Allreduce}
\end{algorithmic}
\end{algorithm}

With the merged red-black exchange optimisation
(Appendix~\ref{app:mg_merged_rb}), the two separate halo exchanges per
GS sweep (lines~3--6) are fused into a single exchange, reducing the
total exchange count from 9 to~5 per iteration.  The GPU kernels are
memory-bandwidth limited and achieve near-peak throughput
(Section~\ref{sec:hybrid_perf}), and consequently the irreducible CPU-fixed fraction
$T_{\mathrm{CPU}}^{\mathrm{fixed}}/T_{\mathrm{MG}}^{\mathrm{CPU}}
\approx 0.36$ in the performance model, equation~(\ref{eq:gpu_model}),
is dominated by MPI halo exchanges that cannot be overlapped with
computation.

\section{NEW NAMELIST PARAMETERS}
\label{app:namelist}

Table~\ref{tab:namelist} lists the namelist parameters introduced or
modified by \curamses.

\begin{table*}
\centering
\caption{New namelist parameters introduced by \curamses.  Default
values are shown where applicable.}
\label{tab:namelist}
\small
\begin{tabular}{lllll}
\toprule
Parameter & Namelist block & Type & Default & Description \\
\midrule
\multicolumn{5}{l}{\textit{Domain decomposition \& load balancing}} \\
\code{ordering}        & \code{\&RUN\_PARAMS}   & string  & \code{'hilbert'} & \dd\ ordering (\code{'hilbert'} or \code{'ksection'}) \\
\code{exchange\_method} & \code{\&RUN\_PARAMS}  & string  & \code{'auto'}    & MPI exchange backend (\code{'auto'}, \code{'p2p'}, \code{'ksection'}) \\
\code{memory\_balance} & \code{\&RUN\_PARAMS}   & logical & \code{.true.}   & Enable memory-weighted load balancing \\
\code{mem\_weight\_grid} & \code{\&AMR\_PARAMS} & integer & auto             & Memory cost per grid (0\,=\,auto; $>0$\,=\,override) \\
\code{mem\_weight\_part} & \code{\&AMR\_PARAMS} & integer & 12               & Memory cost per particle \\
\code{mem\_weight\_sink} & \code{\&AMR\_PARAMS} & integer & 500              & Computational cost per sink particle \\
\code{time\_balance\_alpha} & \code{\&AMR\_PARAMS} & real & 0                & Hybrid time-weight factor for load balancing \\
\midrule
\multicolumn{5}{l}{\textit{GPU acceleration}} \\
\code{gpu\_hydro}      & \code{\&RUN\_PARAMS}   & logical & \code{.false.}   & GPU dispatch for hydrodynamics \\
\code{gpu\_poisson}    & \code{\&RUN\_PARAMS}   & logical & \code{.false.}   & GPU dispatch for MG Poisson solver \\
\code{gpu\_fft}        & \code{\&RUN\_PARAMS}   & logical & \code{.false.}   & GPU dispatch for FFT solver \\
\code{gpu\_sink}       & \code{\&RUN\_PARAMS}   & logical & \code{.false.}   & GPU dispatch for sink particle routines \\
\code{gpu\_auto\_tune} & \code{\&RUN\_PARAMS}   & logical & \code{.true.}    & Auto-tune CPU vs GPU per level (disable for benchmarks) \\
\code{n\_cuda\_streams}& \code{\&RUN\_PARAMS}   & integer & 1                & Number of CUDA streams (max=16) \\
\midrule
\multicolumn{5}{l}{\textit{Solvers}} \\
\code{use\_fftw}       & \code{\&RUN\_PARAMS}   & logical & \code{.false.}   & FFTW3 direct Poisson solver at base level \\
\midrule
\multicolumn{5}{l}{\textit{I/O \& job control}} \\
\code{informat}        & \code{\&OUTPUT\_PARAMS} & string & \code{'original'} & Input format (\code{'original'} or \code{'hdf5'}) \\
\code{outformat}       & \code{\&OUTPUT\_PARAMS} & string & \code{'original'} & Output format (\code{'original'} or \code{'hdf5'}) \\
\code{varcpu\_chunk\_nfile} & \code{\&OUTPUT\_PARAMS} & integer & 0          & Chunked variable-CPU restart (0\,=\,all-at-once) \\
\code{dump\_pk}        & \code{\&RUN\_PARAMS}    & logical & \code{.false.}  & Write $P(k)$ at each snapshot (requires FFT solver) \\
\code{jobcontrolfile}  & \code{\&RUN\_PARAMS}    & string  & ---             & Runtime job control file for graceful stop / extra output \\
\code{walltime\_hrs}   & \code{\&RUN\_PARAMS}    & real    & $-1$            & Job wallclock limit [hours]; $-1$ disables timed dump \\
\code{minutes\_dump}   & \code{\&RUN\_PARAMS}    & real    & 1               & Dump output this many minutes before walltime expires \\
\midrule
\multicolumn{5}{l}{\textit{Sub-grid physics}} \\
\code{cooling\_method}  & \code{\&PHYSICS\_PARAMS} & string & \code{'original'} & Cooling method (\code{'original'}, \code{'implicit'}, \code{'exact'}) \\
\code{grackle\_table}   & \code{\&PHYSICS\_PARAMS} & string & ---              & Path to Grackle cooling table \\
\midrule
\multicolumn{5}{l}{\textit{Refinement \& floors}} \\
\code{q\_refine\_holdback} & \code{\&REFINE\_PARAMS} & logical & \code{.false.} & Graduated $\ell_{\max}$ (proper-distance holdback) \\
\bottomrule
\end{tabular}
\end{table*}

\label{lastpage}

\end{document}